\documentclass[12pt]{article}
\usepackage{amsmath}
\usepackage[left=1in,right=1in,top=1in,bottom=1in]{geometry}
\usepackage{bm}
\usepackage{amsfonts}
\usepackage{amssymb}
\usepackage{graphicx}
\usepackage{xcolor}
\usepackage{array}
\usepackage{soul}
\usepackage{algorithm}
\usepackage{algpseudocode}
\usepackage{appendix}
\usepackage{booktabs}
\usepackage{comment}
\usepackage{setspace}
\usepackage[natbibapa]{apacite}
\usepackage{multibib}
\newcites{S}{References}

\DeclareMathOperator*{\argmax}{arg max}
\DeclareMathOperator*{\argmin}{arg min}

\ifx\theorem\undefined
\newtheorem{theorem}{Theorem}
\fi
\ifx\theorem\undefined
\newtheorem{proposition}{Proposition}
\fi
\ifx\example\undefined
\newtheorem{example}[theorem]{Example}
\fi
\ifx\property\undefined
\newtheorem{property}{Property}
\fi
\ifx\lemma\undefined
\newtheorem{lemma}[theorem]{Lemma}
\fi
\ifx\proposition\undefined
\newtheorem{proposition}[theorem]{Proposition}
\fi
\ifx\remark\undefined
\newtheorem{remark}[theorem]{Remark}
\fi
\ifx\corollary\undefined
\newtheorem{corollary}[theorem]{Corollary}
\fi
\ifx\definition\undefined
\newtheorem{definition}[theorem]{Definition}
\fi
\ifx\conjecture\undefined
\newtheorem{conjecture}[theorem]{Conjecture}
\fi
\ifx\fact\undefined
\newtheorem{fact}[theorem]{Fact}
\fi
\ifx\claim\undefined
\newtheorem{claim}[theorem]{Claim}
\fi
\ifx\assumption\undefined
\newtheorem{assumption}[theorem]{Assumption}
\fi
\ifx\cond\undefined
\newtheorem{cond}[theorem]{Condition}
\fi
\ifx\problem\undefined
\newtheorem{problem}[theorem]{Problem}
\fi
\ifx\BlackBox\undefined
\newcommand{\BlackBox}{\rule{1.5ex}{1.5ex}}
\fi
\ifx\proof\undefined
\newenvironment{proof}{\par\noindent{\bf Proof\ }}{\hfill\BlackBox\\}
\fi
\def\bm{\boldsymbol}

\newcommand{\bx}{\mathbf{x}}

\newcommand{\bw}{\mathbf{w}}
\newcommand{\bU}{\mathbf{U}}
\newcommand{\bv}{\mathbf{v}}
\newcommand{\bV}{\mathbf{V}}

\newcommand{\bmu}{\boldsymbol{\mu}}

\algrenewcommand\algorithmicrequire{\textbf{Input:}}
\algrenewcommand\algorithmicensure{\textbf{Output:}}

\begin{document}
	\doublespacing
	
	\title{\textbf{Statistical Inference for Fuzzy Clustering}}
	\date{}
	\author{Qiuyi Wu\textsuperscript{1}\thanks{Department of Biostatistics \& Bioinformatics, Duke University}, ~ Zihan Zhu\textsuperscript{1}\thanks{Department of Statistics \& Data Science, University of Pennsylvania}, ~ and ~ Anru R. Zhang\textsuperscript{2}\thanks{Department of Biostatistics \& Bioinformatics and Department of Computer Science, Duke University}}
	\maketitle
	
	\def\thefootnote{}\footnotetext{$^1$Equal contribution. $^2$Corresponding author. E-mail: \texttt{anru.zhang@duke.edu}}\def\thefootnote{\arabic{footnote}}
	
	\begin{abstract}
		Clustering is a central tool in biomedical research for discovering heterogeneous patient subpopulations, where group boundaries are often diffuse rather than sharply separated. While soft clustering methods such as fuzzy $c$-means (FCM) allow mixed memberships, principled statistical inference for fuzzy clustering remains limited. 
		
		We propose a weighted fuzzy $c$-means (WFCM) framework for settings with cluster-size imbalance. Cluster-specific weights $\bw$ re-balance the classic FCM criterion to adjust the relative influence of clusters of unequal size, and we construct a normalized working density from the weighted objective with scale parameter $\sigma$ and fuzziness parameter $m$. Estimation is performed via FCM-style alternating initialization followed by numerical likelihood refinement; the intractable normalizing constant is approximated by importance sampling using a data-adaptive Gaussian-mixture as the proposal distribution. We further develop bootstrap-based confidence intervals.
		
		We establish consistency and asymptotic normality of the maximum likelihood estimator, validate the method in simulations, and illustrate it on single-cell RNA-seq, Alzheimer’s Disease Neuroimaging Initiative (ADNI), and TCGA-BRCA data. These applications illustrate uncertainty quantification and interpretable soft memberships, ranging from imbalanced cell populations to graded compatibility with AD-like and non-AD-like prototypes.
		
	\end{abstract}
	
	\begin{sloppypar}
		\section{Introduction}\label{sec:intro}
		
		Clustering is a fundamental task in unsupervised learning, with widespread applications in bioinformatics, social science, and image analysis. In many biomedical and clinical domains, subgroup boundaries are rarely clear-cut. For instance, sepsis can exhibit overlapping phenotypes in which patients display mixed inflammatory and hemodynamic responses \citep{jiang2024soft}. Similarly, neuroimaging studies often suggest disease processes that evolve along a continuum rather than a strict dichotomy, as in Alzheimer's disease and mild cognitive impairment. These examples motivate clustering methods that can explicitly represent uncertainty and gradual transitions between subgroups.
		
		Traditional clustering algorithms such as $k$-means \citep{lloyd1982least} and hierarchical clustering \citep{ward1963hierarchical,sokal1963principles} assign each observation to a single cluster, producing hard partitions of the data. While effective in many contexts, hard clustering can fail to reflect uncertainty and ambiguity in complex datasets. Soft clustering, in contrast, assigns each observation a vector of membership weights that quantify partial association with clusters. Among distance-based soft clustering methods, fuzzy $c$-means (FCM) \citep{bezdek2013pattern} has been particularly influential, jointly estimating cluster centers and membership weights. Despite widespread empirical use, fuzzy clustering is often treated primarily as an optimization procedure, and formal statistical inference and uncertainty quantification remain underdeveloped.
		
		\subsection{Our Contributions}
		
		This paper addresses this gap by developing a framework for statistical inference in soft clustering based on a weighted extension of fuzzy $c$-means. Our main contributions are as follows.
		
		We propose a \emph{weighted} fuzzy $c$-means (WFCM) formulation that adjusts to settings with severe cluster-size imbalance. By introducing cluster-specific weights, the objective re-balances the classic FCM criterion to accommodate unequal cluster contributions. We embed this weighted objective in a normalized working density with scale parameter $\sigma$ and fuzziness parameter $m$, under which fuzzy memberships admit closed-form, weighted FCM-style updates. The resulting log-likelihood separates into an intractable normalizing constant and the WFCM loss, and likelihood estimation accounts for both terms.
		
		To estimate the model parameters under the induced likelihood, we use FCM-style alternating initialization with closed-form membership and centroid updates. The intractable normalizing constant is approximated via importance sampling with a data-adaptive Gaussian-mixture as the proposal distribution, and we use a numerical likelihood refinement step to further reduce the negative log-likelihood. Building on the fitted model, we provide practical inferential tools, including a center-equality likelihood-ratio statistic requiring null-specific calibration and bootstrap-based confidence regions with label alignment. We also adapt the Xie--Beni index to provide a practical criterion for selecting the number of clusters under the proposed formulation.
		
		We establish asymptotic guarantees for the proposed likelihood formulation. In particular, we prove strong consistency of the MLE up to label permutation (Theorem~\ref{thm:mle_consistency}) and asymptotic normality under interior, separated-center conditions (Theorem~\ref{thm:mle-an}), with covariance characterized by the inverse Fisher information in free coordinates. Theorem~\ref{thm:approx-transfer} gives sufficient importance-sampling, grid, and optimization accuracy conditions for transferring these guarantees to numerical fits.
		
		Simulation studies examine the finite-sample behavior predicted by the theory and compare bootstrap and Fisher-information-based uncertainty quantification. We further illustrate the framework on three biomedical datasets. In scRNA-seq PBMC data (memory T cells, B cells, monocytes), the method yields tight uncertainty regions even when clusters are biologically distinct under class imbalance. In ADNI (Alzheimer's Disease Neuroimaging Initiative), soft memberships describe relative proximity to AD-like and non-AD-like prototypes. In TCGA-BRCA breast cancer data (Appendix~\ref{app:brca}), WFCM increases recall of the overlapping LumB subtype at the cost of lower HER2-enriched recall and lower overall agreement.
		
		\subsection{Organization}\label{sec:organization}
		
		Section~\ref{sec:notation-preliminary} reviews classic fuzzy $c$-means, Section~\ref{sec:method} introduces the weighted formulation and its working density, Section~\ref{sec:computation} presents the fitting algorithm, inference procedures, and selection of $k$, Section~\ref{sec:theory} gives the asymptotic theory, Section~\ref{sec:real-data-analysis} analyzes single-cell and ADNI data, and Section~\ref{sec:discussion} concludes. The supplementary materials contain the proofs, properties of the WFCM density, simulation studies, related literature, and a third application to TCGA-BRCA breast cancer subtypes (Appendix~\ref{app:brca}).
		
		\section{Notation and Preliminaries}\label{sec:notation-preliminary}
		
		We use boldface letters, e.g., $\mathbf{x}, \mathbf{y}, \mathbf{z}$, to denote vectors, 
		and regular letters, e.g., $x, y, z$, to denote scalars. $\bx_1,\ldots, \bx_n$ are $n$ samples, and $x_{i1},\ldots, x_{id}$ are $d$ entries of $\bx_i$.
		For any vector $\mathbf{x} \in \mathbb{R}^d$, we write 
		$\mathbf{x} = (x_{1}, \ldots, x_{d})$ for its components and define its $\ell_{2}$-norm as
		$\|\bx\| = \sqrt{\sum_{j=1}^{d} x_j^{2}}.$
		For any non-negative integer $n$, we denote $[n] = \{1, \ldots, n\}$.
		
		Next, we describe the classic fuzzy $c$-means (FCM) method as the benchmark of our work. The FCM algorithm attempts to partition a finite collection of $n$ $d$-dimensional observations $\{\bx_{1},\ldots,\bx_{n}\} \subseteq \mathbb{R}^d$ into a collection of $k$ fuzzy clusters with respect to a given criterion. Given a finite set of data, the algorithm returns an estimate $\bV=(\mathbf{v}_{1},\ldots,\mathbf{v}_{k})$ of the cluster centroids and a partition matrix $\bU = (u_{ij})\in [0,1]^{n\times k}, i = 1,\ldots,n; j = 1,\ldots, k$, where the pairwise distances $d_{ij} = \|\bx_i - \bv_j\|, i\in[n], j\in [k]$ and the membership coefficient of point $\bx_i$ to cluster $j$ is 
		\begin{equation}\label{eq:FCM-member}
			\begin{split}
				\text{if $m>1$,~~} & u_{ij} = \left\{\sum_{l=1}^k \left(\frac{d_{ij}}{d_{il}}\right)^{2/(m-1)}\right\}^{-1};\\
				\text{if $m=1$,~~} & u_{ij} = \left\{\begin{array}{ll}
					\frac{1}{|A|}, & d_{ij} = \min_{l'\in [k]} d_{il'},  A = \{l: d_{il} = \min_{l'\in [k]}d_{il'}\};\\
					0 & d_{ij} \neq \min_{l'\in [k]}d_{il'}.
				\end{array}\right.
			\end{split}
		\end{equation}
		Here, $m\ge 1$ represents the ``fuzziness" or the degree of softness: when $m=1$, the method corresponds to the hard clustering and the $k$-means algorithm. The membership and cluster centers are chosen to minimize the FCM objective function, which can be rewritten as
		\begin{equation}\label{eq:J-FCM}
			\begin{split}
				\mathcal{J}^{FCM}_m(\bV)  = &  \sum_{i=1}^n \sum_{j=1}^k u_{ij}^m \|\bx_i - \bv_j\|^2\\
				= & \sum_{i=1}^n \sum_{j=1}^k \frac{d_{ij}^2}{\left[ \sum_{l=1}^k \left(d_{ij}/d_{il}\right)^{\frac{2}{m-1}}\right]^m } = \sum_{i=1}^n \frac{1 }{\left [  \sum_{j=1}^k (d_{ij}^2)^{-\frac{1}{m-1}}\right]^{m-1}}.
			\end{split}
		\end{equation}
		
		Relations among generalized means \citep{bullen2013handbook} give three special cases. As $m\to1$, $\mathcal{J}^{FCM}_m\to\sum_{i=1}^n\min_{j} d_{ij}^2$, the $k$-means objective, so FCM reduces to $k$-means. As $m\to\infty$, $k^{m-1}\mathcal{J}^{FCM}_m=\sum_{i=1}^n[k^{-1}\sum_{j=1}^k d_{ij}^{-2/(m-1)}]^{-(m-1)}\to\sum_{i=1}^n\prod_{j=1}^k d_{ij}^{2/k}$, a sum of geometric means of the squared distances. At $m=2$, $\mathcal{J}^{FCM}_2=\sum_{i=1}^n(\sum_{j=1}^k d_{ij}^{-2})^{-1}$, so each observation contributes the harmonic mean of its squared distances divided by $k$.
		The procedure of classic FCM follows an iterative optimization scheme \citep{bezdek2013pattern}. It begins by randomly initializing the cluster centers $\bV$. Given the current cluster centers, the membership coefficients $u_{ij}$ are updated using the distance-based formula described above. With the updated memberships, each cluster center $\bv_j$ is recalculated as the weighted average of all data points $\bv_j = \frac{\sum_i u_{ij}^m \bx_i}{\sum_i u_{ij}^m}$. These two steps of updating memberships and updating cluster centers are repeated alternately until the objective function or the membership matrix converges. This alternating optimization ensures that the algorithm monotonically decreases the objective at each iteration and typically converges to a local minimum.
		
		The classic fuzzy $c$-means has many limitations. First, the method does not immediately yield a statistical inference or uncertainty quantification, which can hinder rigorous interpretation of clustering results. Second, it treats all clusters equally in the objective, regardless of their relative sizes or importance, whereas in real datasets some clusters may contain substantially more points than others (see Figure \ref{fig:imbalance}). Third, FCM does not yield a generative model like Gaussian mixture models. It does not define a likelihood function or allow the generation of synthetic samples from the learned clusters. These limitations motivate extensions of FCM, such as weighted fuzzy $c$-means, which incorporate cluster-specific contributions and probabilistic interpretations, offering a more flexible modeling framework and enabling statistical inference.
		
		\section{A Statistical Framework for Weighted Fuzzy $C$-Means}\label{sec:method}
		
		We develop statistical inference for soft clustering in settings where clusters may have markedly different sizes. Let
		\(\bx_1,\ldots,\bx_n \in \mathbb{R}^d\) be observed data. Our goal is to assign them to a small number of (potentially overlapping) latent clusters while retaining uncertainty information through fuzzy memberships.
		
		\paragraph{Weighted fuzzy $c$-means.}
		
\begin{figure}[ht!]
			\centering
			\includegraphics[width=1\linewidth]{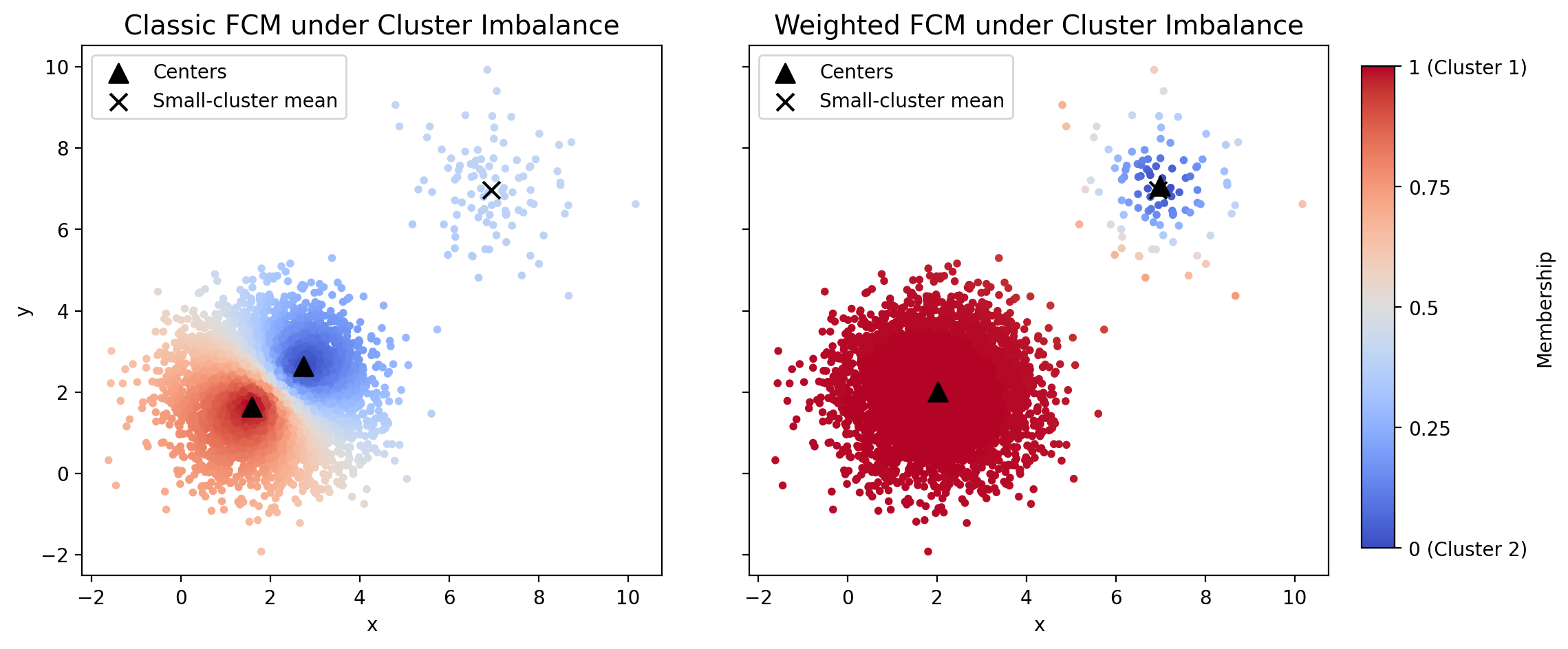}
			\caption{Classic and weighted fuzzy $c$-means under cluster imbalance (5,000 and 100 points; $m=2$). Both methods use the same five starts and keep their best fit. Colors show membership in the cluster whose center is nearest the large-cluster mean. Classic FCM places both centers inside the large cluster, leaving the small cluster without a center. WFCM, with cluster weights and scale estimated by maximum likelihood (Algorithm~\ref{alg:emfuzzy}), places a center 0.13 from the small-cluster mean. Its smaller fitted weight for the large cluster (0.07 versus 0.93) enlarges that cluster's high-membership region, so small-cluster points receive moderate memberships (mean 0.71).}
			\label{fig:imbalance}
		\end{figure}

		A well-known limitation of standard fuzzy $c$-means (FCM) is that all clusters are treated as equally important in the objective. When cluster sizes are imbalanced, large clusters dominate the optimization and can draw all centroids into dense regions, leaving smaller clusters without a prototype (Figure \ref{fig:imbalance}). To mitigate this, we introduce cluster-specific weights and obtain a \emph{weighted} FCM objective. 
		
		Let $\bV = (\bv_1,\ldots,\bv_k)$ denote the cluster centroids. Let
		\[
		\bw = (w_1,\ldots,w_k) \in \mathcal{S}_k := \left\{(w_1,\ldots,w_k): w_j > 0 \ \forall j,\ \sum_{j=1}^k w_j = 1\right\}
		\]
		be a weight vector over clusters, and write $d_{ij} = \|\bx_i - \bv_j\|$. $w_j$ acts as a cluster-specific distance penalty: a smaller $w_j$ enlarges the region in which cluster $j$ receives high membership, so the weights adjust the relative influence of clusters of unequal size. For the fuzziness parameter $m>1$, we define the WFCM loss as
		\begin{equation}\label{eq:WFCM}
			\begin{split}
				\mathcal{J}_{m}^{\mathrm{WFCM}}(\bV,\bw)
				= \sum_{i=1}^n \Bigl[ \sum_{j=1}^k \bigl( w_j d_{ij}^2 \bigr)^{-\frac{1}{m-1}} \Bigr]^{-m+1}.
			\end{split}
		\end{equation}
		The weights $\bw$ directly re-balance the contribution of each cluster to the data-fitting term.
		
		\paragraph{A density induced by WFCM.}
		
		To carry out statistical inference and uncertainty quantification, we embed the WFCM loss into a probabilistic model. Define a family of densities on $\mathbb{R}^d$ by
		\begin{equation}\label{eq:WFCM-density}
			f(\bx; \bV, \bw, m, \sigma)
			= C_{\bV, \bw, m, \sigma}
			\exp\!\left(
			- \frac{1}{\sigma^2 \bigl[ \sum_{j=1}^k (w_j \|\bx - \bv_j\|^2)^{-\frac{1}{m-1}} \bigr]^{m-1} }
			\right), \quad \bx \in \mathbb{R}^d,
		\end{equation}
		where $\bV = (\bv_1,\ldots,\bv_k)$ are the centroids, $\bw \in \mathcal{S}_k$ are the cluster weights, $m>1$ is the fuzziness parameter, and $\sigma>0$ is a scale parameter. The normalizing constant
		\begin{equation}\label{eq:normalizing-constant}
			C_{\bV, \bw, m, \sigma}
			=
			\left\{
			\int_{\mathbb{R}^d}
			\exp\!\left(
			- \frac{1}{\sigma^2 \bigl[ \sum_{j=1}^k (w_j \|\bx - \bv_j\|^2)^{-\frac{1}{m-1}} \bigr]^{m-1} }
			\right) d\bx
			\right\}^{-1}
		\end{equation}
		ensures that $f$ integrates to one.
		
		\begingroup
		\begin{remark}[From a clustering loss to a working density]
			\label{rem:density-derivation}
			The distance-based WFCM criterion does not require a sampling model. Equation~\eqref{eq:WFCM-density} adds a working density for likelihood-based inference. Appendix~\ref{app:density-derivation} establishes the algebraic identity
			\begin{equation}
				\min_{U}\sum_{i=1}^n\sum_{j=1}^k u_{ij}^m\,w_j\,\|x_i-v_j\|^2
				= \sum_{i=1}^n\left[\sum_{j=1}^k
				\bigl(w_j\,\|x_i-v_j\|^2\bigr)^{-1/(m-1)}
				\right]^{-(m-1)}
				= J_m^{\mathrm{WFCM}}(V,w),
				\label{eq:loss-per-obs-remark}
			\end{equation}
			where the minimum is over membership matrices whose rows lie in the probability simplex. Exponentiating the negative loss contribution and normalizing it is a modeling choice. With the convention in~\eqref{eq:normalizing-constant},
			\[
			-\ell_n(\theta) = -n\log C(\theta) + \sigma^{-2}\,J_m^{\mathrm{WFCM}}(V,w).
			\]
			Since $C(\theta)$ generally depends on the fitted parameters, minimizing the WFCM loss alone is not maximum likelihood estimation.
		\end{remark}
		\endgroup
		
		This model has a direct clustering interpretation: the term $\bigl[ \sum_{j=1}^k (w_j \|\bx - \bv_j\|^2)^{-\frac{1}{m-1}} \bigr]^{-(m-1)}$
		plays the same role as the pointwise contribution in \eqref{eq:WFCM}, so that high density regions coincide with locations close to one (or more) centroids, modulated by the cluster weights.
		
		\begin{figure}[ht!]
			\centering
			\includegraphics[width=0.95\linewidth]{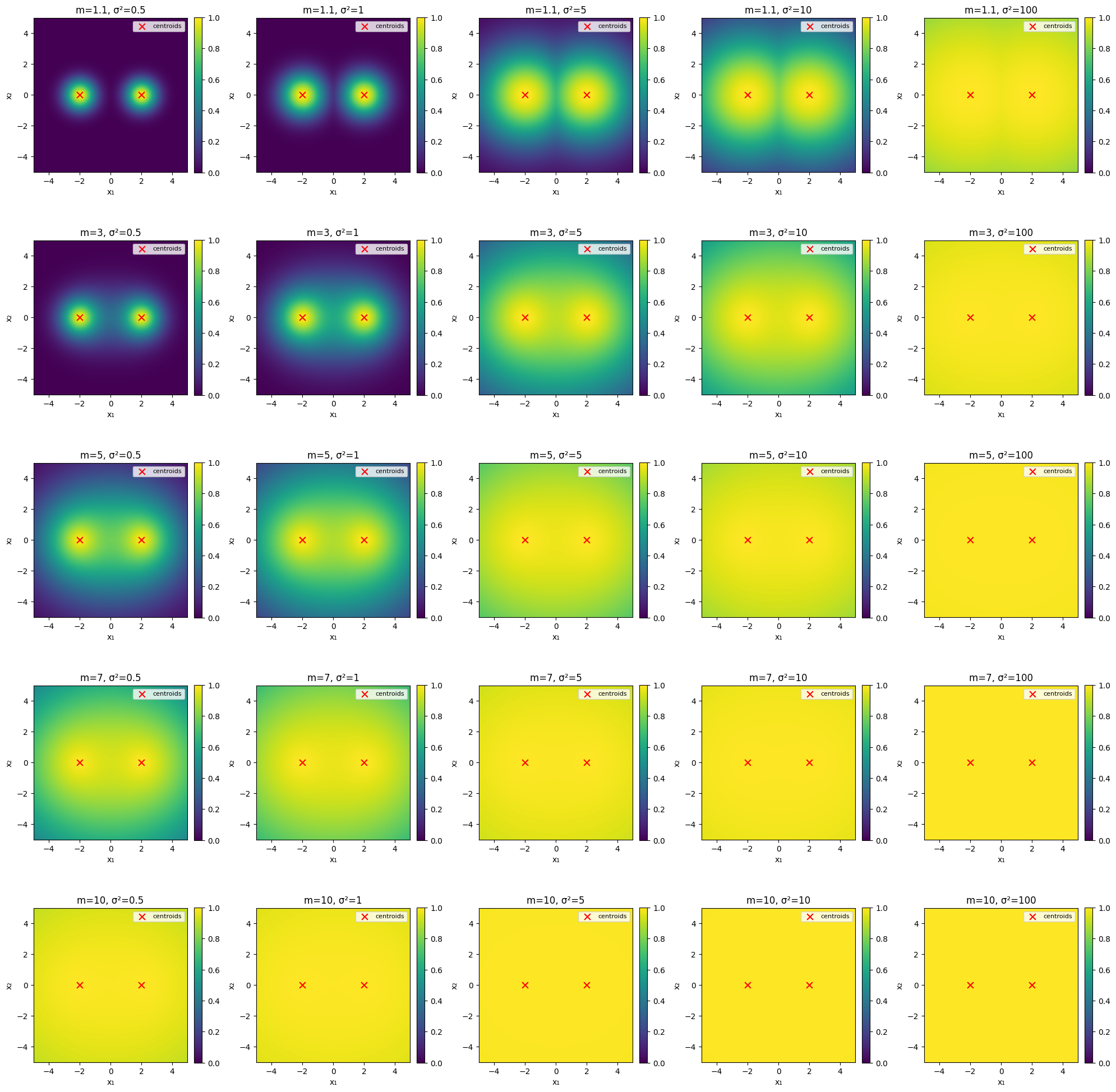}
			\caption{Induced unnormalized density kernel in \eqref{eq:WFCM-density} for different values of $\sigma^2$ and $m$ in two dimensions with two centroids (red crosses) and equal weights $w_1=w_2=1/2$. Smaller $m$ produces sharper, nearly hard-clustering modes; larger $m$ yields smoother, overlapping regions. Increasing $\sigma^2$ broadens the density and flattens the boundaries.}%
			\label{fig:density}
		\end{figure}
		Figure~\ref{fig:density} illustrates the kernel, and Appendix~\ref{densityproof} gives the limits of the WFCM density as $m\to1^+$ and $m\to\infty$ and its form at $m=2$.
		
		\paragraph{Memberships, likelihood, and estimation}
		
		Minimizing the WFCM loss over each row of the membership matrix (Remark~\ref{rem:density-derivation} and Appendix~\ref{app:density-derivation}) gives the fuzzy membership of observation $\bx_i$ in cluster $j$ in closed form: $\bU = (u_{ij})_{i\in [n], j\in[k]}$,
		\begin{equation}\label{eq:membership}
			u_{ij}
			= \frac{ w_j^{-\frac{1}{m-1}} \|\bx_i - \bv_j\|^{-\frac{2}{m-1}} }
			{ \displaystyle \sum_{l=1}^k
				w_l^{-\frac{1}{m-1}} \|\bx_i - \bv_l\|^{-\frac{2}{m-1}} }
			= \frac{1}{\displaystyle \sum_{l=1}^k
				\left( \frac{ w_j \|\bx_i - \bv_j\|^2 }{ w_l \|\bx_i - \bv_l\|^2 } \right)^{\frac{1}{m-1}} } = \frac{ \bigl( w_j d_{ij}^2 \bigr)^{-\frac{1}{m-1}} }{ \displaystyle \sum_{l=1}^k \bigl( w_l d_{il}^2 \bigr)^{-\frac{1}{m-1}} }.
		\end{equation}
		Thus, the weights $\bw$ directly re-balance the contribution of each cluster to the data-fitting term, yielding the usual FCM membership formula \eqref{eq:FCM-member} but now explicitly weighted by $\bw$.
		
		Assume $\bx_1,\ldots,\bx_n$ are independent draws from $f(\cdot; \bV, \bw, m, \sigma)$. The joint likelihood is $L(\bV, \bw, m, \sigma)
		= \prod_{i=1}^n f(\bx_i; \bV, \bw, m, \sigma)$ and the log-likelihood can be written as
		\begin{equation}
			\ell(\bV, \bw, m, \sigma)
			= \sum_{i=1}^n \log f(\bx_i; \bV, \bw, m, \sigma)
			= n \log C_{\bV, \bw, m, \sigma}
			- \frac{1}{\sigma^2} \, \mathcal{J}_{m}^{\mathrm{WFCM}}(\bV,\bw).
			\label{eq:loglik}
		\end{equation}
		The WFCM loss is the data-fitting term in this likelihood; the parameter-dependent normalizing constant must also be included in estimation.
		
		For the asymptotic analysis, we define the exact global MLE on the compact regular parameter set $\Theta_\delta$ specified in \eqref{eq:regular-parameter-set}:
		\begingroup
		\begin{equation}\label{eq:MLE}
			(\hat\bV, \hat\bw, \hat m, \hat\sigma)
			\in \argmax_{(\sigma,\bw,\bV,m)\in\Theta_\delta}
			\ \ell(\bV, \bw, m, \sigma),
		\end{equation}
		\endgroup
		with $\ell$ given in \eqref{eq:loglik}. The fixed bounds and center-separation constraint defining $\Theta_\delta$ also apply in Theorems~\ref{thm:mle_consistency}--\ref{thm:approx-transfer}. Direct optimization, however, is nontrivial because
		(i) the memberships \eqref{eq:membership} depend on \((\bV,\bw,m )\) in a nonlinear way, and
		(ii) the normalizing constant $C_{\bV, \bw, m, \sigma}$ in \eqref{eq:normalizing-constant} also depends on the parameters. To address these coupled dependencies, we adopt FCM-style alternating initialization followed by numerical likelihood refinement, detailed in Section \ref{sec:computation}.
		
		\begin{remark}[WFCM versus Gaussian mixture modeling]
			\label{rem:wfcm-vs-gmm}
			Weighted fuzzy $c$-means (WFCM) and Gaussian mixture models (GMMs) both produce soft cluster assignments but rest on different assumptions. A GMM models the data by the density $\sum_{j=1}^k\pi_j\mathcal{N}(\bx\mid\bmu_j,\Sigma_j)$, and its responsibilities are posterior component probabilities that depend on the component volumes $|\Sigma_j|$ and mixing proportions $\pi_j$. WFCM memberships are proximity-based, $u_{ij}\propto (w_j\|\bx_i-\bv_j\|^2)^{-1/(m-1)}$, although our likelihood formulation imposes a working density. Its Euclidean distances and common scale favor approximately isotropic local geometry, and scalar cluster weights cannot represent cluster-specific orientations. WFCM has $\mathcal{O}(kd)$ parameters, whereas a full-covariance GMM has $\mathcal{O}(kd^2)$ covariance parameters and can accommodate anisotropy.
			
		\end{remark}
		
		\section{Weighted Fuzzy $C$-Means: Algorithm and Statistical Inference}\label{sec:computation}
		
		\subsection{Blockwise Alternating Fuzzy Clustering Algorithm}\label{sec:EM-like}
		
		The procedure estimates the parameters $\bV = (\bv_1, \ldots, \bv_k)$, $\bw = (w_1,\ldots,w_k)$, the scale parameter $\sigma$, and the fuzziness parameter $m$. For each grid value of $m$, membership and centroid updates alternate with numerical scale and weight updates to initialize the subsequent full-likelihood refinement.
		Unlike conventional weighted fuzzy clustering, the proposed likelihood formulation involves a parameter-dependent normalizing constant $C_{\bV, \bw, m, \sigma}$ in \eqref{eq:normalizing-constant}, which must be accounted for in likelihood-based estimation. We approximate this constant using importance sampling with a data-adaptive Gaussian mixture as the proposal distribution. Although importance sampling itself is well established, its approximation error must be controlled to ensure that numerical likelihood estimation retains the statistical properties of the exact MLE. Theorem~\ref{thm:approx-transfer} provides sufficient conditions for this purpose.
		
		For fixed $(\bV,\bw)$ and $m>1$, minimizing the one-point objective $\sum_{j=1}^k w_ju_{ij}^m\|\bx_i-\bv_j\|^2$ over the simplex $\sum_j u_{ij}=1$ (a Lagrange-multiplier calculation) gives the membership update~\eqref{eq:membership}. With memberships fixed, minimizing the WFCM loss over $\bv_j$ gives the centroid update
		\begin{equation}
			\mathbf{v}_j = \frac{\sum_i u_{ij}^m\mathbf{x}_i}{\sum_i  u_{ij}^m},
			\label{eq:centroid-update}
		\end{equation}
		in which $w_j$ cancels. The scale and weights $(\sigma,\bw)$ are then updated by minimizing the negative log-likelihood $\mathrm{NLL}=-\ell$ in~\eqref{eq:loglik} with the L-BFGS algorithm \citep{liu1989limited}, using the importance-sampling approximation of $C_{\bV,\bw,m,\sigma}$ described below. Optimizing the fuzziness parameter $m$ jointly with $(\bV, \bw, \sigma)$ is difficult because the objective is neither smooth nor convex in $m$, so we estimate $m$ by a grid search over a prespecified range, selecting the value that minimizes the negative log-likelihood. Algorithm~\ref{alg:emfuzzy} summarizes the procedure.
		
		\paragraph{Initialization.}
		Numerical fits depend on initialization; Appendix~\ref{app:approx-mle} lists the starting values, numbers of starts, and parameter bounds used in each analysis.
		
		\paragraph{Importance Sampling for the Normalizing Constant.} The constant $C_{\bV,\bw, m, \sigma}$ has no closed form. Given a proposal density $q$ fitted to the data and draws $\mathbf{x}_1,\ldots,\mathbf{x}_M$ from $q$, we use
		\[
		C^{-1}_{\bV,\bw, m, \sigma} = \int \tilde{f}(\mathbf{x};\theta)\,\mathrm{d}\mathbf{x}
		\approx
		\frac{1}{M} \sum_{r=1}^{M}
		\frac{\tilde{f}(\mathbf{x}_r;\theta)}{q(\mathbf{x}_r)},
		\]
		where $\tilde f(\cdot;\theta)$ is the unnormalized kernel in~\eqref{eq:WFCM-density}. Lemma~\ref{lem:is-defensive} shows that the uniform importance-sampling conditions of Theorem~\ref{thm:approx-transfer} hold when $q$ mixes a Gaussian mixture model (GMM) fitted to the data with a Student-$t$ density at a fixed positive weight. The benchmark in Section~\ref{sec:revision-benchmark} uses such a defensive proposal, and Section~\ref{sec:revision-computation} compares defensive and GMM-only proposals.
		
		In the real-data analyses (Section~\ref{sec:real-data-analysis} and Appendix~\ref{app:brca}), $q$ is a defensive mixture of a fitted GMM, isotropic Gaussian components at the fitted means, and a heavy-tailed Student-$t$ component with a fixed weight, as in Lemma~\ref{lem:is-defensive}; Appendix~\ref{app:approx-mle} gives its exact form and the Monte Carlo sizes. Following Theorem~\ref{thm:approx-transfer}, the proposal and the integration cloud are held fixed throughout optimization, including across candidate values of $m$, and are rebuilt for each bootstrap refit.
		
		\begin{algorithm}[htbp]
			\caption{Blockwise Alternating Fuzzy Clustering Algorithm}
			\label{alg:emfuzzy}
			\begin{algorithmic}[1]
				\Require Observed data $\{\mathbf{x}_i\}_{i=1}^n$, number of clusters $k$, initial parameters $\sigma^{(0)}, \bm{\bV}^{(0)}=(\mathbf{v}_1^{(0)},\ldots,\mathbf{v}_k^{(0)}), \mathbf{w}^{(0)}$, tolerances $\varepsilon_\theta,\varepsilon_{\mathrm{NLL}}$, $m$-grid $\mathcal{M}=\{m_1,m_2,\ldots,m_T\}$, max iters of likelihood refinement $K_{\text{post}}$
				
				\State\textbf{For} $m\in\mathcal{M}$:
				\State \hspace{\algorithmicindent}\textbf{Initialize:} Set $t\gets 0$ and $\theta^{(0,m)}\gets(\sigma^{(0,m)},\bV^{(0,m)},\mathbf{w}^{(0,m)})$
				\State \hspace{\algorithmicindent}\textbf{Repeat}
				\State \hspace{\algorithmicindent}\hspace{\algorithmicindent} $t \gets t+1$
				\State \hspace{\algorithmicindent}\hspace{\algorithmicindent} \textbf{Membership update:} compute $u_{ij}^{(t,m)}$ by~\eqref{eq:membership} at $(\bV^{(t-1,m)},\mathbf{w}^{(t-1,m)})$ for all $i\in[n]$, $j\in[k]$.
				\State \hspace{\algorithmicindent}\hspace{\algorithmicindent} \textbf{Centroid update:} compute $\mathbf{v}_j^{(t,m)}$ by~\eqref{eq:centroid-update} with memberships $u_{ij}^{(t,m)}$ for all $j\in[k]$.
				\State \hspace{\algorithmicindent}\hspace{\algorithmicindent} \textbf{Scale/weight update  via L-BFGS}
				\State \hspace{\algorithmicindent}\hspace{\algorithmicindent} Numerically minimize $\mathrm{NLL}(\sigma,\bV^{(t,m)},\mathbf{w};m)$ over $\sigma>0$, $\mathbf{w}\in\mathcal S_k$ to obtain $(\sigma^{(t,m)},\mathbf{w}^{(t,m)})$.
				\State\hspace{\algorithmicindent}\textbf{Until} $\|\theta^{(t,m)}-\theta^{(t-1,m)}\|_2\leq \varepsilon_\theta$ \textbf{ or } $\left|\mathrm{NLL}^{(t-1,m)}-\mathrm{NLL}^{(t,m)}\right|\leq \varepsilon_{\mathrm{NLL}}$
				\State \hspace{\algorithmicindent}\textbf{Numerical likelihood refinement:}
				\State \hspace{\algorithmicindent} Numerically minimize $\mathrm{NLL}(\theta;m)$ starting from $(\sigma^{(t,m)},\bV^{(t,m)},\mathbf{w}^{(t,m)})$, for at most $K_{\text{post}}$ iterations, to obtain $\theta^{(m)}$.
				\State \hspace{\algorithmicindent}\textbf{Store} $\theta^{(m)},\ \mathrm{NLL}\bigl(\theta^{(m)}; m\bigr)$
				\State \textbf{End for}
				
				\Ensure $m^* = \argmin_{m\in\mathcal{M}}\mathrm{NLL}\bigl(\theta^{(m)}; m\bigr)$,\quad $\theta^{*}=(\sigma^{(m^* )},\bV^{(m^* )},\mathbf{w}^{(m^* )})$
			\end{algorithmic}
		\end{algorithm}
		
		For fixed cluster weights $\mathbf{w}$ and fuzziness parameter $m$, the membership and centroid updates alternately minimize the WFCM loss with respect to $\mathbf{U}$ and $\bV$ in closed form. This blockwise organization combines two practical features: (1) computational simplicity, as membership and centroid updates do not require numerical optimization; and (2) likelihood-based estimation, as the scale and cluster weights are updated numerically using the full negative log-likelihood, including the parameter-dependent normalizing constant. The resulting estimates serve as initial values for subsequent joint numerical refinement of $(\bV,\mathbf{w},\sigma)$ under the full negative log-likelihood, with $m$ fixed.
		
		\begin{algorithm}[!t]
			\setstretch{1.15}
			\caption{Parametric Bootstrap Inference for WFCM at Fixed $k$ and $m$}
			\label{alg:inference}
			\begin{algorithmic}[1]
				\Require Fitted parameters $\hat\theta=(\hat\sigma,\hat{\bV},\hat{\bw})$, fixed $k,m$, sample size $n$, bootstrap replicates $B\ge\max\{d+1,k\}$, confidence level $1-\alpha$
				\For{$b=1,\dots,B$}
				\State Draw $X^{(b)}=\{\bx_i^{(b)}\}_{i=1}^n$ independently from $f(\cdot;\hat\theta,m)$.
				\State Refit by Algorithm~\ref{alg:emfuzzy} with $k,m$ fixed, initialized at $\hat\theta$ and rebuilding the proposal and importance-sampling cloud, to obtain $\hat\theta^{(b)}$.
				\State Find the permutation that matches the bootstrap centers to the observed centers:
				\[
				\pi^{(b)}\in\argmin_{\pi\in\mathfrak S_k}\sum_{a=1}^k
				\bigl\|\hat{\bv}_{\pi(a)}^{(b)}-\hat{\bv}_a\bigr\|^2.
				\]
				\State Apply $\pi^{(b)}$ to both $\hat{\bV}^{(b)}$ and $\hat{\bw}^{(b)}$.
				\EndFor
				\State For each scalar component $\theta_j$, compute the aligned bootstrap mean and standard deviation:
				\[
				\bar\theta_j^*=\frac1B\sum_{b=1}^B\hat\theta_j^{(b)},\qquad
				s_j^*=\left\{\frac1B\sum_{b=1}^B
				(\hat\theta_j^{(b)}-\bar\theta_j^*)^2\right\}^{1/2}.
				\]
				\State Form the interval using the standard normal quantile $z_{1-\alpha/2}$:
				\[
				\mathrm{CI}(\theta_j)=
				\bigl[\bar\theta_j^*-z_{1-\alpha/2}s_j^*,\ \bar\theta_j^*+z_{1-\alpha/2}s_j^*\bigr].
				\]
				\State Intersect scale and weight intervals with their parameter ranges.
				\State Compute the sample covariances $\hat\Sigma_a$ of $\{\hat{\bv}_a^{(b)}\}_{b=1}^B$ and $\hat\Lambda$ of $\{\hat{\bw}^{(b)}\}_{b=1}^B$, using denominator $B-1$.
				\State Form the regions in \eqref{eq:bootstrap-regions} using the bootstrap means, $\hat\Sigma_a^{-1}$ and $\chi^2_{d,1-\alpha}$ for each center, and $\hat\Lambda^\dagger$ and $\chi^2_{k-1,1-\alpha}$ for the weights, restricting the weight region to the simplex.
				\Ensure Scalar confidence intervals and center and weight confidence regions.
			\end{algorithmic}
		\end{algorithm}
		
		\subsection{Statistical Inference}\label{sec:inference}
		
		For fixed $k$ and $m$, the inferential targets are the parameters of the working density (the centers, weights, and scale), not the memberships $u_{ij}$, which measure relative geometric compatibility rather than posterior class probabilities. A center confidence region describes repeated-sample uncertainty in a prototype's location (in the applications, in the retained principal-component coordinates), conditional on the selected clustering specification and preprocessing. The parametric bootstrap and model-Fisher intervals below assume a correctly specified working density; otherwise the estimator targets the working-density parameter closest to the data distribution in Kullback--Leibler divergence, whose sandwich covariance (Appendix~\ref{app:approx-mle}) a nonparametric bootstrap estimates instead. Below we distinguish center confidence intervals from a center-equality statistic, whose null requires separate calibration.
		
		\paragraph{Cluster-center comparison.}
		
		A center comparison asks whether a fitted solution contains
		redundant prototypes. Unlike a conventional two-sample test,
		the two centers are estimated from the same data, and cluster
		labels are identifiable only up to permutation.
		
		To test $H_0:\bv_a=\bv_b$ against $H_1:\bv_a\neq\bv_b$ for some $1\le a<b\le k$, let $\tilde\theta$ and $\hat\theta$ be the MLEs with and without the constraint that the two components share a common center, on nested parameter spaces that allow coincident centers rather than the separated set $\Theta_\delta$; numerical fits approximate these targets. With $\ell(\theta)=\sum_{i=1}^n\log f(\bx_i;\theta)$, the likelihood-ratio statistic is $\Lambda=-2\{\ell(\tilde\theta)-\ell(\hat\theta)\}$.
		Large values of $\Lambda$ indicate improvement over
		the common-center fit. Because coincident centers
		violate the regularity conditions of Theorem~\ref{thm:mle-an},
		we calibrate $\Lambda$ with a fitted-null parametric
		bootstrap rather than a $\chi^2_d$ reference.
		Section~\ref{sec:revision-lrt} evaluates this procedure
		for one versus two centers.
		
		\paragraph{Confidence interval construction.} We quantify uncertainty via a parametric bootstrap at fixed $k$ and $m$. Let $\hat\theta=(\hat\sigma,\hat{\bV},\hat{\bw})$ be the estimate returned by Algorithm~\ref{alg:emfuzzy}. For $b=1,\dots,B$, generate $n$ observations from the fitted density $f(\cdot;\hat\theta,m)$, refit the same estimation routine on the bootstrap data, initialized at $\hat\theta$, and obtain $\hat\theta^{(b)}$. The one-dimensional simulations draw these observations by numerical inverse-CDF sampling. In the applications the draws are exact rejection samples: because $\bigl[\sum_{j}(w_j\|\bx-\bv_j\|^2)^{-1/(m-1)}\bigr]^{-(m-1)}\ge k^{-(m-1)}\min_j w_j\|\bx-\bv_j\|^2$, the kernel in~\eqref{eq:WFCM-density} is bounded by $\sum_j\exp\{-w_j\|\bx-\bv_j\|^2/(\sigma^2k^{m-1})\}$, a sum of isotropic Gaussian kernels that serves as the envelope.
		
		Section~\ref{sec:revision-coverage} compares normal-approximation bootstrap intervals with model-Fisher intervals. Nonparametric percentile intervals are also reported there as a supplementary comparison. In that comparison the two have similar coverage and length, and the Fisher intervals are about 60--70 times faster. The bootstrap, however, rebuilds the proposal and importance-sampling cloud for each replicate, so it also reflects importance-sampling variability and local optimization error that the model-Fisher intervals omit. Because each refit starts at $\hat\theta$, it does not reflect variability from convergence to other local optima.
		
		Because cluster labels are identifiable only up to permutation, Algorithm~\ref{alg:inference} aligns each replicate to $\hat\theta$ by the center permutation that minimizes the total squared distance to $\hat{\bV}$. For each scalar parameter $\theta_j$, it forms the normal-approximation interval $\bar\theta_j^*\pm z_{1-\alpha/2}s_j^*$ from the mean $\bar\theta_j^*$ and standard deviation $s_j^*$ of the aligned replicates. Centering at $\hat\theta_j$ instead changes the interval by the bootstrap bias estimate $\bar\theta_j^*-\hat\theta_j$, which is $O_p(n^{-1})$ under the conditions of Theorem~\ref{thm:mle-an}, so both centerings have the same first-order limit. Scale and weight intervals are intersected with their parameter ranges.
		For vector parameters, let $\bar{\bv}_a$ and $\bar{\bw}$ be the bootstrap means and $\hat\Sigma_a$ and $\hat\Lambda$ the sample covariances of the aligned center and weight replicates. With positive definite $\hat\Sigma_a$ and $\operatorname{rank}(\hat\Lambda)=k-1$, the normal-approximation regions are
		\begin{equation}\label{eq:bootstrap-regions}
			\begin{split}
				\mathcal C_{1-\alpha}(\mathbf v_a)
				&=\bigl\{\mathbf v:\ (\mathbf v-\bar{\mathbf v}_a)^\top\hat\Sigma_a^{-1}(\mathbf v-\bar{\mathbf v}_a)\le\chi^2_{d,1-\alpha}\bigr\},\\
				\mathcal C_{1-\alpha}(\bw)
				&=\bigl\{\bw\in\mathcal S_k:\ (\bw-\bar{\bw})^\top\hat\Lambda^\dagger(\bw-\bar{\bw})\le\chi^2_{k-1,1-\alpha}\bigr\},
			\end{split}
		\end{equation}
		where $\hat\Lambda^\dagger$ is the Moore--Penrose pseudoinverse.
		
		\subsection{Number of Clusters via Weighted Xie-Beni Index}
		
		To determine an appropriate number of clusters in our weighted fuzzy $c$-means framework, we use a heuristic weighted modification of the Xie--Beni index  (XBI), a widely used validity measure for fuzzy clustering \citep{xie1991new}. The XBI evaluates both the compactness and separation of fuzzy partitions, balancing within-cluster cohesion against between-cluster separation.
		
		The classical XBI divides the fuzzy within-cluster sum of squares $\sum_{i=1}^n\sum_{j=1}^k u_{ij}^m\|\bx_i-\bv_j\|^2$, computed with the FCM memberships~\eqref{eq:FCM-member}, by $n\min_{j\neq l}\|\bv_j-\bv_l\|^2$; smaller values indicate compact, well-separated clusters. With the WFCM memberships~\eqref{eq:membership} and fitted weights, we use
		\[
		\text{XBI}_{\text{weighted}}(k) = \frac{\sum_{i=1}^n \sum_{j=1}^k w_{j}\, u_{ij}^m \|\bx_i - \bv_j\|^2}{\min_{j \neq l} \|\bv_j - \bv_l\|^2}.
		\]
		
		We evaluate the weighted XBI over a range of candidate
		cluster numbers, with smaller values indicating a more
		favorable compactness--separation trade-off. For systematic
		comparisons across candidate cluster numbers, we use the
		normalized criterion
		\[
		\mathrm{WXBI}_{\mathrm{rel}}(k)
		= \frac{k}{n}\,\mathrm{XBI}_{\mathrm{weighted}}(k),
		\qquad
		\hat{k} = \argmin_{k\in\mathcal K}
		\mathrm{WXBI}_{\mathrm{rel}}(k).
		\]
		The factor $k$ compensates for the $1/k$ scaling
		associated with uniform simplex weights, so that
		the normalized criterion reduces to the conventional
		sample-averaged XBI when $w_j=1/k$.
		
		In the single-cell application, the only real-data
		analysis in which $k$ is selected, we minimize
		$\mathrm{WXBI}_{\mathrm{rel}}$ and additionally consider the shape
		of the index curves and the interpretability of the
		resulting clustering solutions. Fits containing a nearly
		empty cluster can produce artificially small index values, so
		such fits should be screened (Section~\ref{sec:real-data-analysis}). These approaches provide
		practical guidance for selecting the number of clusters.
		
		The subsequent statistical inference is conditional on the selected clustering specification. Bayesian clustering, by contrast, places a posterior on partitions and, in suitable models, on the number of clusters; credible balls summarize uncertainty over those partitions \citep{wade2018bayesian}.
		
		\section{Theory}\label{sec:theory}
		\begingroup
We first establish consistency and asymptotic normality of the exact global MLE, and then give conditions under which these results extend to the computed estimator. The number of clusters $k\geq2$ and dimension $d$ are fixed. All statements allow a common permutation of centers and weights.

\paragraph{Model and parameter space.}
Write $t=m-1$, $q=2/t$, and
\[
 E_\theta(x)=\sigma^{-2}\left\{\sum_{j=1}^k
 w_j^{-1/t}\|x-v_j\|^{-q}\right\}^{-t},\qquad
 g_\theta(x)=e^{-E_\theta(x)},\qquad f_\theta(x)=g_\theta(x)/Z(\theta),
\]
where $Z(\theta)=\int g_\theta(x)\,dx=1/C(\theta)$ and $E_\theta(v_j)=0$ by continuity. We use the compact, nonempty parameter set
\begin{equation}\label{eq:regular-parameter-set}
\begin{split}
 \Theta_\delta={}&[\sigma_{\min},\sigma_{\max}]
 \times\left\{w\in[\varepsilon_w,1]^k:\sum_jw_j=1\right\}\\
 &\times\left\{V\in\mathcal V^k:\min_{j\ne l}\|v_j-v_l\|\geq\delta\right\}
 \times[m_{\min},m_{\max}],
\end{split}
\end{equation}
where $0<\sigma_{\min}<\sigma_{\max}<\infty$, $0<\varepsilon_w<1/k$, $\delta>0$, $\mathcal V\subset\mathbb R^d$ is compact with nonempty interior, and $1<m_{\min}<m_{\max}<\infty$. The separation condition excludes coincident-center singularities. Positive weights ensure a proper density: $w_j\downarrow0$ makes $w_j^{-1/t}$ diverge, rather than removing a mixture component.

\setcounter{theorem}{0}
\begin{theorem}[Exact MLE consistency]\label{thm:mle_consistency}
Let $X_1,\ldots,X_n$ be i.i.d. from $f_{\theta_0}$ with $\theta_0\in\Theta_\delta$, and let $\hat\theta_n$ maximize the exact log-likelihood over $\Theta_\delta$. Then
\[
 \min_\pi\|\pi\hat\theta_n-\theta_0\|\longrightarrow0
 \quad\text{almost surely}.
\]
Here $\pi$ ranges over the common permutations of centers and weights.
\end{theorem}

For local inference, use free coordinates $\vartheta=(\sigma,w_1,\ldots,w_{k-1},\operatorname{vec}V,m)$ and the map $\theta=T(\vartheta)$ with $w_k=1-\sum_{j<k}w_j$. Let $\Xi=T^{-1}(\Theta_\delta)$. Interior below means strict slack in all bounds and separation constraints, with each center in the interior of $\mathcal V$.

\begin{theorem}[Exact MLE asymptotic normality]\label{thm:mle-an}
Under Theorem~\ref{thm:mle_consistency}, suppose that $\theta_0$ is interior in the above sense. After local label matching,
\[
 \sqrt n(\hat\vartheta_n-\vartheta_0)
 \xrightarrow{d}\mathcal N\{0,I_r(\vartheta_0)^{-1}\},\qquad
 I_r(\vartheta_0)=\operatorname{Var}_{\theta_0}
 \{\nabla_\vartheta E_{\theta_0}(X)\}>0.
\]
In the original coordinates the covariance is
$DT(\vartheta_0)\,I_r(\vartheta_0)^{-1}DT(\vartheta_0)^{\mathsf T}$, where $DT$ is the Jacobian of $T$.
\end{theorem}

Appendices~\ref{sec:pftheory}--\ref{app:normality} establish identifiability, including $m$, differentiability at the centers, and positive definite reduced information. The covariance is obtained in free coordinates and mapped back to the simplex. When $m$ is fixed in advance, the active-coordinate information must instead be inverted.

\paragraph{Importance sampling, the $m$ grid, and numerical optimization.}
Let $Q_n(\vartheta)=n^{-1}\sum_iE_{T(\vartheta)}(X_i)+\log Z(\vartheta)$. Conditional on the data, draw $M_n$ independent points $Y_r$ from a fitted proposal $\hat p_n$, and hold both the proposal and this cloud fixed throughout optimization, including across $m$ values. Define
\[
 \widehat Z_n(\vartheta)=\frac1{M_n}\sum_{r=1}^{M_n}
 \frac{g_{T(\vartheta)}(Y_r)}{\hat p_n(Y_r)},\qquad
 \widehat Q_n(\vartheta)=\frac1n\sum_iE_{T(\vartheta)}(X_i)+\log\widehat Z_n(\vartheta).
\]
For a grid $\mathcal G_n\subset[m_{\min},m_{\max}]$, let $h_n=\sup_m\min_{a\in\mathcal G_n}|m-a|$ and $\Xi_{h_n}=\{\vartheta\in\Xi:m\in\mathcal G_n\}$. The numerical output $\tilde\vartheta_n\in\Xi_{h_n}$ has average negative-log-likelihood gap
\[
 r_n=\widehat Q_n(\tilde\vartheta_n)-\inf_{\vartheta\in\Xi_{h_n}}\widehat Q_n(\vartheta)\geq0.
\]

\begin{theorem}[Conditional approximation transfer]\label{thm:approx-transfer}
Assume the interior setting of Theorem~\ref{thm:mle-an} and the uniform importance-sampling conditions in Appendix~\ref{app:approx-mle}. If $M_n\to\infty$, $h_n\to0$, and $r_n=o_p(1)$, then $\tilde\vartheta_n$ is consistent after label matching, and
\begin{equation}\label{eq:approx-transfer-rate}
 \|\tilde\vartheta_n-\hat\vartheta_n\|
 =O_p\{M_n^{-1/2}+h_n+r_n^{1/2}\}.
\end{equation}
Consequently, $M_n/n\to\infty$, $h_n=o(n^{-1/2})$, and $r_n=o_p(n^{-1})$ suffice for the same first-order limit as the exact MLE. If instead $n/M_n\to\rho\in[0,\infty)$, the conditional Monte Carlo central limit condition in Appendix~\ref{app:approx-mle} and the same grid and optimization rates give
\begin{equation}\label{eq:approx-transfer-clt}
 \sqrt n(\tilde\vartheta_n-\vartheta_0)
 \xrightarrow{d}\mathcal N\left(0,
 I_r^{-1}\{I_r+\rho\Omega_*\}I_r^{-1}\right),
\end{equation}
where $\Omega_*$ is the limiting covariance of the importance-sampling score error defined in \eqref{eq:mc-score-covariance}.
\end{theorem}

Thus $M_n$ proportional to $n$ generally contributes additional first-order variance. The result applies to numerical outputs in the basin of the global maximizer that attain the stated accuracy. Appendix~\ref{app:approx-mle} also treats continuous refinement, fixed finite $m$ grids, and misspecified working densities. A fixed grid containing $m_0$ selects it with probability tending to one, and an off-grid truth leads to a grid-restricted target.
\endgroup

		\section{Real Data Analysis}\label{sec:real-data-analysis}
		\subsection{Single-Cell RNA Sequencing Data}
		Single-cell RNA sequencing (scRNA-seq) measures gene expression in individual cells, enabling exploration of cellular heterogeneity. Standard workflows often rely on point-estimate cluster assignments and ignore uncertainty \citep{Laehnemann2020,Luecken2019}. We analyze a PBMC scRNA-seq dataset pre-classified by \cite{zheng2017massively} to illustrate geometric memberships and center-estimation uncertainty.
		
		\subsubsection{Data Pre-processing}
		
		We focus on a subset of
		PBMCs consisting of memory T cells, B cells, and monocytes. Following standard preprocessing \citep{duo2018}, we filter low-quality cells, normalize for sequencing depth, and log-transform counts using $\log_2(1+\text{count})$. We then select the top 500 genes by average expression, yielding a combined dataset of 3,000 cells (1,500 T cells, 1,000 B cells, 500 monocytes). These well-separated lineages provide a baseline for evaluating soft clustering, allowing us to describe geometric ambiguity arising from technical noise, biological variability, and class imbalance. We retain two principal components based on the scree plot: PC1 and PC2 explain 12.51\% and 8.62\% of the variance, respectively, whereas PC3 explains only 1.45\%, indicating a clear elbow after the second component. We standardized the two retained PC scores before clustering.
		
		\subsubsection{Data Analysis}
\begin{figure}
			\centering
			\includegraphics[width=1\linewidth]{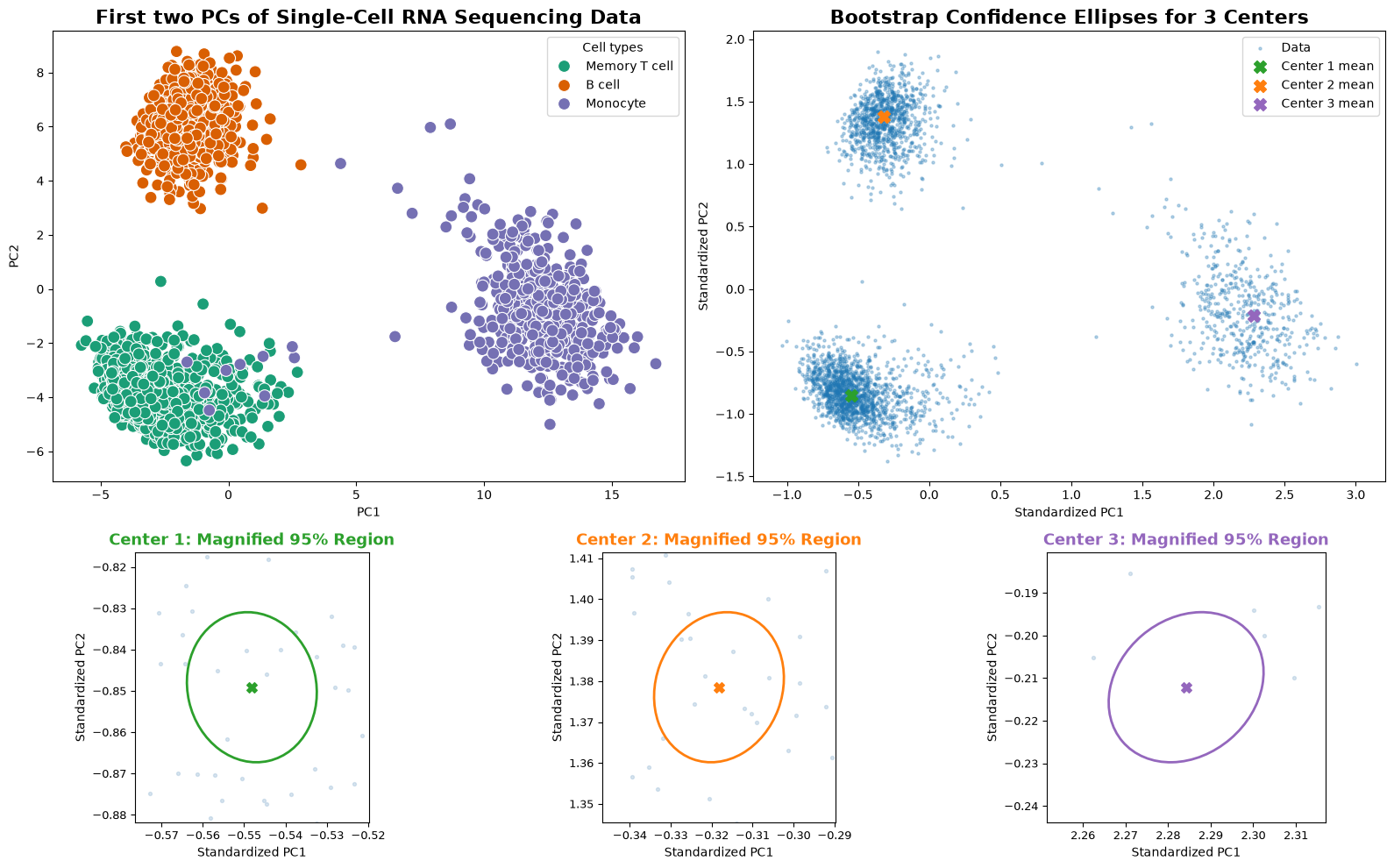}
			\caption{Top left: first two PCs of Single-Cell RNA-Seq data consisting of memory T cells, B cells, and monocytes. Top right: The estimated centers with joint 95\% confidence ellipses. Bottom: Magnified views of each region. 
			} \label{fig:cell-center1}
		\end{figure}

		Figure \ref{fig:cell-center1} displays joint confidence ellipses for the estimated cluster centers, which capture the joint variability across bootstrap resamples. The relatively small and non-overlapping ellipses suggest that the three centers are well separated and estimated with reasonable precision. Table \ref{tb:cell} demonstrates bootstrap summary statistics and 95\% confidence intervals for the parameters, including the cluster centers $(v_j^x, v_j^y)$, the scale parameter $\sigma$, and the weights $w_j$. The narrow intervals indicate reproducible parameter estimates across bootstrap resamples, supporting the stability of the fitted
		WFCM representation.
		
		\begin{table}[htbp]
			\centering
			\caption{scRNA-seq Bootstrap Summary Statistics and 95\% Confidence Intervals}\label{tb:cell}
			\begin{tabular}{lcc}
				\toprule
				\textbf{Parameter} & \textbf{Mean $\pm$ SD} & \textbf{95\% CI} \\
				\midrule
				$\sigma$ & $ 0.064 \pm 0.001$ & $[ 0.063,\  0.065]$ \\
				$v_1^x$  & $-0.548 \pm 0.006$ & $[-0.561,\ -0.536]$ \\
				$v_1^y$  & $-0.849 \pm 0.007$ & $[-0.864,\ -0.835]$ \\
				$v_2^x$  & $-0.318 \pm 0.006$ & $[-0.331,\ -0.306]$ \\
				$v_2^y$  & $ 1.379 \pm 0.007$ & $[ 1.364,\  1.393]$ \\
				$v_3^x$  & $ 2.284 \pm 0.007$ & $[ 2.270,\  2.299]$ \\
				$v_3^y$  & $-0.212 \pm 0.007$ & $[-0.226,\ -0.198]$ \\
				$w_1$    & $ 0.240 \pm 0.010$ & $[ 0.220,\  0.260]$ \\
				$w_2$    & $ 0.384 \pm 0.014$ & $[ 0.357,\  0.412]$ \\
				$w_3$    & $ 0.376 \pm 0.013$ & $[ 0.351,\  0.400]$ \\
				\bottomrule
			\end{tabular}
		\end{table}
		
\begin{figure}
			\centering
			\includegraphics[width=1\linewidth]{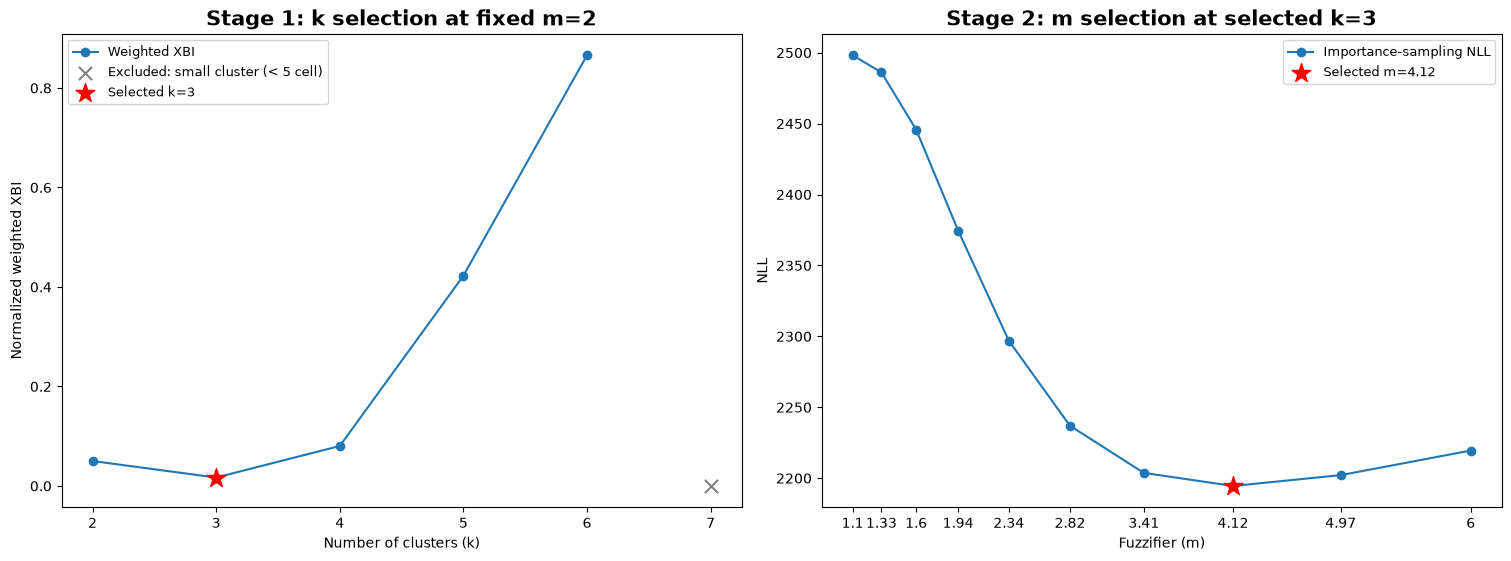}
			\caption{Two-stage parameter selection for the single-cell RNA-seq analysis. Left: Normalized weighted XBI across candidate numbers of clusters at a fixed fuzziness parameter. Right: Importance-sampling NLL across candidate fuzziness parameters at the selected number of clusters.
			} \label{fig:cell-xbi}
		\end{figure}

		We used a two-stage procedure to select the number of clusters and the
		fuzzifier. First, we fixed $m=2$ and selected the number of clusters by
		minimizing the normalized weighted Xie--Beni index, which balances
		within-cluster compactness and between-center separation. We restricted
		this comparison to fits with at least five cells assigned to each cluster
		by highest membership. Among the eligible fits, the index was minimized
		at $K=3$ (left panel of Figure~\ref{fig:cell-xbi}). This screening rule was
		added after we observed that, without it, the index is minimized by a $k=7$
		fit containing a single-cell cluster (the excluded point in
		Figure~\ref{fig:cell-xbi}).
		
		We then selected the fuzziness parameter by minimizing the negative
		log-likelihood over ten geometrically spaced values between 1.1 and 6. The minimum occurred at
		$m=4.12$ (right panel of Figure~\ref{fig:cell-xbi}). Below, we examine how the fitted memberships are distributed across the three cell types and summarize uncertainty in the estimated cluster centers and weights using the parametric bootstrap.
		
		\begin{figure}
			\centering
			\includegraphics[width=0.5\linewidth]{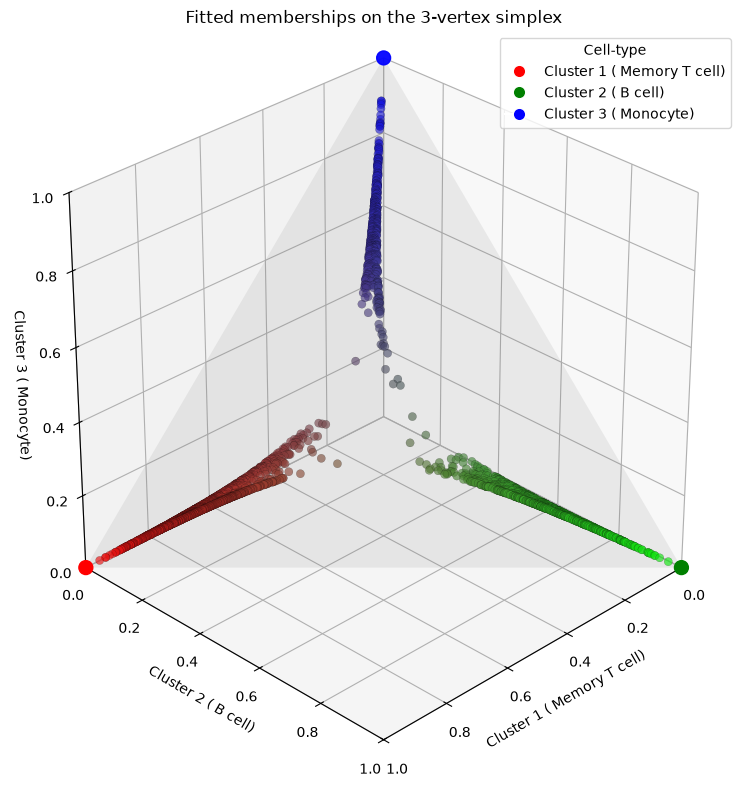}
			\caption{Estimated cluster memberships on the 3-vertex simplex}
			\label{fig:cell-weight1}
		\end{figure}
		
		Figure \ref{fig:cell-weight1} shows the estimated cluster memberships on the three-dimensional simplex. As expected given the biological distinctness of memory T cells, B cells, and monocytes, most cells concentrate near the simplex vertices, reflecting near-crisp memberships; however, the soft framework still describes relative compatibility and highlights the few borderline cases.
		
\begin{figure}
			\centering
			\includegraphics[width=0.8\linewidth]{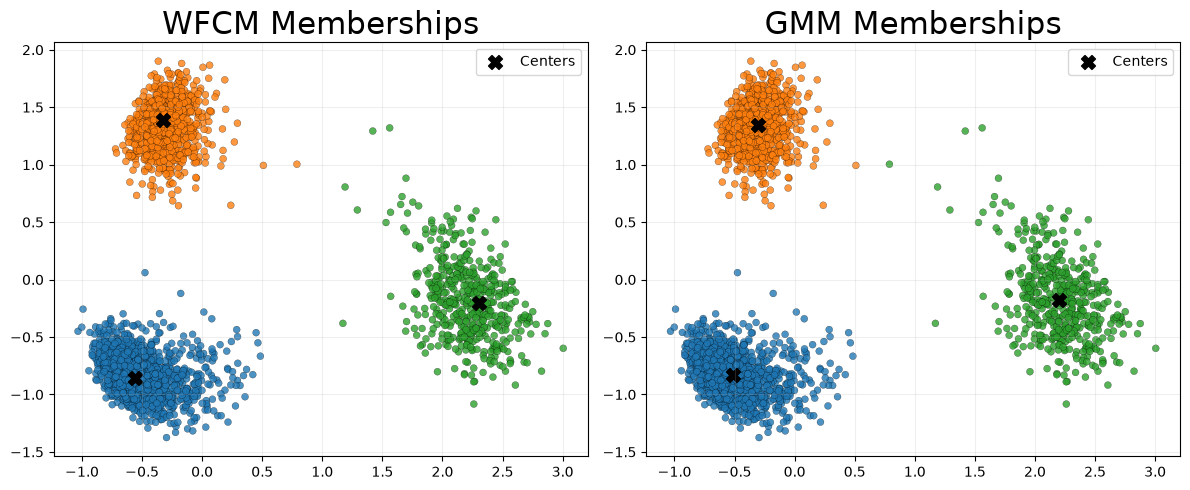}
			\caption{Cluster Membership Comparison between WFCM and GMM}\label{fig:cell.gmm}
		\end{figure}

		We compared the proposed WFCM with a GMM in Figure \ref{fig:cell.gmm}. Both methods achieved nearly identical cluster structures, as indicated by a high Adjusted Rand Index (ARI = 0.999), where ARI was computed by first assigning each observation to the cluster with the highest fuzzy membership (WFCM) or posterior responsibility (GMM). WFCM achieved this result without assuming Gaussianity or estimating covariance matrices, using distance-based memberships within its working density, whereas GMM required full-covariance estimation and likelihood optimization (BIC = 3925.96).
		
		From a biological perspective, these results confirm that the three immune cell populations are transcriptionally distinct. Center intervals quantify estimation uncertainty; cell-level memberships describe geometric compatibility.
		
		\subsection{Alzheimer's Disease Neuroimaging Data}
		
		The Alzheimer's Disease Neuroimaging Initiative (ADNI) is a longitudinal multicenter study of cognitive aging and Alzheimer's disease (AD). It enrolls cognitively normal (CN) adults, individuals with mild cognitive impairment (MCI), and AD patients, collecting repeated neuroimaging, genetic, cognitive, and clinical measures. For our analysis, we used baseline visits only, retaining the earliest record per subject, resulting in a cohort of 2,436 subjects.
		
		\subsubsection{Data Pre-processing}
		
		For each subject, we selected the earliest baseline visit (VISCODE = ``bl''). We extracted demographic (age, education), genetic (APOE4 status), cognitive and functional (CDR-SB, ADAS11/13/Q4, MMSE, RAVLT percent forgetting, Logical Memory delayed recall (LDELTOTAL), TRABSCOR, FAQ) variables associated with AD progression, 12 predictors in total. Missing numeric values were median-imputed, and all features were standardized to mean 0 and unit variance. For analysis, dementia cases were labeled ``AD,'' while cognitively normal (CN) and mild cognitive impairment (MCI) were grouped as ``non-AD.'' We retained the first three principal components of the standardized features for the ADNI analysis, allowing us to visualize the fitted memberships and cluster-center confidence ellipsoids in three dimensions. Together, these components explained 70.3\% of the total variance. The first two eigenvalues exceed one; we also retain the third (eigenvalue 0.97) to allow three-dimensional display. The three PC scores were re-standardized before clustering. Fourteen subjects without a recorded diagnosis are included in the fit but excluded from label-based summaries.
		
		\subsubsection{Data Analysis}
\begin{figure}
			\centering
			\includegraphics[width=0.8\linewidth]{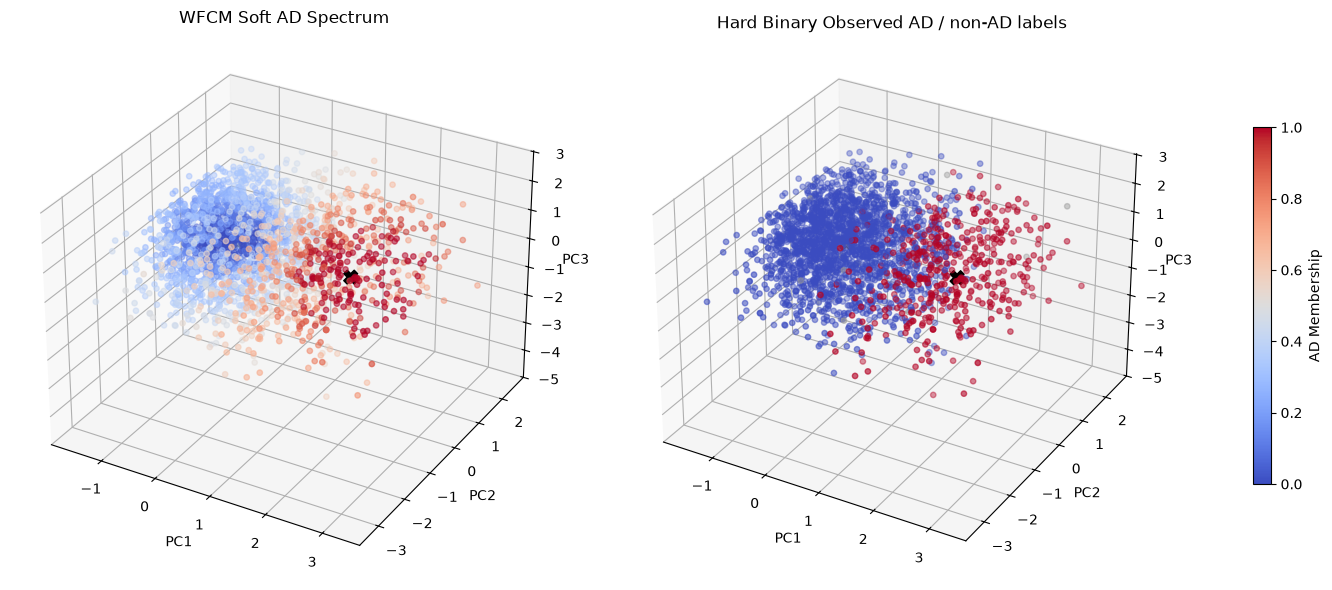}
			\caption{Left: membership weights from WFCM, representing the membership or degree of belonging to the "AD-like" cluster; colors are stretched between the 5th and 95th percentiles of the fitted memberships (raw values in Figure~\ref{fig:adni2}). Right: the binary clinical labels (AD vs. non-AD).}\label{fig:adni1}
		\end{figure}

		We fixed $k = 2$ to represent AD-like and non-AD-like prototypes, and selected the fuzzifier by minimizing the numerical negative log-likelihood over eight geometrically spaced values between 1.1 and 6; the minimum occurred at $m = 4.71$. We projected ADNI data onto the first three principal components and compared two perspectives as shown in Figure \ref{fig:adni1}. On the right, the hard labels split subjects into AD versus non-AD, creating a rigid boundary. But notice how the two groups overlap heavily---many individuals diagnosed as non-AD lie close to the AD cluster, and vice versa.
		
\begin{figure}
			\centering
			\includegraphics[width=0.8\linewidth]{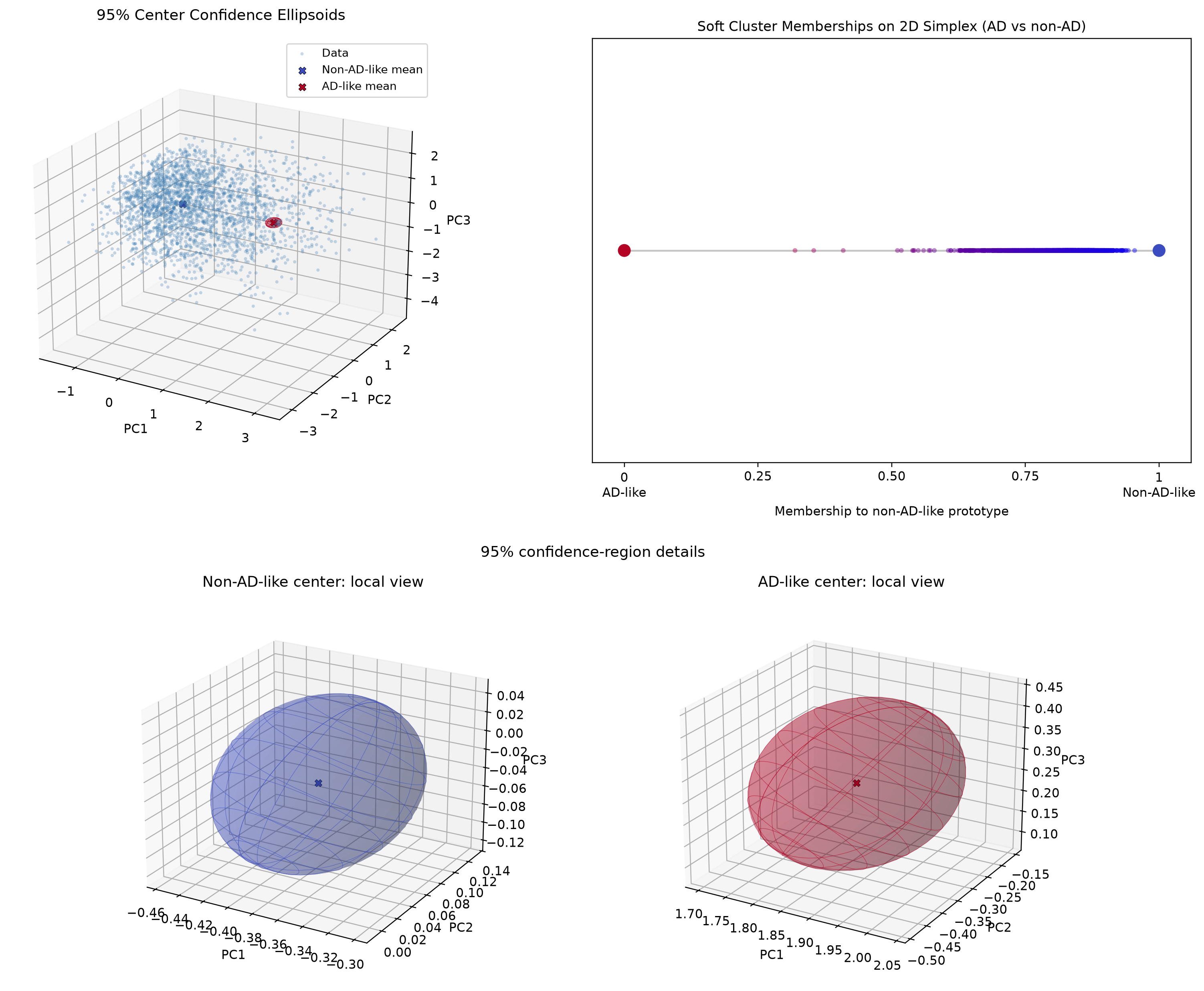}
			\caption{Left: The estimated centers with 95\% confidence regions. Right: Estimated cluster memberships for ADNI subjects on the 2-vertex simplex. Bottom: Magnified views of each region.}\label{fig:adni2}
		\end{figure}

		On the left, our WFCM approach produces a soft spectrum of AD membership. Here, instead of forcing each person into a binary label, we quantify the degree of belonging to the AD-like cluster. The WFCM memberships form a smooth gradient from blue (non-AD-like) to red (AD-like); the color scale in the left panel is stretched between the 5th and 95th percentiles of the fitted AD-like memberships, and raw values appear in Figure~\ref{fig:adni2}. The soft assignments identify a ``spectrum zone" where individuals have intermediate membership values (e.g., 0.3--0.5). These values describe geometric compatibility with the two prototypes.
		
		Table~\ref{tab:bootstrap} summarizes the bootstrap distribution of parameter estimates. For numerical stability, the ADNI fits constrain $\sigma \geq 0.05$ and each weight $w_j \geq 0.01$, and weight intervals are intersected with this range (Algorithm 2, step 9). The estimated weight of the non-AD-like cluster, $w_1$, lies close to this bound, so its lower endpoint in Table~\ref{tab:bootstrap} equals the constraint value.
		Because a smaller $w_j$ enlarges the high-membership region of cluster $j$, the small $w_1$ gives the larger non-AD-like cluster a wide basin. The results show that variance estimates are relatively stable, while some cluster center coordinates and weights exhibit wider confidence intervals, reflecting greater uncertainty in those directions.
		
		\begin{table}[ht]
			\centering
			\caption{ADNI Bootstrap Summary Statistics with 95\% Confidence Intervals}
			\label{tab:bootstrap}
			\begin{tabular}{lcc}
				\hline
				\textbf{Parameter} & \textbf{Mean $\pm$ SD} & \textbf{95\% CI} \\
				\hline
				$\sigma$              & $ 0.092 \pm 0.006$ & $[ 0.079,\phantom{-}0.104]$ \\
				$v_{1}^{x}$ & $-0.379 \pm 0.027$ & $[-0.432,-0.326]$ \\
				$v_{1}^{y}$ & $ 0.061 \pm 0.025$ & $[ 0.013,\phantom{-}0.110]$ \\
				$v_{1}^{z}$ & $-0.038 \pm 0.029$ & $[-0.094,\phantom{-}0.018]$ \\
				$v_{2}^{x}$ & $ 1.870 \pm 0.060$ & $[ 1.753,\phantom{-}1.987]$ \\
				$v_{2}^{y}$ & $-0.323 \pm 0.059$ & $[-0.439,-0.207]$ \\
				$v_{2}^{z}$ & $ 0.258 \pm 0.063$ & $[ 0.136,\phantom{-}0.381]$ \\
				$w_1$                 & $ 0.012 \pm 0.002$ & $[ 0.010,\phantom{-}0.016]$ \\
				$w_2$                 & $ 0.988 \pm 0.002$ & $[ 0.984,\phantom{-}0.990]$ \\
				\hline
			\end{tabular}
		\end{table}
		
		To further illustrate the soft clustering results, we project the estimated memberships for each ADNI subject onto a 2-vertex simplex, which in the case of two clusters reduces to a line segment between the Alzheimer's disease (AD) and non-AD corners (right plot of Figure \ref{fig:adni2}). Each subject is represented as a point whose position reflects its estimated cluster membership. Positions closer to the red corner indicate higher AD-like membership. In this fit, almost all subjects have AD-like membership below 0.5, so the spectrum lies mainly on the non-AD-like half of the segment. Points located in the middle of the line indicate subjects with intermediate geometric memberships.
		
\begin{figure}
			\centering
			\includegraphics[width=0.8\linewidth]{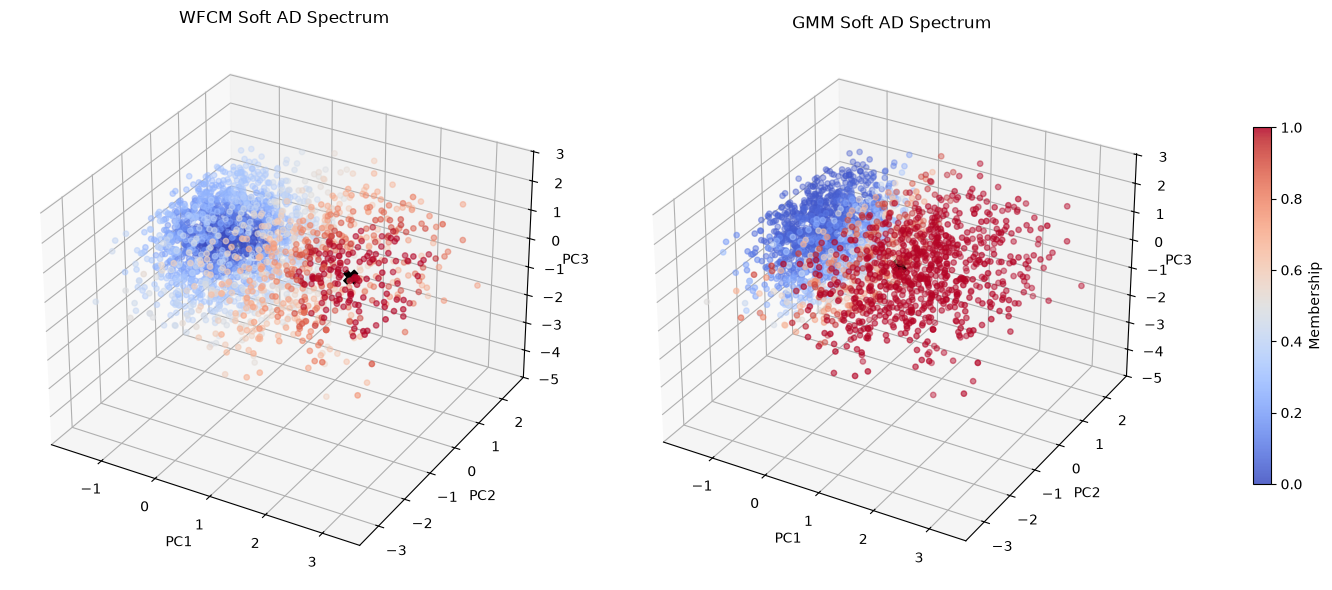}
			\caption{Soft AD-like cluster memberships estimated by WFCM and GMM. The WFCM panel uses the percentile-stretched color scale of Figure~\ref{fig:adni1}; the GMM panel shows raw responsibilities on $[0,1]$.}\label{fig:adni.gmm}
		\end{figure}

		We compared the proposed WFCM with a Gaussian mixture model (GMM) in Figure~\ref{fig:adni.gmm}. Against the clinical labels, which were not used for fitting, WFCM achieved a threshold-based adjusted Rand index (ARI)\footnote{Subjects are labeled AD-like when their AD-like membership (WFCM) or responsibility (GMM) exceeds the midpoint of the 5th and 95th percentiles of that method's scores.} of 0.485, compared with 0.185 for a full-covariance GMM (BIC = 20052.39).
		
		\section{Discussion}\label{sec:discussion}
		
		The weighted fuzzy $c$-means framework proposed here provides a principled way to incorporate heterogeneity into soft clustering. Our formulation combines geometric memberships with likelihood estimation: membership updates retain the classical FCM form, while cluster weights adapt to imbalance and support a likelihood-based probabilistic interpretation. Simulations are consistent with the theory, with estimation errors decaying at approximately the $n^{-1/2}$ rate and Fisher-standardized QQ diagnostics broadly consistent with asymptotic normality, apart from two degenerate numerical fits (Appendix~\ref{sec:sup_simulation}). This enables uncertainty quantification via both bootstrap and Fisher-information-based methods. In the real-data analyses, WFCM produces well-separated clusters with tight confidence regions in scRNA-seq; in ADNI it describes graded compatibility with AD-like and non-AD-like prototypes; and in TCGA-BRCA it raises LumB recall at the cost of HER2-enriched recall and overall agreement.
		
		Several limitations warrant discussion. First, all reported results are conditional on
		the data-dependent preprocessing choices made
		for each application, including selection of
		highly expressed genes for the single-cell analysis,
		selection of highly variable genes for TCGA-BRCA,
		and PCA dimension selection for all three applications. 
		In addition, likelihood optimization and importance sampling incur additional computation and can be sensitive to dimension and proposal choice; the iterative routine can also converge to local optima, so initialization matters. Future work may extend the framework by incorporating structured prior information (e.g., spatial structure, network constraints, or biological pathways) into the weighting scheme, and by developing dynamic or longitudinal extensions where memberships evolve over time.
		
		Additional simulations in Sections~\ref{sec:revision-benchmark}--\ref{sec:revision-computation} compare methods across geometry, imbalance, and overlap; distinguish fixed-$k$ coverage from selection uncertainty; and quantify computational sensitivity. They show that WFCM has higher ARI than FCM but lower ARI than a full-covariance GMM, and performs poorly on nonconvex clusters; that selecting $k$ lowers joint coverage; and that importance-sampling efficiency declines with dimension.
		
		Two computational extensions warrant further investigation. Particle swarm optimization (PSO), used for constrained mixture designs \citep{wong2015modifiedpso}, minimax optimal designs \citep{chen2015minimaxpso}, and fuzzy $c$-means \citep{izakian2011fuzzyswarm,silvafilho2015hybridpso}, could generate diverse starting centers and weights for local likelihood refinement, although each evaluation of the normalizing constant adds computation and importance-sampling variability. Quasi-Newton acceleration \citep{zhou2011quasinewton} and majorization--minimization \citep{lange2021mmnonconvex} may make refinement under parameter constraints faster or more reliable; establishing descent for the normalized likelihood and controlling importance-sampling error within such algorithms remain open.
		
		
		\section*{Data Availability Statement} 
		Upon acceptance, we will release all code needed to reproduce the simulations and all three application analyses in a public GitHub repository. The single-cell and TCGA-BRCA data are publicly available; ADNI data are subject to third-party access restrictions. None of the application datasets are redistributed with this article. Our processed file \texttt{processed\_mix3\_df.rds} is derived from the PBMC scRNA-seq dataset of \citet{zheng2017massively} using the preprocessing pipeline in \citet{gao2024selective}. The repository documentation gives step-by-step instructions, using the upstream script of \citet{gao2024selective}, to download the raw data and recreate the processed file. TCGA-BRCA STAR TPM expression values and PAM50 annotations are available from the UCSC Xena data portal; the repository code performs the cohort selection, gene filtering, and PCA. ADNI data are available to qualified researchers via the ADNI Image and Data Archive (IDA) under the ADNI Data Use Agreement. The repository code performs the median imputation, standardization, and PCA from a baseline ADNIMERGE extract containing the 12 listed predictors, so approved users can reproduce the derived features; a synthetic ADNI-format facsimile allows the analysis to be run without restricted data. Computations use NumPy \citep{harris2020numpy}, SciPy \citep{virtanen2020scipy}, and scikit-learn \citep{pedregosa2011scikit}.
		
		\bibliographystyle{apacite}
		\bibliography{ref}

\begin{thebibliography}{}

\bibitem [\protect \citeauthoryear {%
Bezdek%
}{%
Bezdek%
}{%
{\protect \APACyear {2013}}%
}]{%
bezdek2013pattern}
\APACinsertmetastar {%
bezdek2013pattern}%
\begin{APACrefauthors}%
Bezdek, J\BPBI C.%
\end{APACrefauthors}%
\unskip\
\newblock
\APACrefYear{2013}.
\newblock
\APACrefbtitle {Pattern recognition with fuzzy objective function algorithms}
  {Pattern recognition with fuzzy objective function algorithms}.
\newblock
\APACaddressPublisher{}{Springer Science \& Business Media}.
\PrintBackRefs{\CurrentBib}

\bibitem [\protect \citeauthoryear {%
Bullen%
}{%
Bullen%
}{%
{\protect \APACyear {2013}}%
}]{%
bullen2013handbook}
\APACinsertmetastar {%
bullen2013handbook}%
\begin{APACrefauthors}%
Bullen, P\BPBI S.%
\end{APACrefauthors}%
\unskip\
\newblock
\APACrefYear{2013}.
\newblock
\APACrefbtitle {Handbook of means and their inequalities} {Handbook of means
  and their inequalities}\ (\BVOL~560).
\newblock
\APACaddressPublisher{}{Springer Science \& Business Media}.
\PrintBackRefs{\CurrentBib}

\bibitem [\protect \citeauthoryear {%
Chen%
, Chang%
, Wang%
, Tung%
\BCBL {}\ \BBA {} Wong%
}{%
Chen%
\ \protect \BOthers {.}}{%
{\protect \APACyear {2015}}%
}]{%
chen2015minimaxpso}
\APACinsertmetastar {%
chen2015minimaxpso}%
\begin{APACrefauthors}%
Chen, R\BHBI B.%
, Chang, S\BHBI P.%
, Wang, W.%
, Tung, H\BHBI C.%
\BCBL {}\ \BBA {} Wong, W\BPBI K.%
\end{APACrefauthors}%
\unskip\
\newblock
\APACrefYearMonthDay{2015}{}{}.
\newblock
{\BBOQ}\APACrefatitle {Minimax Optimal Designs via Particle Swarm Optimization
  Methods} {Minimax optimal designs via particle swarm optimization
  methods}.{\BBCQ}
\newblock
\APACjournalVolNumPages{Statistics and Computing}{25}{5}{975--988}.
\newblock
\begin{APACrefDOI} \doi{10.1007/s11222-014-9466-0} \end{APACrefDOI}
\PrintBackRefs{\CurrentBib}

\bibitem [\protect \citeauthoryear {%
Du{\`o}%
, Robinson%
\BCBL {}\ \BBA {} Soneson%
}{%
Du{\`o}%
\ \protect \BOthers {.}}{%
{\protect \APACyear {2018}}%
}]{%
duo2018}
\APACinsertmetastar {%
duo2018}%
\begin{APACrefauthors}%
Du{\`o}, A.%
, Robinson, M\BPBI D.%
\BCBL {}\ \BBA {} Soneson, C.%
\end{APACrefauthors}%
\unskip\
\newblock
\APACrefYearMonthDay{2018}{}{}.
\newblock
{\BBOQ}\APACrefatitle {A systematic performance evaluation of clustering
  methods for single-cell {RNA}-seq data} {A systematic performance evaluation
  of clustering methods for single-cell {RNA}-seq data}.{\BBCQ}
\newblock
\APACjournalVolNumPages{F1000Research}{7}{}{1141}.
\newblock
\begin{APACrefDOI} \doi{10.12688/f1000research.15666.3} \end{APACrefDOI}
\PrintBackRefs{\CurrentBib}

\bibitem [\protect \citeauthoryear {%
Gao%
, Bien%
\BCBL {}\ \BBA {} Witten%
}{%
Gao%
\ \protect \BOthers {.}}{%
{\protect \APACyear {2024}}%
}]{%
gao2024selective}
\APACinsertmetastar {%
gao2024selective}%
\begin{APACrefauthors}%
Gao, L\BPBI L.%
, Bien, J.%
\BCBL {}\ \BBA {} Witten, D.%
\end{APACrefauthors}%
\unskip\
\newblock
\APACrefYearMonthDay{2024}{}{}.
\newblock
{\BBOQ}\APACrefatitle {Selective inference for hierarchical clustering}
  {Selective inference for hierarchical clustering}.{\BBCQ}
\newblock
\APACjournalVolNumPages{Journal of the American Statistical
  Association}{119}{545}{332--342}.
\PrintBackRefs{\CurrentBib}

\bibitem [\protect \citeauthoryear {%
Harris%
, Millman%
, van~der Walt%
\BCBL {}\ \protect \BOthers {.}}{%
Harris%
\ \protect \BOthers {.}}{%
{\protect \APACyear {2020}}%
}]{%
harris2020numpy}
\APACinsertmetastar {%
harris2020numpy}%
\begin{APACrefauthors}%
Harris, C\BPBI R.%
, Millman, K\BPBI J.%
, van~der Walt, S\BPBI J.%
\BCBL {}\ \BOthersPeriod {.}\end{APACrefauthors}%
\unskip\
\newblock
\APACrefYearMonthDay{2020}{}{}.
\newblock
{\BBOQ}\APACrefatitle {Array programming with {NumPy}} {Array programming with
  {NumPy}}.{\BBCQ}
\newblock
\APACjournalVolNumPages{Nature}{585}{7825}{357--362}.
\PrintBackRefs{\CurrentBib}

\bibitem [\protect \citeauthoryear {%
Izakian%
\ \BBA {} Abraham%
}{%
Izakian%
\ \BBA {} Abraham%
}{%
{\protect \APACyear {2011}}%
}]{%
izakian2011fuzzyswarm}
\APACinsertmetastar {%
izakian2011fuzzyswarm}%
\begin{APACrefauthors}%
Izakian, H.%
\BCBT {}\ \BBA {} Abraham, A.%
\end{APACrefauthors}%
\unskip\
\newblock
\APACrefYearMonthDay{2011}{}{}.
\newblock
{\BBOQ}\APACrefatitle {Fuzzy C-means and Fuzzy Swarm for Fuzzy Clustering
  Problem} {Fuzzy c-means and fuzzy swarm for fuzzy clustering problem}.{\BBCQ}
\newblock
\APACjournalVolNumPages{Expert Systems with Applications}{38}{3}{1835--1838}.
\newblock
\begin{APACrefDOI} \doi{10.1016/j.eswa.2010.07.112} \end{APACrefDOI}
\PrintBackRefs{\CurrentBib}

\bibitem [\protect \citeauthoryear {%
Jiang%
\ \protect \BOthers {.}}{%
Jiang%
\ \protect \BOthers {.}}{%
{\protect \APACyear {2024}}%
}]{%
jiang2024soft}
\APACinsertmetastar {%
jiang2024soft}%
\begin{APACrefauthors}%
Jiang, S.%
, Gai, X.%
, Treggiari, M\BPBI M.%
, Stead, W\BPBI W.%
, Zhao, Y.%
, Page, C\BPBI D.%
\BCBL {}\ \BBA {} Zhang, A\BPBI R.%
\end{APACrefauthors}%
\unskip\
\newblock
\APACrefYearMonthDay{2024}{}{}.
\newblock
{\BBOQ}\APACrefatitle {Soft phenotyping for sepsis via EHR time-aware soft
  clustering} {Soft phenotyping for sepsis via ehr time-aware soft
  clustering}.{\BBCQ}
\newblock
\APACjournalVolNumPages{Journal of biomedical informatics}{152}{}{104615}.
\PrintBackRefs{\CurrentBib}

\bibitem [\protect \citeauthoryear {%
L{\"a}hnemann%
\ \protect \BOthers {.}}{%
L{\"a}hnemann%
\ \protect \BOthers {.}}{%
{\protect \APACyear {2020}}%
}]{%
Laehnemann2020}
\APACinsertmetastar {%
Laehnemann2020}%
\begin{APACrefauthors}%
L{\"a}hnemann, D.%
, K{\"o}ster, J.%
, Szczurek, E.%
, McCarthy, D\BPBI J.%
, Hicks, S\BPBI C.%
, Robinson, M\BPBI D.%
\BDBL {}others%
\end{APACrefauthors}%
\unskip\
\newblock
\APACrefYearMonthDay{2020}{}{}.
\newblock
{\BBOQ}\APACrefatitle {Eleven grand challenges in single-cell data science}
  {Eleven grand challenges in single-cell data science}.{\BBCQ}
\newblock
\APACjournalVolNumPages{Genome Biology}{21}{}{31}.
\newblock
\begin{APACrefDOI} \doi{10.1186/s13059-020-1926-6} \end{APACrefDOI}
\PrintBackRefs{\CurrentBib}

\bibitem [\protect \citeauthoryear {%
Lange%
, Won%
, Landeros%
\BCBL {}\ \BBA {} Zhou%
}{%
Lange%
\ \protect \BOthers {.}}{%
{\protect \APACyear {2021}}%
}]{%
lange2021mmnonconvex}
\APACinsertmetastar {%
lange2021mmnonconvex}%
\begin{APACrefauthors}%
Lange, K.%
, Won, J\BHBI H.%
, Landeros, A.%
\BCBL {}\ \BBA {} Zhou, H.%
\end{APACrefauthors}%
\unskip\
\newblock
\APACrefYearMonthDay{2021}{}{}.
\newblock
{\BBOQ}\APACrefatitle {Nonconvex Optimization via {MM} Algorithms: Convergence
  Theory} {Nonconvex optimization via {MM} algorithms: Convergence
  theory}.{\BBCQ}
\newblock
\BIn{} \APACrefbtitle {Wiley StatsRef: Statistics Reference Online.} {Wiley
  statsref: Statistics reference online.}
\newblock
\APACaddressPublisher{}{John Wiley \& Sons}.
\newblock
\begin{APACrefDOI} \doi{10.1002/9781118445112.stat08295} \end{APACrefDOI}
\PrintBackRefs{\CurrentBib}

\bibitem [\protect \citeauthoryear {%
Liu%
\ \BBA {} Nocedal%
}{%
Liu%
\ \BBA {} Nocedal%
}{%
{\protect \APACyear {1989}}%
}]{%
liu1989limited}
\APACinsertmetastar {%
liu1989limited}%
\begin{APACrefauthors}%
Liu, D\BPBI C.%
\BCBT {}\ \BBA {} Nocedal, J.%
\end{APACrefauthors}%
\unskip\
\newblock
\APACrefYearMonthDay{1989}{}{}.
\newblock
{\BBOQ}\APACrefatitle {On the limited memory BFGS method for large scale
  optimization} {On the limited memory bfgs method for large scale
  optimization}.{\BBCQ}
\newblock
\APACjournalVolNumPages{Mathematical programming}{45}{1}{503--528}.
\PrintBackRefs{\CurrentBib}

\bibitem [\protect \citeauthoryear {%
Lloyd%
}{%
Lloyd%
}{%
{\protect \APACyear {1982}}%
}]{%
lloyd1982least}
\APACinsertmetastar {%
lloyd1982least}%
\begin{APACrefauthors}%
Lloyd, S.%
\end{APACrefauthors}%
\unskip\
\newblock
\APACrefYearMonthDay{1982}{}{}.
\newblock
{\BBOQ}\APACrefatitle {Least squares quantization in PCM} {Least squares
  quantization in pcm}.{\BBCQ}
\newblock
\APACjournalVolNumPages{IEEE transactions on information
  theory}{28}{2}{129--137}.
\PrintBackRefs{\CurrentBib}

\bibitem [\protect \citeauthoryear {%
Luecken%
\ \BBA {} Theis%
}{%
Luecken%
\ \BBA {} Theis%
}{%
{\protect \APACyear {2019}}%
}]{%
Luecken2019}
\APACinsertmetastar {%
Luecken2019}%
\begin{APACrefauthors}%
Luecken, M\BPBI D.%
\BCBT {}\ \BBA {} Theis, F\BPBI J.%
\end{APACrefauthors}%
\unskip\
\newblock
\APACrefYearMonthDay{2019}{}{}.
\newblock
{\BBOQ}\APACrefatitle {Current best practices in single-cell {RNA}-seq
  analysis: a tutorial} {Current best practices in single-cell {RNA}-seq
  analysis: a tutorial}.{\BBCQ}
\newblock
\APACjournalVolNumPages{Molecular Systems Biology}{15}{6}{e8746}.
\newblock
\begin{APACrefDOI} \doi{10.15252/msb.20188746} \end{APACrefDOI}
\PrintBackRefs{\CurrentBib}

\bibitem [\protect \citeauthoryear {%
Pedregosa%
, Varoquaux%
, Gramfort%
\BCBL {}\ \protect \BOthers {.}}{%
Pedregosa%
\ \protect \BOthers {.}}{%
{\protect \APACyear {2011}}%
}]{%
pedregosa2011scikit}
\APACinsertmetastar {%
pedregosa2011scikit}%
\begin{APACrefauthors}%
Pedregosa, F.%
, Varoquaux, G.%
, Gramfort, A.%
\BCBL {}\ \BOthersPeriod {.}\end{APACrefauthors}%
\unskip\
\newblock
\APACrefYearMonthDay{2011}{}{}.
\newblock
{\BBOQ}\APACrefatitle {Scikit-learn: Machine learning in {Python}}
  {Scikit-learn: Machine learning in {Python}}.{\BBCQ}
\newblock
\APACjournalVolNumPages{Journal of Machine Learning
  Research}{12}{}{2825--2830}.
\PrintBackRefs{\CurrentBib}

\bibitem [\protect \citeauthoryear {%
Silva~Filho%
, Pimentel%
, de Souza%
\BCBL {}\ \BBA {} Oliveira%
}{%
Silva~Filho%
\ \protect \BOthers {.}}{%
{\protect \APACyear {2015}}%
}]{%
silvafilho2015hybridpso}
\APACinsertmetastar {%
silvafilho2015hybridpso}%
\begin{APACrefauthors}%
Silva~Filho, T\BPBI M.%
, Pimentel, B\BPBI A.%
, de Souza, R\BPBI M\BPBI C\BPBI R.%
\BCBL {}\ \BBA {} Oliveira, A\BPBI L\BPBI I.%
\end{APACrefauthors}%
\unskip\
\newblock
\APACrefYearMonthDay{2015}{}{}.
\newblock
{\BBOQ}\APACrefatitle {Hybrid Methods for Fuzzy Clustering Based on Fuzzy
  C-means and Improved Particle Swarm Optimization} {Hybrid methods for fuzzy
  clustering based on fuzzy c-means and improved particle swarm
  optimization}.{\BBCQ}
\newblock
\APACjournalVolNumPages{Expert Systems with
  Applications}{42}{17--18}{6315--6328}.
\newblock
\begin{APACrefDOI} \doi{10.1016/j.eswa.2015.04.032} \end{APACrefDOI}
\PrintBackRefs{\CurrentBib}

\bibitem [\protect \citeauthoryear {%
Sokal%
\ \BBA {} Sneath%
}{%
Sokal%
\ \BBA {} Sneath%
}{%
{\protect \APACyear {1963}}%
}]{%
sokal1963principles}
\APACinsertmetastar {%
sokal1963principles}%
\begin{APACrefauthors}%
Sokal, R\BPBI R.%
\BCBT {}\ \BBA {} Sneath, P\BPBI H\BPBI A.%
\end{APACrefauthors}%
\unskip\
\newblock
\APACrefYear{1963}.
\newblock
\APACrefbtitle {Principles of numerical taxonomy.} {Principles of numerical
  taxonomy.}
\newblock
\APACaddressPublisher{}{San Francisco: W. H. Freeman}.
\PrintBackRefs{\CurrentBib}

\bibitem [\protect \citeauthoryear {%
Virtanen%
, Gommers%
, Oliphant%
\BCBL {}\ \protect \BOthers {.}}{%
Virtanen%
\ \protect \BOthers {.}}{%
{\protect \APACyear {2020}}%
}]{%
virtanen2020scipy}
\APACinsertmetastar {%
virtanen2020scipy}%
\begin{APACrefauthors}%
Virtanen, P.%
, Gommers, R.%
, Oliphant, T\BPBI E.%
\BCBL {}\ \BOthersPeriod {.}\end{APACrefauthors}%
\unskip\
\newblock
\APACrefYearMonthDay{2020}{}{}.
\newblock
{\BBOQ}\APACrefatitle {{SciPy} 1.0: Fundamental algorithms for scientific
  computing in {Python}} {{SciPy} 1.0: Fundamental algorithms for scientific
  computing in {Python}}.{\BBCQ}
\newblock
\APACjournalVolNumPages{Nature Methods}{17}{3}{261--272}.
\PrintBackRefs{\CurrentBib}

\bibitem [\protect \citeauthoryear {%
Wade%
\ \BBA {} Ghahramani%
}{%
Wade%
\ \BBA {} Ghahramani%
}{%
{\protect \APACyear {2018}}%
}]{%
wade2018bayesian}
\APACinsertmetastar {%
wade2018bayesian}%
\begin{APACrefauthors}%
Wade, S.%
\BCBT {}\ \BBA {} Ghahramani, Z.%
\end{APACrefauthors}%
\unskip\
\newblock
\APACrefYearMonthDay{2018}{}{}.
\newblock
{\BBOQ}\APACrefatitle {Bayesian Cluster Analysis: Point Estimation and Credible
  Balls (with Discussion)} {Bayesian cluster analysis: Point estimation and
  credible balls (with discussion)}.{\BBCQ}
\newblock
\APACjournalVolNumPages{Bayesian Analysis}{13}{2}{559--626}.
\PrintBackRefs{\CurrentBib}

\bibitem [\protect \citeauthoryear {%
Ward~Jr%
}{%
Ward~Jr%
}{%
{\protect \APACyear {1963}}%
}]{%
ward1963hierarchical}
\APACinsertmetastar {%
ward1963hierarchical}%
\begin{APACrefauthors}%
Ward~Jr, J\BPBI H.%
\end{APACrefauthors}%
\unskip\
\newblock
\APACrefYearMonthDay{1963}{}{}.
\newblock
{\BBOQ}\APACrefatitle {Hierarchical grouping to optimize an objective function}
  {Hierarchical grouping to optimize an objective function}.{\BBCQ}
\newblock
\APACjournalVolNumPages{Journal of the American statistical
  association}{58}{301}{236--244}.
\PrintBackRefs{\CurrentBib}

\bibitem [\protect \citeauthoryear {%
Wong%
, Chen%
, Huang%
\BCBL {}\ \BBA {} Wang%
}{%
Wong%
\ \protect \BOthers {.}}{%
{\protect \APACyear {2015}}%
}]{%
wong2015modifiedpso}
\APACinsertmetastar {%
wong2015modifiedpso}%
\begin{APACrefauthors}%
Wong, W\BPBI K.%
, Chen, R\BHBI B.%
, Huang, C\BHBI C.%
\BCBL {}\ \BBA {} Wang, W.%
\end{APACrefauthors}%
\unskip\
\newblock
\APACrefYearMonthDay{2015}{}{}.
\newblock
{\BBOQ}\APACrefatitle {A Modified Particle Swarm Optimization Technique for
  Finding Optimal Designs for Mixture Models} {A modified particle swarm
  optimization technique for finding optimal designs for mixture
  models}.{\BBCQ}
\newblock
\APACjournalVolNumPages{PLOS ONE}{10}{6}{e0124720}.
\newblock
\begin{APACrefDOI} \doi{10.1371/journal.pone.0124720} \end{APACrefDOI}
\PrintBackRefs{\CurrentBib}

\bibitem [\protect \citeauthoryear {%
Xie%
\ \BBA {} Beni%
}{%
Xie%
\ \BBA {} Beni%
}{%
{\protect \APACyear {1991}}%
}]{%
xie1991new}
\APACinsertmetastar {%
xie1991new}%
\begin{APACrefauthors}%
Xie, X\BPBI L.%
\BCBT {}\ \BBA {} Beni, G.%
\end{APACrefauthors}%
\unskip\
\newblock
\APACrefYearMonthDay{1991}{}{}.
\newblock
{\BBOQ}\APACrefatitle {A new fuzzy clustering validity criterion and its
  application to color image segmentation} {A new fuzzy clustering validity
  criterion and its application to color image segmentation}.{\BBCQ}
\newblock
\BIn{} \APACrefbtitle {Proceedings of the 1991 IEEE International Symposium on
  Intelligent Control} {Proceedings of the 1991 ieee international symposium on
  intelligent control}\ (\BPGS\ 463--468).
\PrintBackRefs{\CurrentBib}

\bibitem [\protect \citeauthoryear {%
Zheng%
\ \protect \BOthers {.}}{%
Zheng%
\ \protect \BOthers {.}}{%
{\protect \APACyear {2017}}%
}]{%
zheng2017massively}
\APACinsertmetastar {%
zheng2017massively}%
\begin{APACrefauthors}%
Zheng, G\BPBI X\BPBI Y.%
, Terry, J\BPBI M.%
, Belgrader, P.%
, Ryvkin, P.%
, Bent, Z\BPBI W.%
, Wilson, R.%
\BDBL {}others%
\end{APACrefauthors}%
\unskip\
\newblock
\APACrefYearMonthDay{2017}{}{}.
\newblock
{\BBOQ}\APACrefatitle {Massively parallel digital transcriptional profiling of
  single cells} {Massively parallel digital transcriptional profiling of single
  cells}.{\BBCQ}
\newblock
\APACjournalVolNumPages{Nature Communications}{8}{1}{14049}.
\newblock
\begin{APACrefDOI} \doi{10.1038/ncomms14049} \end{APACrefDOI}
\PrintBackRefs{\CurrentBib}

\bibitem [\protect \citeauthoryear {%
Zhou%
, Alexander%
\BCBL {}\ \BBA {} Lange%
}{%
Zhou%
\ \protect \BOthers {.}}{%
{\protect \APACyear {2011}}%
}]{%
zhou2011quasinewton}
\APACinsertmetastar {%
zhou2011quasinewton}%
\begin{APACrefauthors}%
Zhou, H.%
, Alexander, D.%
\BCBL {}\ \BBA {} Lange, K.%
\end{APACrefauthors}%
\unskip\
\newblock
\APACrefYearMonthDay{2011}{}{}.
\newblock
{\BBOQ}\APACrefatitle {A Quasi-Newton Acceleration for High-Dimensional
  Optimization Algorithms} {A quasi-newton acceleration for high-dimensional
  optimization algorithms}.{\BBCQ}
\newblock
\APACjournalVolNumPages{Statistics and Computing}{21}{2}{261--273}.
\newblock
\begin{APACrefDOI} \doi{10.1007/s11222-009-9166-3} \end{APACrefDOI}
\PrintBackRefs{\CurrentBib}

\end{thebibliography}


\begin{thebibliography}{}

\bibitem [\protect \citeauthoryear {%
Antoniak%
}{%
Antoniak%
}{%
{\protect \APACyear {1974}}%
}]{%
antoniak1974mixtures}
\APACinsertmetastar {%
antoniak1974mixtures}%
\begin{APACrefauthors}%
Antoniak, C\BPBI E.%
\end{APACrefauthors}%
\unskip\
\newblock
\APACrefYearMonthDay{1974}{}{}.
\newblock
{\BBOQ}\APACrefatitle {Mixtures of Dirichlet processes with applications to
  Bayesian nonparametric problems} {Mixtures of dirichlet processes with
  applications to bayesian nonparametric problems}.{\BBCQ}
\newblock
\APACjournalVolNumPages{The Annals of Statistics}{}{}{1152--1174}.
\PrintBackRefs{\CurrentBib}

\bibitem [\protect \citeauthoryear {%
Ball%
\ \BBA {} Hall%
}{%
Ball%
\ \BBA {} Hall%
}{%
{\protect \APACyear {1965}}%
}]{%
ball1965isodata}
\APACinsertmetastar {%
ball1965isodata}%
\begin{APACrefauthors}%
Ball, G\BPBI H.%
\BCBT {}\ \BBA {} Hall, D\BPBI J.%
\end{APACrefauthors}%
\unskip\
\newblock
\APACrefYearMonthDay{1965}{{\APACmonth{05}}}{}.
\newblock
\APACrefbtitle {ISODATA: A Novel Method of Data Analysis and Pattern
  Classification} {Isodata: A novel method of data analysis and pattern
  classification}\ \APACbVolEdTR {}{Technical Report\ \BNUM\ AD699616}.
\newblock
\APACaddressInstitution{Menlo Park, CA}{Stanford Research Institute}.
\newblock
\APACrefnote{SRI Project 5533}
\PrintBackRefs{\CurrentBib}

\bibitem [\protect \citeauthoryear {%
Banerjee%
, Krumpelman%
, Ghosh%
, Basu%
\BCBL {}\ \BBA {} Mooney%
}{%
Banerjee%
\ \protect \BOthers {.}}{%
{\protect \APACyear {2005}}%
}]{%
banerjee2005model}
\APACinsertmetastar {%
banerjee2005model}%
\begin{APACrefauthors}%
Banerjee, A.%
, Krumpelman, C.%
, Ghosh, J.%
, Basu, S.%
\BCBL {}\ \BBA {} Mooney, R\BPBI J.%
\end{APACrefauthors}%
\unskip\
\newblock
\APACrefYearMonthDay{2005}{}{}.
\newblock
{\BBOQ}\APACrefatitle {Model-based overlapping clustering} {Model-based
  overlapping clustering}.{\BBCQ}
\newblock
\BIn{} \APACrefbtitle {Proceedings of the eleventh ACM SIGKDD international
  conference on Knowledge discovery in data mining} {Proceedings of the
  eleventh acm sigkdd international conference on knowledge discovery in data
  mining}\ (\BPGS\ 532--537).
\PrintBackRefs{\CurrentBib}

\bibitem [\protect \citeauthoryear {%
Basu%
, Davidson%
\BCBL {}\ \BBA {} Wagstaff%
}{%
Basu%
\ \protect \BOthers {.}}{%
{\protect \APACyear {2008}}%
}]{%
basu2008constrained}
\APACinsertmetastar {%
basu2008constrained}%
\begin{APACrefauthors}%
Basu, S.%
, Davidson, I.%
\BCBL {}\ \BBA {} Wagstaff, K.%
\end{APACrefauthors}%
\unskip\
\newblock
\APACrefYear{2008}.
\newblock
\APACrefbtitle {Constrained clustering: Advances in algorithms, theory, and
  applications} {Constrained clustering: Advances in algorithms, theory, and
  applications}.
\newblock
\APACaddressPublisher{}{Chapman and Hall/CRC}.
\PrintBackRefs{\CurrentBib}

\bibitem [\protect \citeauthoryear {%
Bezdek%
, Ehrlich%
\BCBL {}\ \BBA {} Full%
}{%
Bezdek%
\ \protect \BOthers {.}}{%
{\protect \APACyear {1984}}%
}]{%
bezdek1984fcm}
\APACinsertmetastar {%
bezdek1984fcm}%
\begin{APACrefauthors}%
Bezdek, J\BPBI C.%
, Ehrlich, R.%
\BCBL {}\ \BBA {} Full, W.%
\end{APACrefauthors}%
\unskip\
\newblock
\APACrefYearMonthDay{1984}{}{}.
\newblock
{\BBOQ}\APACrefatitle {FCM: The fuzzy c-means clustering algorithm} {Fcm: The
  fuzzy c-means clustering algorithm}.{\BBCQ}
\newblock
\APACjournalVolNumPages{Computers \& geosciences}{10}{2-3}{191--203}.
\PrintBackRefs{\CurrentBib}

\bibitem [\protect \citeauthoryear {%
Campello%
, Moulavi%
\BCBL {}\ \BBA {} Sander%
}{%
Campello%
\ \protect \BOthers {.}}{%
{\protect \APACyear {2013}}%
}]{%
campello2013density}
\APACinsertmetastar {%
campello2013density}%
\begin{APACrefauthors}%
Campello, R\BPBI J\BPBI G\BPBI B.%
, Moulavi, D.%
\BCBL {}\ \BBA {} Sander, J.%
\end{APACrefauthors}%
\unskip\
\newblock
\APACrefYearMonthDay{2013}{}{}.
\newblock
{\BBOQ}\APACrefatitle {Density-based clustering based on hierarchical density
  estimates} {Density-based clustering based on hierarchical density
  estimates}.{\BBCQ}
\newblock
\BIn{} \APACrefbtitle {Advances in Knowledge Discovery and Data Mining (PAKDD
  2013)} {Advances in knowledge discovery and data mining (pakdd 2013)}\
  (\BVOL\ 7819, \BPGS\ 160--172).
\newblock
\APACaddressPublisher{}{Springer}.
\PrintBackRefs{\CurrentBib}

\bibitem [\protect \citeauthoryear {%
{Cancer Genome Atlas Network}%
}{%
{Cancer Genome Atlas Network}%
}{%
{\protect \APACyear {2012}}%
}]{%
tcga2012}
\APACinsertmetastar {%
tcga2012}%
\begin{APACrefauthors}%
{Cancer Genome Atlas Network}.%
\end{APACrefauthors}%
\unskip\
\newblock
\APACrefYearMonthDay{2012}{}{}.
\newblock
{\BBOQ}\APACrefatitle {Comprehensive molecular portraits of human breast
  tumours} {Comprehensive molecular portraits of human breast tumours}.{\BBCQ}
\newblock
\APACjournalVolNumPages{Nature}{490}{}{61--70}.
\newblock
\begin{APACrefDOI} \doi{10.1038/nature11412} \end{APACrefDOI}
\PrintBackRefs{\CurrentBib}

\bibitem [\protect \citeauthoryear {%
Dave%
}{%
Dave%
}{%
{\protect \APACyear {1991}}%
}]{%
dave1991characterization}
\APACinsertmetastar {%
dave1991characterization}%
\begin{APACrefauthors}%
Dave, R\BPBI N.%
\end{APACrefauthors}%
\unskip\
\newblock
\APACrefYearMonthDay{1991}{}{}.
\newblock
{\BBOQ}\APACrefatitle {Characterization and detection of noise in clustering}
  {Characterization and detection of noise in clustering}.{\BBCQ}
\newblock
\APACjournalVolNumPages{Pattern Recognition Letters}{12}{11}{657--664}.
\PrintBackRefs{\CurrentBib}

\bibitem [\protect \citeauthoryear {%
Dempster%
, Laird%
\BCBL {}\ \BBA {} Rubin%
}{%
Dempster%
\ \protect \BOthers {.}}{%
{\protect \APACyear {1977}}%
}]{%
dempster1977maximum}
\APACinsertmetastar {%
dempster1977maximum}%
\begin{APACrefauthors}%
Dempster, A\BPBI P.%
, Laird, N\BPBI M.%
\BCBL {}\ \BBA {} Rubin, D\BPBI B.%
\end{APACrefauthors}%
\unskip\
\newblock
\APACrefYearMonthDay{1977}{}{}.
\newblock
{\BBOQ}\APACrefatitle {Maximum likelihood from incomplete data via the EM
  algorithm} {Maximum likelihood from incomplete data via the em
  algorithm}.{\BBCQ}
\newblock
\APACjournalVolNumPages{Journal of the royal statistical society: series B
  (methodological)}{39}{1}{1--22}.
\PrintBackRefs{\CurrentBib}

\bibitem [\protect \citeauthoryear {%
Duda%
, Hart%
\BCBL {}\ \protect \BOthers {.}}{%
Duda%
\ \protect \BOthers {.}}{%
{\protect \APACyear {2006}}%
}]{%
duda2006pattern}
\APACinsertmetastar {%
duda2006pattern}%
\begin{APACrefauthors}%
Duda, R\BPBI O.%
, Hart, P\BPBI E.%
\BCBL {}\ \BOthersPeriod {.}\end{APACrefauthors}%
\unskip\
\newblock
\APACrefYear{2006}.
\newblock
\APACrefbtitle {Pattern classification} {Pattern classification}.
\newblock
\APACaddressPublisher{}{John Wiley \& Sons}.
\PrintBackRefs{\CurrentBib}

\bibitem [\protect \citeauthoryear {%
Fan%
, Zhen%
\BCBL {}\ \BBA {} Xie%
}{%
Fan%
\ \protect \BOthers {.}}{%
{\protect \APACyear {2003}}%
}]{%
fan2003suppressed}
\APACinsertmetastar {%
fan2003suppressed}%
\begin{APACrefauthors}%
Fan, J\BHBI L.%
, Zhen, W\BHBI Z.%
\BCBL {}\ \BBA {} Xie, W\BHBI X.%
\end{APACrefauthors}%
\unskip\
\newblock
\APACrefYearMonthDay{2003}{}{}.
\newblock
{\BBOQ}\APACrefatitle {Suppressed fuzzy c-means clustering algorithm}
  {Suppressed fuzzy c-means clustering algorithm}.{\BBCQ}
\newblock
\APACjournalVolNumPages{Pattern Recognition Letters}{24}{9-10}{1607--1612}.
\PrintBackRefs{\CurrentBib}

\bibitem [\protect \citeauthoryear {%
Gao%
, Bien%
\BCBL {}\ \BBA {} Witten%
}{%
Gao%
\ \protect \BOthers {.}}{%
{\protect \APACyear {2024}}%
}]{%
gao2024selective}
\APACinsertmetastar {%
gao2024selective}%
\begin{APACrefauthors}%
Gao, L\BPBI L.%
, Bien, J.%
\BCBL {}\ \BBA {} Witten, D.%
\end{APACrefauthors}%
\unskip\
\newblock
\APACrefYearMonthDay{2024}{}{}.
\newblock
{\BBOQ}\APACrefatitle {Selective inference for hierarchical clustering}
  {Selective inference for hierarchical clustering}.{\BBCQ}
\newblock
\APACjournalVolNumPages{Journal of the American Statistical
  Association}{119}{545}{332--342}.
\PrintBackRefs{\CurrentBib}

\bibitem [\protect \citeauthoryear {%
Gath%
\ \BBA {} Geva%
}{%
Gath%
\ \BBA {} Geva%
}{%
{\protect \APACyear {1989}}%
}]{%
gath1989unsupervised}
\APACinsertmetastar {%
gath1989unsupervised}%
\begin{APACrefauthors}%
Gath, I.%
\BCBT {}\ \BBA {} Geva, A\BPBI B.%
\end{APACrefauthors}%
\unskip\
\newblock
\APACrefYearMonthDay{1989}{}{}.
\newblock
{\BBOQ}\APACrefatitle {Unsupervised optimal fuzzy clustering} {Unsupervised
  optimal fuzzy clustering}.{\BBCQ}
\newblock
\APACjournalVolNumPages{IEEE Transactions on Pattern Analysis and Machine
  Intelligence}{11}{7}{773--780}.
\newblock
\begin{APACrefDOI} \doi{10.1109/34.192473} \end{APACrefDOI}
\PrintBackRefs{\CurrentBib}

\bibitem [\protect \citeauthoryear {%
Griffiths%
\ \BBA {} Ghahramani%
}{%
Griffiths%
\ \BBA {} Ghahramani%
}{%
{\protect \APACyear {2011}}%
}]{%
griffiths2011indian}
\APACinsertmetastar {%
griffiths2011indian}%
\begin{APACrefauthors}%
Griffiths, T\BPBI L.%
\BCBT {}\ \BBA {} Ghahramani, Z.%
\end{APACrefauthors}%
\unskip\
\newblock
\APACrefYearMonthDay{2011}{}{}.
\newblock
{\BBOQ}\APACrefatitle {The Indian buffet process: An introduction and review}
  {The indian buffet process: An introduction and review}.{\BBCQ}
\newblock
\APACjournalVolNumPages{Journal of Machine Learning
  Research}{12}{}{1185--1224}.
\PrintBackRefs{\CurrentBib}

\bibitem [\protect \citeauthoryear {%
Gustafson%
\ \BBA {} Kessel%
}{%
Gustafson%
\ \BBA {} Kessel%
}{%
{\protect \APACyear {1979}}%
}]{%
gustafson1979fuzzy}
\APACinsertmetastar {%
gustafson1979fuzzy}%
\begin{APACrefauthors}%
Gustafson, D\BPBI E.%
\BCBT {}\ \BBA {} Kessel, W\BPBI C.%
\end{APACrefauthors}%
\unskip\
\newblock
\APACrefYearMonthDay{1979}{}{}.
\newblock
{\BBOQ}\APACrefatitle {Fuzzy clustering with a fuzzy covariance matrix} {Fuzzy
  clustering with a fuzzy covariance matrix}.{\BBCQ}
\newblock
\BIn{} \APACrefbtitle {1978 IEEE conference on decision and control including
  the 17th symposium on adaptive processes} {1978 ieee conference on decision
  and control including the 17th symposium on adaptive processes}\ (\BPGS\
  761--766).
\PrintBackRefs{\CurrentBib}

\bibitem [\protect \citeauthoryear {%
Hathaway%
\ \BBA {} Bezdek%
}{%
Hathaway%
\ \BBA {} Bezdek%
}{%
{\protect \APACyear {2006}}%
}]{%
hathaway2006extending}
\APACinsertmetastar {%
hathaway2006extending}%
\begin{APACrefauthors}%
Hathaway, R\BPBI J.%
\BCBT {}\ \BBA {} Bezdek, J\BPBI C.%
\end{APACrefauthors}%
\unskip\
\newblock
\APACrefYearMonthDay{2006}{}{}.
\newblock
{\BBOQ}\APACrefatitle {Extending fuzzy and probabilistic clustering to very
  large data sets} {Extending fuzzy and probabilistic clustering to very large
  data sets}.{\BBCQ}
\newblock
\APACjournalVolNumPages{Computational Statistics \& Data
  Analysis}{51}{1}{215--234}.
\PrintBackRefs{\CurrentBib}

\bibitem [\protect \citeauthoryear {%
He%
}{%
He%
}{%
{\protect \APACyear {2024}}%
}]{%
he2024imbalanced}
\APACinsertmetastar {%
he2024imbalanced}%
\begin{APACrefauthors}%
He, Y.%
\end{APACrefauthors}%
\unskip\
\newblock
\APACrefYearMonthDay{2024}{}{}.
\newblock
{\BBOQ}\APACrefatitle {Imbalanced Data Clustering Using Equilibrium K-Means}
  {Imbalanced data clustering using equilibrium k-means}.{\BBCQ}
\newblock
\APACjournalVolNumPages{arXiv preprint arXiv:2402.14490}{}{}{}.
\newblock
\begin{APACrefURL} \url{https://arxiv.org/abs/2402.14490} \end{APACrefURL}
\PrintBackRefs{\CurrentBib}

\bibitem [\protect \citeauthoryear {%
Huang%
, Ng%
, Rong%
\BCBL {}\ \BBA {} Li%
}{%
Huang%
\ \protect \BOthers {.}}{%
{\protect \APACyear {2005}}%
}]{%
huang2005automated}
\APACinsertmetastar {%
huang2005automated}%
\begin{APACrefauthors}%
Huang, J\BPBI Z.%
, Ng, M\BPBI K.%
, Rong, H.%
\BCBL {}\ \BBA {} Li, Z.%
\end{APACrefauthors}%
\unskip\
\newblock
\APACrefYearMonthDay{2005}{}{}.
\newblock
{\BBOQ}\APACrefatitle {Automated variable weighting in $k$-means type
  clustering} {Automated variable weighting in $k$-means type
  clustering}.{\BBCQ}
\newblock
\APACjournalVolNumPages{IEEE Transactions on Pattern Analysis and Machine
  Intelligence}{27}{5}{657--668}.
\newblock
\begin{APACrefDOI} \doi{10.1109/TPAMI.2005.95} \end{APACrefDOI}
\PrintBackRefs{\CurrentBib}

\bibitem [\protect \citeauthoryear {%
Jain%
}{%
Jain%
}{%
{\protect \APACyear {2010}}%
}]{%
jain2010data}
\APACinsertmetastar {%
jain2010data}%
\begin{APACrefauthors}%
Jain, A\BPBI K.%
\end{APACrefauthors}%
\unskip\
\newblock
\APACrefYearMonthDay{2010}{}{}.
\newblock
{\BBOQ}\APACrefatitle {Data clustering: 50 years beyond K-means} {Data
  clustering: 50 years beyond k-means}.{\BBCQ}
\newblock
\APACjournalVolNumPages{Pattern recognition letters}{31}{8}{651--666}.
\PrintBackRefs{\CurrentBib}

\bibitem [\protect \citeauthoryear {%
Jain%
\ \BBA {} Dubes%
}{%
Jain%
\ \BBA {} Dubes%
}{%
{\protect \APACyear {1988}}%
}]{%
jain1988algorithms}
\APACinsertmetastar {%
jain1988algorithms}%
\begin{APACrefauthors}%
Jain, A\BPBI K.%
\BCBT {}\ \BBA {} Dubes, R\BPBI C.%
\end{APACrefauthors}%
\unskip\
\newblock
\APACrefYear{1988}.
\newblock
\APACrefbtitle {Algorithms for clustering data} {Algorithms for clustering
  data}.
\newblock
\APACaddressPublisher{}{Prentice-Hall, Inc.}
\PrintBackRefs{\CurrentBib}

\bibitem [\protect \citeauthoryear {%
Ji%
, Sun%
\BCBL {}\ \BBA {} Xia%
}{%
Ji%
\ \protect \BOthers {.}}{%
{\protect \APACyear {2011}}%
}]{%
ji2011modified}
\APACinsertmetastar {%
ji2011modified}%
\begin{APACrefauthors}%
Ji, Z\BHBI X.%
, Sun, Q\BHBI S.%
\BCBL {}\ \BBA {} Xia, D\BHBI S.%
\end{APACrefauthors}%
\unskip\
\newblock
\APACrefYearMonthDay{2011}{}{}.
\newblock
{\BBOQ}\APACrefatitle {A modified possibilistic fuzzy c-means clustering
  algorithm for bias field estimation and segmentation of brain MR image} {A
  modified possibilistic fuzzy c-means clustering algorithm for bias field
  estimation and segmentation of brain mr image}.{\BBCQ}
\newblock
\APACjournalVolNumPages{Computerized Medical Imaging and
  Graphics}{35}{5}{383--397}.
\PrintBackRefs{\CurrentBib}

\bibitem [\protect \citeauthoryear {%
Jing%
, Ng%
\BCBL {}\ \BBA {} Huang%
}{%
Jing%
\ \protect \BOthers {.}}{%
{\protect \APACyear {2007}}%
}]{%
jing2007entropy}
\APACinsertmetastar {%
jing2007entropy}%
\begin{APACrefauthors}%
Jing, L.%
, Ng, M\BPBI K.%
\BCBL {}\ \BBA {} Huang, J\BPBI Z.%
\end{APACrefauthors}%
\unskip\
\newblock
\APACrefYearMonthDay{2007}{}{}.
\newblock
{\BBOQ}\APACrefatitle {An entropy weighting $k$-means algorithm for subspace
  clustering of high-dimensional sparse data} {An entropy weighting $k$-means
  algorithm for subspace clustering of high-dimensional sparse data}.{\BBCQ}
\newblock
\APACjournalVolNumPages{IEEE Transactions on Knowledge and Data
  Engineering}{19}{8}{1026--1041}.
\newblock
\begin{APACrefDOI} \doi{10.1109/TKDE.2007.1048} \end{APACrefDOI}
\PrintBackRefs{\CurrentBib}

\bibitem [\protect \citeauthoryear {%
Klawonn%
\ \BBA {} H{\"o}ppner%
}{%
Klawonn%
\ \BBA {} H{\"o}ppner%
}{%
{\protect \APACyear {2003}}%
}]{%
klawonn2003alternative}
\APACinsertmetastar {%
klawonn2003alternative}%
\begin{APACrefauthors}%
Klawonn, F.%
\BCBT {}\ \BBA {} H{\"o}ppner, F.%
\end{APACrefauthors}%
\unskip\
\newblock
\APACrefYearMonthDay{2003}{}{}.
\newblock
{\BBOQ}\APACrefatitle {An alternative approach to the fuzzifier in fuzzy
  clustering to obtain better clustering results} {An alternative approach to
  the fuzzifier in fuzzy clustering to obtain better clustering
  results}.{\BBCQ}
\newblock
\BIn{} \APACrefbtitle {Proceedings of Eusflat Conference} {Proceedings of
  eusflat conference}\ (\BPGS\ 730--734).
\PrintBackRefs{\CurrentBib}

\bibitem [\protect \citeauthoryear {%
Krishnapuram%
\ \BBA {} Keller%
}{%
Krishnapuram%
\ \BBA {} Keller%
}{%
{\protect \APACyear {1993}}%
}]{%
krishnapuram1993possibilistic}
\APACinsertmetastar {%
krishnapuram1993possibilistic}%
\begin{APACrefauthors}%
Krishnapuram, R.%
\BCBT {}\ \BBA {} Keller, J\BPBI M.%
\end{APACrefauthors}%
\unskip\
\newblock
\APACrefYearMonthDay{1993}{}{}.
\newblock
{\BBOQ}\APACrefatitle {A possibilistic approach to clustering} {A possibilistic
  approach to clustering}.{\BBCQ}
\newblock
\APACjournalVolNumPages{IEEE Transactions on Fuzzy Systems}{1}{2}{98--110}.
\PrintBackRefs{\CurrentBib}

\bibitem [\protect \citeauthoryear {%
Lin%
, Huang%
, Kuo%
\BCBL {}\ \BBA {} Lai%
}{%
Lin%
\ \protect \BOthers {.}}{%
{\protect \APACyear {2014}}%
}]{%
lin2014size}
\APACinsertmetastar {%
lin2014size}%
\begin{APACrefauthors}%
Lin, P\BHBI L.%
, Huang, P\BHBI W.%
, Kuo, C\BPBI H.%
\BCBL {}\ \BBA {} Lai, Y\BPBI H.%
\end{APACrefauthors}%
\unskip\
\newblock
\APACrefYearMonthDay{2014}{}{}.
\newblock
{\BBOQ}\APACrefatitle {A size-insensitive integrity-based fuzzy c-means method
  for data clustering} {A size-insensitive integrity-based fuzzy c-means method
  for data clustering}.{\BBCQ}
\newblock
\APACjournalVolNumPages{Pattern Recognition}{47}{5}{2042--2056}.
\PrintBackRefs{\CurrentBib}

\bibitem [\protect \citeauthoryear {%
McInnes%
, Healy%
\BCBL {}\ \BBA {} Melville%
}{%
McInnes%
\ \protect \BOthers {.}}{%
{\protect \APACyear {2018}}%
}]{%
mcinnes2018umap}
\APACinsertmetastar {%
mcinnes2018umap}%
\begin{APACrefauthors}%
McInnes, L.%
, Healy, J.%
\BCBL {}\ \BBA {} Melville, J.%
\end{APACrefauthors}%
\unskip\
\newblock
\APACrefYearMonthDay{2018}{}{}.
\newblock
{\BBOQ}\APACrefatitle {{UMAP}: Uniform manifold approximation and projection
  for dimension reduction} {{UMAP}: Uniform manifold approximation and
  projection for dimension reduction}.{\BBCQ}
\newblock
\APACjournalVolNumPages{arXiv preprint arXiv:1802.03426}{}{}{}.
\PrintBackRefs{\CurrentBib}

\bibitem [\protect \citeauthoryear {%
McLachlan%
\ \BBA {} Peel%
}{%
McLachlan%
\ \BBA {} Peel%
}{%
{\protect \APACyear {2000}}%
}]{%
mclachlan2000finite}
\APACinsertmetastar {%
mclachlan2000finite}%
\begin{APACrefauthors}%
McLachlan, G\BPBI J.%
\BCBT {}\ \BBA {} Peel, D.%
\end{APACrefauthors}%
\unskip\
\newblock
\APACrefYear{2000}.
\newblock
\APACrefbtitle {Finite mixture models} {Finite mixture models}.
\newblock
\APACaddressPublisher{}{John Wiley \& Sons}.
\PrintBackRefs{\CurrentBib}

\bibitem [\protect \citeauthoryear {%
Meil{\u{a}}%
}{%
Meil{\u{a}}%
}{%
{\protect \APACyear {2006}}%
}]{%
meilua2006uniqueness}
\APACinsertmetastar {%
meilua2006uniqueness}%
\begin{APACrefauthors}%
Meil{\u{a}}, M.%
\end{APACrefauthors}%
\unskip\
\newblock
\APACrefYearMonthDay{2006}{}{}.
\newblock
{\BBOQ}\APACrefatitle {The uniqueness of a good optimum for k-means} {The
  uniqueness of a good optimum for k-means}.{\BBCQ}
\newblock
\BIn{} \APACrefbtitle {Proceedings of the 23rd international conference on
  Machine learning} {Proceedings of the 23rd international conference on
  machine learning}\ (\BPGS\ 625--632).
\PrintBackRefs{\CurrentBib}

\bibitem [\protect \citeauthoryear {%
Miyamoto%
\ \BBA {} Mukaidono%
}{%
Miyamoto%
\ \BBA {} Mukaidono%
}{%
{\protect \APACyear {1997}}%
}]{%
miyamoto1997fuzzy}
\APACinsertmetastar {%
miyamoto1997fuzzy}%
\begin{APACrefauthors}%
Miyamoto, S.%
\BCBT {}\ \BBA {} Mukaidono, M.%
\end{APACrefauthors}%
\unskip\
\newblock
\APACrefYearMonthDay{1997}{}{}.
\newblock
{\BBOQ}\APACrefatitle {Fuzzy $c$-means as a regularization and maximum entropy
  approach} {Fuzzy $c$-means as a regularization and maximum entropy
  approach}.{\BBCQ}
\newblock
\BIn{} \APACrefbtitle {Proceedings of the 7th International Fuzzy Systems
  Association World Congress (IFSA)} {Proceedings of the 7th international
  fuzzy systems association world congress (ifsa)}\ (\BPGS\ 86--92).
\PrintBackRefs{\CurrentBib}

\bibitem [\protect \citeauthoryear {%
Ng%
, Jordan%
\BCBL {}\ \BBA {} Weiss%
}{%
Ng%
\ \protect \BOthers {.}}{%
{\protect \APACyear {2002}}%
}]{%
ng2002spectral}
\APACinsertmetastar {%
ng2002spectral}%
\begin{APACrefauthors}%
Ng, A\BPBI Y.%
, Jordan, M\BPBI I.%
\BCBL {}\ \BBA {} Weiss, Y.%
\end{APACrefauthors}%
\unskip\
\newblock
\APACrefYearMonthDay{2002}{}{}.
\newblock
{\BBOQ}\APACrefatitle {On spectral clustering: Analysis and an algorithm} {On
  spectral clustering: Analysis and an algorithm}.{\BBCQ}
\newblock
\BIn{} \APACrefbtitle {Advances in Neural Information Processing Systems 14}
  {Advances in neural information processing systems 14}\ (\BPGS\ 849--856).
\PrintBackRefs{\CurrentBib}

\bibitem [\protect \citeauthoryear {%
Noordam%
, van~den Broek%
\BCBL {}\ \BBA {} Buydens%
}{%
Noordam%
\ \protect \BOthers {.}}{%
{\protect \APACyear {2002}}%
}]{%
noordam2002multivariate}
\APACinsertmetastar {%
noordam2002multivariate}%
\begin{APACrefauthors}%
Noordam, J\BPBI C.%
, van~den Broek, W\BPBI H\BPBI A\BPBI M.%
\BCBL {}\ \BBA {} Buydens, L\BPBI M\BPBI C.%
\end{APACrefauthors}%
\unskip\
\newblock
\APACrefYearMonthDay{2002}{}{}.
\newblock
{\BBOQ}\APACrefatitle {Multivariate image segmentation with cluster size
  insensitive fuzzy {C}-means} {Multivariate image segmentation with cluster
  size insensitive fuzzy {C}-means}.{\BBCQ}
\newblock
\APACjournalVolNumPages{Chemometrics and Intelligent Laboratory
  Systems}{64}{1}{65--78}.
\PrintBackRefs{\CurrentBib}

\bibitem [\protect \citeauthoryear {%
Parker%
\ \protect \BOthers {.}}{%
Parker%
\ \protect \BOthers {.}}{%
{\protect \APACyear {2009}}%
}]{%
Parker2009}
\APACinsertmetastar {%
Parker2009}%
\begin{APACrefauthors}%
Parker, J\BPBI S.%
, Mullins, M.%
, Cheang, M\BPBI C\BPBI U.%
, Leung, S.%
, Voduc, D.%
, Vickery, T.%
\BDBL {}Bernard, P\BPBI S.%
\end{APACrefauthors}%
\unskip\
\newblock
\APACrefYearMonthDay{2009}{}{}.
\newblock
{\BBOQ}\APACrefatitle {Supervised risk predictor of breast cancer based on
  intrinsic subtypes} {Supervised risk predictor of breast cancer based on
  intrinsic subtypes}.{\BBCQ}
\newblock
\APACjournalVolNumPages{Journal of Clinical Oncology}{27}{8}{1160--1167}.
\newblock
\begin{APACrefDOI} \doi{10.1200/JCO.2008.18.1370} \end{APACrefDOI}
\PrintBackRefs{\CurrentBib}

\bibitem [\protect \citeauthoryear {%
Pedrycz%
}{%
Pedrycz%
}{%
{\protect \APACyear {2005}}%
}]{%
pedrycz2005knowledge}
\APACinsertmetastar {%
pedrycz2005knowledge}%
\begin{APACrefauthors}%
Pedrycz, W.%
\end{APACrefauthors}%
\unskip\
\newblock
\APACrefYear{2005}.
\newblock
\APACrefbtitle {Knowledge-Based Clustering: From Data to Information Granules}
  {Knowledge-based clustering: From data to information granules}.
\newblock
\APACaddressPublisher{}{Wiley}.
\PrintBackRefs{\CurrentBib}

\bibitem [\protect \citeauthoryear {%
Pham%
\ \BBA {} Prince%
}{%
Pham%
\ \BBA {} Prince%
}{%
{\protect \APACyear {1999}}%
}]{%
pham1999adaptive}
\APACinsertmetastar {%
pham1999adaptive}%
\begin{APACrefauthors}%
Pham, D\BPBI L.%
\BCBT {}\ \BBA {} Prince, J\BPBI L.%
\end{APACrefauthors}%
\unskip\
\newblock
\APACrefYearMonthDay{1999}{}{}.
\newblock
{\BBOQ}\APACrefatitle {An adaptive fuzzy C-means algorithm for image
  segmentation in the presence of intensity inhomogeneities} {An adaptive fuzzy
  c-means algorithm for image segmentation in the presence of intensity
  inhomogeneities}.{\BBCQ}
\newblock
\APACjournalVolNumPages{Pattern recognition letters}{20}{1}{57--68}.
\PrintBackRefs{\CurrentBib}

\bibitem [\protect \citeauthoryear {%
S{\o}rlie%
\ \protect \BOthers {.}}{%
S{\o}rlie%
\ \protect \BOthers {.}}{%
{\protect \APACyear {2003}}%
}]{%
Sorlie2003}
\APACinsertmetastar {%
Sorlie2003}%
\begin{APACrefauthors}%
S{\o}rlie, T.%
, Tibshirani, R.%
, Parker, J.%
, Hastie, T.%
, Marron, J\BPBI S.%
, Nobel, A.%
\BDBL {}Botstein, D.%
\end{APACrefauthors}%
\unskip\
\newblock
\APACrefYearMonthDay{2003}{}{}.
\newblock
{\BBOQ}\APACrefatitle {Repeated observation of breast tumor subtypes in
  independent gene expression data sets} {Repeated observation of breast tumor
  subtypes in independent gene expression data sets}.{\BBCQ}
\newblock
\APACjournalVolNumPages{Proceedings of the National Academy of
  Sciences}{100}{14}{8418--8423}.
\newblock
\begin{APACrefDOI} \doi{10.1073/pnas.0932692100} \end{APACrefDOI}
\PrintBackRefs{\CurrentBib}

\bibitem [\protect \citeauthoryear {%
Taylor%
\ \BBA {} Tibshirani%
}{%
Taylor%
\ \BBA {} Tibshirani%
}{%
{\protect \APACyear {2015}}%
}]{%
taylor2015statistical}
\APACinsertmetastar {%
taylor2015statistical}%
\begin{APACrefauthors}%
Taylor, J.%
\BCBT {}\ \BBA {} Tibshirani, R\BPBI J.%
\end{APACrefauthors}%
\unskip\
\newblock
\APACrefYearMonthDay{2015}{}{}.
\newblock
{\BBOQ}\APACrefatitle {Statistical learning and selective inference}
  {Statistical learning and selective inference}.{\BBCQ}
\newblock
\APACjournalVolNumPages{Proceedings of the National Academy of
  Sciences}{112}{25}{7629--7634}.
\PrintBackRefs{\CurrentBib}

\bibitem [\protect \citeauthoryear {%
van~der Maaten%
\ \BBA {} Hinton%
}{%
van~der Maaten%
\ \BBA {} Hinton%
}{%
{\protect \APACyear {2008}}%
}]{%
vanDerMaaten2008}
\APACinsertmetastar {%
vanDerMaaten2008}%
\begin{APACrefauthors}%
van~der Maaten, L.%
\BCBT {}\ \BBA {} Hinton, G.%
\end{APACrefauthors}%
\unskip\
\newblock
\APACrefYearMonthDay{2008}{}{}.
\newblock
{\BBOQ}\APACrefatitle {Visualizing data using {t-SNE}} {Visualizing data using
  {t-SNE}}.{\BBCQ}
\newblock
\APACjournalVolNumPages{Journal of Machine Learning Research}{9}{}{2579--2605}.
\PrintBackRefs{\CurrentBib}

\bibitem [\protect \citeauthoryear {%
Winkler%
, Klawonn%
\BCBL {}\ \BBA {} Kruse%
}{%
Winkler%
\ \protect \BOthers {.}}{%
{\protect \APACyear {2011}}%
}]{%
winkler2011fuzzy}
\APACinsertmetastar {%
winkler2011fuzzy}%
\begin{APACrefauthors}%
Winkler, R.%
, Klawonn, F.%
\BCBL {}\ \BBA {} Kruse, R.%
\end{APACrefauthors}%
\unskip\
\newblock
\APACrefYearMonthDay{2011}{}{}.
\newblock
{\BBOQ}\APACrefatitle {Fuzzy clustering with polynomial fuzzifier function in
  connection with $m$-estimators} {Fuzzy clustering with polynomial fuzzifier
  function in connection with $m$-estimators}.{\BBCQ}
\newblock
\APACjournalVolNumPages{Applied and Computational Mathematics}{10}{}{146--163}.
\PrintBackRefs{\CurrentBib}

\bibitem [\protect \citeauthoryear {%
J.~Xie%
, Kelley%
\BCBL {}\ \BBA {} Szymanski%
}{%
J.~Xie%
\ \protect \BOthers {.}}{%
{\protect \APACyear {2013}}%
}]{%
xie2013overlapping}
\APACinsertmetastar {%
xie2013overlapping}%
\begin{APACrefauthors}%
Xie, J.%
, Kelley, S.%
\BCBL {}\ \BBA {} Szymanski, B\BPBI K.%
\end{APACrefauthors}%
\unskip\
\newblock
\APACrefYearMonthDay{2013}{}{}.
\newblock
{\BBOQ}\APACrefatitle {Overlapping community detection in networks: The
  state-of-the-art and comparative study} {Overlapping community detection in
  networks: The state-of-the-art and comparative study}.{\BBCQ}
\newblock
\APACjournalVolNumPages{Acm computing surveys (csur)}{45}{4}{1--35}.
\PrintBackRefs{\CurrentBib}

\bibitem [\protect \citeauthoryear {%
X\BPBI L.~Xie%
\ \BBA {} Beni%
}{%
X\BPBI L.~Xie%
\ \BBA {} Beni%
}{%
{\protect \APACyear {1991}}%
}]{%
xie1991new}
\APACinsertmetastar {%
xie1991new}%
\begin{APACrefauthors}%
Xie, X\BPBI L.%
\BCBT {}\ \BBA {} Beni, G.%
\end{APACrefauthors}%
\unskip\
\newblock
\APACrefYearMonthDay{1991}{}{}.
\newblock
{\BBOQ}\APACrefatitle {A new fuzzy clustering validity criterion and its
  application to color image segmentation} {A new fuzzy clustering validity
  criterion and its application to color image segmentation}.{\BBCQ}
\newblock
\BIn{} \APACrefbtitle {Proceedings of the 1991 IEEE International Symposium on
  Intelligent Control} {Proceedings of the 1991 ieee international symposium on
  intelligent control}\ (\BPGS\ 463--468).
\PrintBackRefs{\CurrentBib}

\bibitem [\protect \citeauthoryear {%
Yun%
\ \BBA {} Foygel~Barber%
}{%
Yun%
\ \BBA {} Foygel~Barber%
}{%
{\protect \APACyear {2023}}%
}]{%
yun2023selective}
\APACinsertmetastar {%
yun2023selective}%
\begin{APACrefauthors}%
Yun, Y\BHBI J.%
\BCBT {}\ \BBA {} Foygel~Barber, R.%
\end{APACrefauthors}%
\unskip\
\newblock
\APACrefYearMonthDay{2023}{}{}.
\newblock
{\BBOQ}\APACrefatitle {Selective inference for clustering with unknown
  variance} {Selective inference for clustering with unknown variance}.{\BBCQ}
\newblock
\APACjournalVolNumPages{Electronic Journal of Statistics}{17}{2}{1923--1946}.
\PrintBackRefs{\CurrentBib}

\bibitem [\protect \citeauthoryear {%
Zhang%
\ \BBA {} Chen%
}{%
Zhang%
\ \BBA {} Chen%
}{%
{\protect \APACyear {2003}}%
}]{%
zhang2003clustering}
\APACinsertmetastar {%
zhang2003clustering}%
\begin{APACrefauthors}%
Zhang, D\BHBI Q.%
\BCBT {}\ \BBA {} Chen, S\BHBI C.%
\end{APACrefauthors}%
\unskip\
\newblock
\APACrefYearMonthDay{2003}{}{}.
\newblock
{\BBOQ}\APACrefatitle {Clustering incomplete data using kernel-based fuzzy
  c-means algorithm} {Clustering incomplete data using kernel-based fuzzy
  c-means algorithm}.{\BBCQ}
\newblock
\APACjournalVolNumPages{Neural processing letters}{18}{3}{155--162}.
\PrintBackRefs{\CurrentBib}

\bibitem [\protect \citeauthoryear {%
Zhang%
\ \BBA {} Chen%
}{%
Zhang%
\ \BBA {} Chen%
}{%
{\protect \APACyear {2004}}%
}]{%
zhang2004novel}
\APACinsertmetastar {%
zhang2004novel}%
\begin{APACrefauthors}%
Zhang, D\BHBI Q.%
\BCBT {}\ \BBA {} Chen, S\BHBI C.%
\end{APACrefauthors}%
\unskip\
\newblock
\APACrefYearMonthDay{2004}{}{}.
\newblock
{\BBOQ}\APACrefatitle {A novel kernelized fuzzy c-means algorithm with
  application in medical image segmentation} {A novel kernelized fuzzy c-means
  algorithm with application in medical image segmentation}.{\BBCQ}
\newblock
\APACjournalVolNumPages{Artificial intelligence in medicine}{32}{1}{37--50}.
\PrintBackRefs{\CurrentBib}

\bibitem [\protect \citeauthoryear {%
Zhu%
, Gai%
\BCBL {}\ \BBA {} Zhang%
}{%
Zhu%
\ \protect \BOthers {.}}{%
{\protect \APACyear {2026}}%
}]{%
zhu2026functional}
\APACinsertmetastar {%
zhu2026functional}%
\begin{APACrefauthors}%
Zhu, Z.%
, Gai, X.%
\BCBL {}\ \BBA {} Zhang, A\BPBI R.%
\end{APACrefauthors}%
\unskip\
\newblock
\APACrefYearMonthDay{2026}{}{}.
\newblock
{\BBOQ}\APACrefatitle {Functional post-clustering selective inference with
  applications to ehr data analysis} {Functional post-clustering selective
  inference with applications to ehr data analysis}.{\BBCQ}
\newblock
\APACjournalVolNumPages{Electronic Journal of Statistics}{20}{2}{4134--4178}.
\PrintBackRefs{\CurrentBib}

\end{thebibliography}
		
		\clearpage
		\appendix
		
		\begin{center}
			\LARGE \textbf{Supplementary Materials for}\\[0.5em]
			\textbf{``Statistical Inference for Fuzzy Clustering"}\\[1.5em]
		\end{center}
		
		\section{Proof for Consistency}\label{sec:pftheory}
\begingroup
We work on the compact set \eqref{eq:regular-parameter-set}, retaining all label permutations. The distance between two label orbits is $d_{\rm lab}(\theta,\theta')=\min_\pi\|\pi\theta-\theta'\|$. No globally ordered parameter chart is needed.

\begin{theorem}[Compact likelihood consistency]\label{thm:wald}
Suppose the parameter space is compact, the log-density is pointwise continuous with an integrable envelope under the true distribution, and the model is identifiable modulo a finite label group. For i.i.d. observations from the model, every measurable exact likelihood maximizer is strongly consistent in label-orbit distance.
\end{theorem}
The proof is the uniform-law and separated-population-maximum argument given below. We verify each ingredient for the WFCM density, including its joint identification in $m$.

\paragraph{Normalization and uniform convergence.}
Let $R$ bound the norms of all allowed centers. For $\|x\|>2R$, the ratios $\|x-v_j\|/\|x\|$ lie between $1/2$ and $3/2$. Since $w_j\geq\varepsilon_w$, $t\in[m_{\min}-1,m_{\max}-1]$, and $qt=2$, constants independent of $\theta$ give
\[
 c_S\|x\|^{-q}\leq\sum_jw_j^{-1/t}\|x-v_j\|^{-q}
 \leq C_S\|x\|^{-q},\qquad
 c_E\|x\|^2\leq E_\theta(x)\leq C_E\|x\|^2.
\]
The continuous extension of $E$ at each center is jointly continuous in $(\theta,x)$. Hence $g_\theta$ has a common integrable Gaussian-tail envelope, $Z$ is continuous, and compactness gives $0<Z_{\min}\leq Z(\theta)\leq Z_{\max}<\infty$. Moreover,
\[
 \sup_{\theta\in\Theta_\delta}|\log f_\theta(x)|\leq C(1+\|x\|^2).
\]
The true density has Gaussian tails and therefore all polynomial moments. Pointwise continuity, compactness, and the integrable envelope imply the uniform strong law
\begin{equation}\label{eq:mle-uniform-lln}
 \sup_{\theta\in\Theta_\delta}\left|
 \frac1n\sum_{i=1}^n\log f_\theta(X_i)
 -\mathbb E_{\theta_0}\log f_\theta(X)\right|\longrightarrow0
 \quad\text{almost surely}.
\end{equation}
For completeness, continuity makes the supremum of the local oscillation over a shrinking parameter neighborhood tend pointwise to zero. Dominated convergence makes its expectation tend to zero; a finite cover and the ordinary strong law then give \eqref{eq:mle-uniform-lln}.

\paragraph{Identifiability, including the fuzziness parameter.}
At a center $v_j$, put
\[
 c_j=w_j^{-1/t},\quad S_j=\sum_{l\ne j}w_l^{-1/t}\|v_j-v_l\|^{-q},\quad
 A_j=\sigma^{-2}w_j,\quad B_j=A_jtS_j/c_j>0.
\]
For a unit vector $u$, separation gives
\[
 \sum_lw_l^{-1/t}\|v_j+ru-v_l\|^{-q}
 =c_jr^{-q}+S_j+O(r).
\]
The binomial expansion, $c_j^{-t}=w_j$, and $qt=2$ yield
\begin{equation}\label{eq:mle-center-expansion}
 E_\theta(v_j+ru)=A_jr^2-B_jr^{2+q}+o(r^{2+q}),\qquad r\downarrow0,
\end{equation}
uniformly in $u$ and locally uniformly in regular parameters.

If two densities agree almost everywhere, continuity makes them agree everywhere. Their common maximum is $1/Z$, attained exactly at their centers. Thus their unordered centers and normalizers agree. After matching the centers, their energies agree. Equation~\eqref{eq:mle-center-expansion} first identifies every $A_j$. If their values of $q$ differed, subtracting the common quadratic term and dividing by the smaller power $r^{2+q}$ would give a strictly negative limit for one energy and zero for the other, since $B_j>0$. Hence $q$, and therefore $m$, agree. Finally, $\sum_j A_j=\sigma^{-2}$ identifies $\sigma$, and $w_j=\sigma^2A_j$ identifies the weights. This proves identifiability modulo labels. Coincident centers and $k=1$, for which this argument fails, are excluded.

\paragraph{Completion of Theorem~\ref{thm:mle_consistency}.}
Let $L(\theta)=\mathbb E_{\theta_0}\log f_\theta(X)$. The integrable envelope makes $L$ continuous. The identity
\[
 L(\theta_0)-L(\theta)
 =\operatorname{KL}(f_{\theta_0},f_\theta)\geq0
\]
and identifiability show that only the finite label orbit of $\theta_0$ maximizes $L$. For every $\epsilon>0$, compactness consequently gives
\[
 L(\theta_0)-\sup_{d_{\rm lab}(\theta,\theta_0)\geq\epsilon}L(\theta)>0.
\]
An exact maximizer exists by compactness and continuity of the sample criterion. The uniform strong law and this strict population gap place every such maximizer within orbit distance $\epsilon$ eventually almost surely. Taking a countable sequence $\epsilon\downarrow0$ proves the theorem. Once this distance is small, separation gives an unambiguous local matching of centers to the truth; this is the chart used for asymptotic normality.
\endgroup

		\section{Proof for Asymptotic Normality}\label{app:normality}
\begingroup
All derivatives in this section are in the free coordinates $\vartheta$, with labels locally matched. Work in a neighborhood of the interior truth whose closure has positive weights, separated centers, and $m>1$.

\paragraph{Twice differentiable likelihood at and away from the centers.}
Near $v_j$, let $r=\|x-v_j\|$ and
$K_\theta(x)=c_j^{-1}\sum_{l\ne j}w_l^{-1/t}\|x-v_l\|^{-q}$.
The other distances are bounded away from zero, and the exact factorization is
\begin{equation}\label{eq:mle-center-factorization}
 E_\theta(x)=A_jr^2\{1+K_\theta(x)r^q\}^{-t}.
\end{equation}
Since $\{(1+s)^{-t}-1\}/s$ has a smooth extension at $s=0$, the difference from $A_jr^2$ is a bounded smooth factor times $r^{2+q}$. Let $q_->0$ bound $q$ below in the neighborhood. Up to two parameter derivatives of this remainder are controlled by $r^{q_-}$ for two center derivatives, $r^{1+q_-}(1+|\log r|)$ for a center--$m$ derivative, and $r^{2+q_-}(1+|\log r|^2)$ for two $m$ derivatives. All tend to zero. Thus the energy is jointly twice continuously differentiable, including at $x=v_j$; no restriction $m<3$ is required.

On bounded sets away from the centers, ordinary differentiation suffices. In the tails, write $x=ru$, $\|u\|=1$, and
\[
 E_\theta(ru)=\sigma^{-2}r^2 H_\theta(u,r)^{-t},\qquad
 H_\theta(u,r)=\sum_jw_j^{-1/t}\|u-v_j/r\|^{-q}.
\]
Here $H$ is bounded above and away from zero, and its first two parameter derivatives are bounded uniformly for large $r$. In particular, $qt=2$ cancels the radial powers before differentiation in $m$. Combining the center, compact-region, and tail bounds gives
\begin{equation}\label{eq:mle-derivative-envelopes}
 \|D^a E_\theta(x)\|\leq C(1+\|x\|^2),\qquad
 \|D^a g_\theta(x)\|\leq C(1+\|x\|^4)e^{-c\|x\|^2},\quad |a|\leq2,
\end{equation}
with an enlarged constant on bounded sets. The same argument is uniform on a slightly enlarged regular neighborhood of the compact parameter set. Differentiation under the integral for $Z$ is valid through order two. The score and log-likelihood Hessian have integrable quadratic envelopes, and the finite-cover argument used in Appendix~\ref{sec:pftheory} gives their local uniform laws of large numbers.

\paragraph{Score and information.}
Because $C=1/Z$ and $Dg=-g\,DE$, the correct identities are
\begin{align}
 \nabla\log C(\vartheta)&=\mathbb E_\vartheta\{\nabla E_\vartheta(X)\},\label{eq:constant}\\
 s_\vartheta(x):=\nabla\log f_\vartheta(x)
 &=-\nabla E_\vartheta(x)+\mathbb E_\vartheta\{\nabla E_\vartheta(X)\},\label{eq:score}\\
 \nabla^2\log f_\vartheta(x)
 &=-\nabla^2E_\vartheta(x)+\mathbb E_\vartheta\{\nabla^2E_\vartheta(X)\}
 -\operatorname{Var}_\vartheta\{\nabla E_\vartheta(X)\}.\nonumber
\end{align}
It follows that $\mathbb E_\vartheta s_\vartheta=0$ and
$I_r(\vartheta)=\mathbb E_\vartheta s_\vartheta s_\vartheta^{\mathsf T}
=-\mathbb E_\vartheta\nabla^2\log f_\vartheta
=\operatorname{Var}_\vartheta(\nabla E_\vartheta)$.

We verify nonsingularity directly. If a free-coordinate direction $a$ has $a^{\mathsf T}I_ra=0$, the directional derivative $DE$ is constant almost everywhere and hence everywhere by continuity and positivity of the density. At a center $DE=0$, so the constant is zero. Differentiating the exact factorization \eqref{eq:mle-center-factorization} at $x=v_j+ru$, with $x$ held fixed, first gives
\[
 0=DE(v_j+ru)=-2A_jr\,u^{\mathsf T}\dot v_j+o(r).
\]
Varying $u$ implies $\dot v_j=0$ for each $j$. With all these directions zero, the next term is $\dot A_jr^2+o(r^2)$, so $\dot w_j=2w_j\dot\sigma/\sigma$. The simplex tangent constraint $\sum_j\dot w_j=0$ then forces $\dot\sigma=0$ and all $\dot w_j=0$. Only the $m$ direction remains. Differentiating \eqref{eq:mle-center-factorization} now yields
\[
 0=DE(v_j+ru)=\dot m\,
 \frac{2B_j}{(m-1)^2}r^{2+q}\log r+O(r^{2+q}).
\]
Since $B_j>0$, division by $r^{2+q}\log r$ gives $\dot m=0$. Thus $a=0$, proving $I_r>0$. These directional expansions follow from the exact factorization~\eqref{eq:mle-center-factorization}.

\paragraph{Completion of Theorem~\ref{thm:mle-an}.}
Consistency and interiority put the matched exact MLE in this neighborhood with probability tending to one. For $Q_n=-n^{-1}\ell_n$, the score equation and integral Taylor expansion give
\[
 0=\nabla Q_n(\vartheta_0)+
 \left\{\int_0^1\nabla^2Q_n\bigl(\vartheta_0+s(\hat\vartheta_n-\vartheta_0)\bigr)\,ds\right\}
 (\hat\vartheta_n-\vartheta_0).
\]
The matrix in braces tends in probability to $I_r(\vartheta_0)>0$. The score central limit theorem and the derivative envelopes therefore imply
\[
 \sqrt n(\hat\vartheta_n-\vartheta_0)
 =I_r(\vartheta_0)^{-1}\frac1{\sqrt n}\sum_i s_{\vartheta_0}(X_i)+o_p(1).
\]
The delta method gives the stated ambient covariance. If $m=m_0$ is fixed in advance and $\eta$ contains the remaining free coordinates, the same proof uses the principal information block $I_{\eta\eta}$ and covariance $I_{\eta\eta}^{-1}$, not the corresponding block of $I_r^{-1}$. Third derivatives are not needed for either result.
\endgroup

		\section{Properties of WFCM Density}\label{densityproof}
		
		\begin{theorem}\label{th:limit-density}
			Let $k\geq2$ and fix $\bV=(\bv_1,\ldots,\bv_k)$, $\bw=(w_1,\ldots,w_k)$ with $w_j>0$ and
			$\sum_{j=1}^k w_j=1$, and $\sigma>0$. Define the unnormalized WFCM kernel
			\[
			h_m(\bx):=\ \exp\!\left(-\frac{1}{\sigma^2\, g_m(\bx)}\right),
			\qquad
			g_m(\bx):=\left[\sum_{j=1}^k (w_j\|\bx-\bv_j\|^2)^{-\frac{1}{m-1}}\right]^{m-1},
			\]
			for $\bx\in\mathbb R^d$ with $\bx\neq\bv_j$ for all $j$.
			Set $h_m(\bv_j)=1$ by continuity, $Z_m=\int_{\mathbb R^d}h_m(\bx)\,d\bx$, and $f_m=h_m/Z_m$ for the normalized density.
			Then the following limiting properties hold.
			\begin{enumerate}
				\item[(i)] As $m\to 1^+$,
				\[
				g_m(\bx)\ \longrightarrow\ \max_{1\le j\le k}(w_j\|\bx-\bv_j\|^2)^{-1}
				=\frac{1}{\min_{1\le j\le k} w_j\|\bx-\bv_j\|^2},
				\]
				and hence
				\[
				h_m(\bx)\ \longrightarrow\ \exp\!\left(-\frac{\min_{1\le j\le k} \, w_j\|\bx-\bv_j\|^2}{\sigma^2}\right)
				=\max_{1\le j\le k}\exp\!\left(-\frac{w_j\|\bx-\bv_j\|^2}{\sigma^2}\right).
				\]
				In particular, the dominant contribution comes from the centroid(s) minimizing
				$w_j\|\bx-\bv_j\|^2$ (and if $w_j$ are equal, this reduces to the nearest centroid(s)).
				
				\item[(ii)] For each fixed $\bx$ with $\bx\neq\bv_j$ for all $j$,
				\[
				g_m(\bx)\ \longrightarrow\ \infty,
				\quad\text{and hence}\quad
				h_m(\bx)\ \longrightarrow\ 1
				\]
				as $m\to\infty$. The normalizing integral satisfies $Z_m\to\infty$, so $f_m(\bx)\to0$ for each fixed $\bx$; there is no limiting uniform probability density on $\mathbb R^d$. On any bounded measurable set of positive volume, the normalized restriction converges to the uniform distribution.
				
				\item[(iii)] When $m=2$,
				\[f_2(\bx)\ \propto \,
				\exp\!\left(-\frac{1}{\sigma^2\sum_{j=1}^k (w_j\|\bx-\bv_j\|^2)^{-1}}\right).
				\]
			\end{enumerate}
		\end{theorem}
		
		\begin{proof}
			Throughout, fix $\bx\in\mathbb R^d$ with $\bx\neq\bv_j$ for all $j$.
			
			\textbf{(i) Limit as $m\to 1^+$.}
			Apply Lemma~\ref{lm:softmax} with $a_j=(w_j\|\bx-\bv_j\|^2)^{-1}$ to get
			\[
			\lim_{m\to 1^+} g_m(\bx)=\max_{1\le j\le k}(w_j\|\bx-\bv_j\|^2)^{-1}
			=\frac{1}{\min_{1\le j\le k} w_j\|\bx-\bv_j\|^2}.
			\]
			Therefore,
			\[
			\begin{aligned}
				h_m(\bx)\ &\longrightarrow\ \exp\!\left(
				-\frac{1}{\sigma^2\,\max_j(w_j\|\bx-\bv_j\|^2)^{-1}}
				\right)\\
				&=
				\exp\!\left(-\frac{\min_j w_j\|\bx-\bv_j\|^2}{\sigma^2}\right)
				=
				\max_j \exp\!\left(-\frac{w_j\|\bx-\bv_j\|^2}{\sigma^2}\right).
			\end{aligned}
			\]
			
			\medskip
			\textbf{(ii) Limit as $m\to\infty$.}
			Let $\alpha_m:=\frac{1}{m-1}\downarrow 0$. Then, for each $j$,
			\[
			(w_j\|\bx-\bv_j\|^2)^{-\alpha_m}
			=\exp\!\Big(-\alpha_m\log(w_j\|\bx-\bv_j\|^2)\Big)\longrightarrow 1.
			\]
			Since the sum has finitely many terms,
			\[
			\sum_{j=1}^k (w_j\|\bx-\bv_j\|^2)^{-\alpha_m}\ \longrightarrow\ k,
			\]
			For sufficiently large $m$, this sum is at least $(k+1)/2>1$. Therefore
			\[
			g_m(\bx)\geq\left(\frac{k+1}{2}\right)^{m-1}\longrightarrow\infty,
			\qquad h_m(\bx)\longrightarrow1.
			\]
			The kernel limit also holds at the centers by its continuous extension. Fatou's lemma gives
			$\liminf_{m\to\infty}Z_m\geq\int_{\mathbb R^d}1\,d\bx=\infty$.
			Since $0<h_m\leq1$, it follows that $f_m(\bx)=h_m(\bx)/Z_m\to0$.
			For a bounded measurable set $A$ with $0<|A|<\infty$, dominated convergence gives
			$\int_A h_m(\bx)\,d\bx\to|A|$; hence the density restricted and renormalized on $A$ converges to $\mathbf1_A/|A|$.
			
			\medskip
			\textbf{(iii) Case $m=2$.}
			Substitute $m=2$ into the definition of $g_m(\bx)$ to obtain
			\[
			g_2(\bx)=\sum_{j=1}^k (w_j\|\bx-\bv_j\|^2)^{-1}
			=\sum_{j=1}^k w_j^{-1}\|\bx-\bv_j\|^{-2},
			\]
			and plug into $h_2(\bx)$ and normalize to obtain $f_2(\bx)$.
		\end{proof}
		
		\begin{lemma}\label{lm:softmax}
			Let $\{a_j\}_{j=1}^k$ be positive numbers and let $\{w_j\}_{j=1}^k$ satisfy
			$w_j> 0$ and $\sum_{j=1}^k w_j=1$. Then
			\[
			\lim_{m\to 1^+}\left(\sum_{j=1}^k a_j^{\frac{1}{m-1}}\right)^{m-1}
			=\max_{1\le j\le k} a_j,
			\qquad
			\lim_{m\to 1^+}\left(\sum_{j=1}^k w_j\, a_j^{\frac{1}{m-1}}\right)^{m-1}
			=\max_{1\le j\le k} a_j.
			\]
		\end{lemma}
		
		\begin{proof}
			Let $a_{\max}=\max_{1\le j\le k}a_j$ and
			$J^*=\{j:a_j=a_{\max}\}$.
			Set $t=m-1\to0^+$ and
			$r_j=a_j/a_{\max}\in(0,1]$.
			Define
			\[
			S_t:=\sum_{j=1}^k w_j r_j^{1/t}.
			\]
			Then
			\[
			\left(\sum_{j=1}^k w_j a_j^{1/t}\right)^t
			=\left(a_{\max}^{1/t}\sum_{j=1}^k w_j r_j^{1/t}\right)^t
			= a_{\max}\, (S_t)^t.
			\]
			It remains to show $(S_t)^t\to 1$.
			
			Let $s_*:=\sum_{j\in J^*} w_j\in(0,1]$. For $j\in J^*$, $r_j=1$.
			For $j\notin J^*$, we have $r_j<1$; define
			\[
			r:=\max_{j\notin J^*} r_j \in [0,1).
			\]
			(If $J^*=[k]$, set $r:=0$; then the argument below still holds.)
			
			\emph{Two-sided bound for $S_t$.}
			Since $r_j^{1/t}=1$ for $j\in J^*$ and $0\le r_j^{1/t}\le r^{1/t}$ for
			$j\notin J^*$, we obtain
			\[
			s_*
			=\sum_{j\in J^*} w_j
			\le
			\sum_{j=1}^k w_j r_j^{1/t}
			\le
			\sum_{j\in J^*} w_j
			+
			\sum_{j\notin J^*} w_j\, r^{1/t}
			=
			s_*+(1-s_*)r^{1/t}.
			\]
			Hence, for all $t>0$,
			\begin{equation}\label{eq:St-bounds}
				s_* \le S_t \le s_*+(1-s_*)r^{1/t}.
			\end{equation}
			
			\emph{Two-sided bound for $(S_t)^t$.}
			Raising \eqref{eq:St-bounds} to the power $t$ (monotonicity since all terms are positive) gives
			\[
			(s_*)^t \le (S_t)^t \le \bigl(s_*+(1-s_*)r^{1/t}\bigr)^t.
			\]
			As $t\to 0^+$, we have $(s_*)^t\to 1$.
			Also $r^{1/t}\to 0$ because $0\le r<1$, so $s_*+(1-s_*)r^{1/t}\to s_*$.
			Therefore,
			\[
			\lim_{t\to 0^+}\bigl(s_*+(1-s_*)r^{1/t}\bigr)^t
			=
			\exp\!\left(\lim_{t\to 0^+} t\log\bigl(s_*+(1-s_*)r^{1/t}\bigr)\right)
			=\exp(0)=1,
			\]
			since $\log\bigl(s_*+(1-s_*)r^{1/t}\bigr)\to \log s_*$ is finite.
			By the squeeze theorem, $(S_t)^t\to 1$.
			
			Finally,
			\[
			\lim_{t\to 0^+}\left(\sum_{j=1}^k w_j a_j^{1/t}\right)^t
			=
			\lim_{t\to 0^+} a_{\max}(S_t)^t
			=
			a_{\max}.
			\]
			This proves the weighted claim. Taking $w_j\equiv 1/k$ gives the unweighted claim, because $\bigl(\sum_{j} a_j^{1/t}\bigr)^t=k^{t}\bigl(\sum_{j} k^{-1}a_j^{1/t}\bigr)^t$ and $k^{t}\to1$.
		\end{proof}
		
		\section{Simulation}\label{sec:sup_simulation}
		
		We now conduct a suite of simulation experiments to investigate the finite-sample and asymptotic properties of the proposed WFCM estimator. Specifically, we aim to (i) examine finite-sample consistency, (ii) assess the empirical distribution of the estimator under increasing sample size, and (iii) compare different approaches to uncertainty quantification, including empirical replicates, bootstrap resampling, and Fisher-information-based approximations.
		
		\paragraph{Data-Generating Mechanism.}
		We fix $d=1$, $k=2$, and
		\[
		\theta_0=(\sigma_0=0.9,\ v_{0,1}=-2,\ v_{0,2}=2,\ w_{0,1}=0.35,\ w_{0,2}=0.65,\ m_0=2).
		\]
		Each dataset contains independent draws from a numerical inverse-CDF approximation to the normalized WFCM density in~\eqref{eq:WFCM-density}. We construct the CDF by numerical integration on a 200,001-point grid with extended tails and invert it for independent uniform draws. Each Monte Carlo dataset uses a separate recorded seed. Here the weights are WFCM distance penalties.
		
		\paragraph{Estimation Procedure.}
		Given a dataset of size $n$, we proceed in three steps:
		\begin{enumerate}
			\item Fit the defensive proposal of Section~\ref{sec:revision-computation} (90\% two-component Gaussian mixture, 10\% Student-$t_5$) and draw $M=\max\{2000,\lceil n^{3/2}\rceil\}$ proposal samples. This proposal is used for importance sampling in the approximate likelihood evaluation.
			\item Run the WFCM fitting procedure (Algorithm~\ref{alg:emfuzzy}) from three random starts, keeping the fit with the smallest NLL, to obtain parameter estimates
			\[
			\hat\theta=(\hat\sigma,\hat{\bv}_1,\ldots,\hat{\bv}_k,\hat \bw).
			\]
			\item Resolve label switching by aligning the estimated centers $\hat{\bv}_j$ with the true centers $\bv_{0,j}$. We adopt the Hungarian assignment method, which minimizes the total Euclidean distance between estimated and true centers. The estimated weights $\hat \bw$ are permuted consistently with the center assignment.
		\end{enumerate}
		This alignment step is essential because the underlying density is invariant under permutations of the cluster labels, and therefore the likelihood does not distinguish among labelings. It corresponds to the local label matching in Theorem~\ref{thm:mle-an}.
		
\begin{figure}
			\centering
			\includegraphics[width=\linewidth]{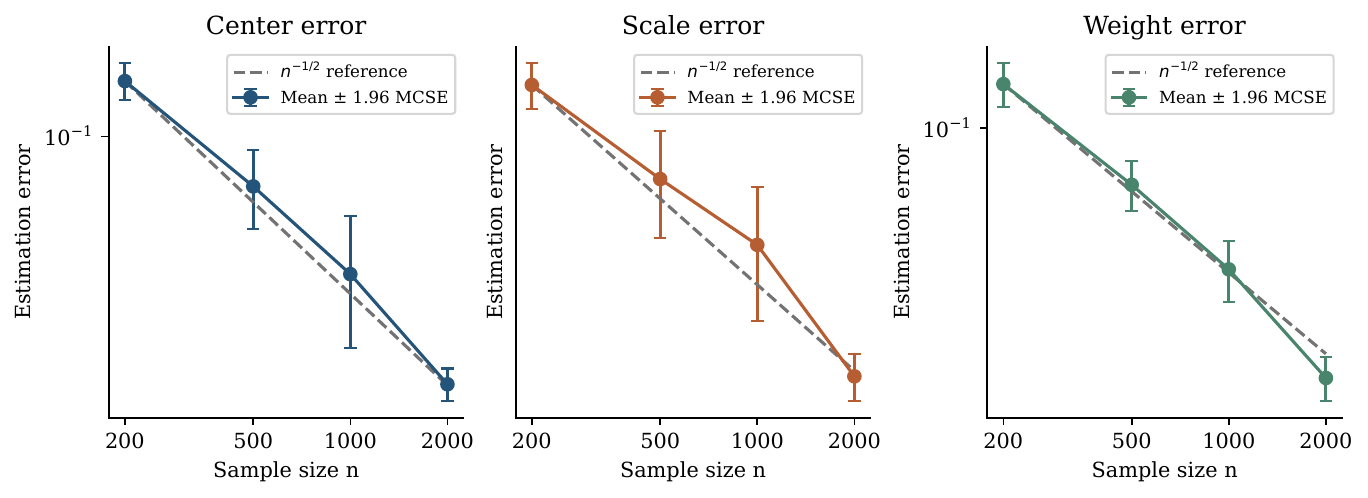}
			\caption{Finite-sample consistency of the WFCM estimator. Errors are shown as functions of sample size $n$ on a log--log scale. Dashed curves have slope $-1/2$.}
			\label{fig:con}
		\end{figure}

		\paragraph{Consistency.}
		To verify consistency, we compute the label-invariant errors between the estimates and the truth across sample sizes $n\in\{200,500,1000,2000\}$, with 250 datasets per $n$. Figure~\ref{fig:con} plots the empirical root mean squared error of the centers, the absolute error of $\hat\sigma$, and the $\ell_1$ error of the weights on a log--log scale. In each case, the empirical error decreases at approximately the $n^{-1/2}$ rate, consistent with the theoretical guarantees in Section~\ref{sec:theory}. The means include the two degenerate fits described under Asymptotic Normality below. Confidence intervals are constructed across independent Monte Carlo replicates.  
		\begin{figure}
			\centering
			\includegraphics[width=\linewidth]{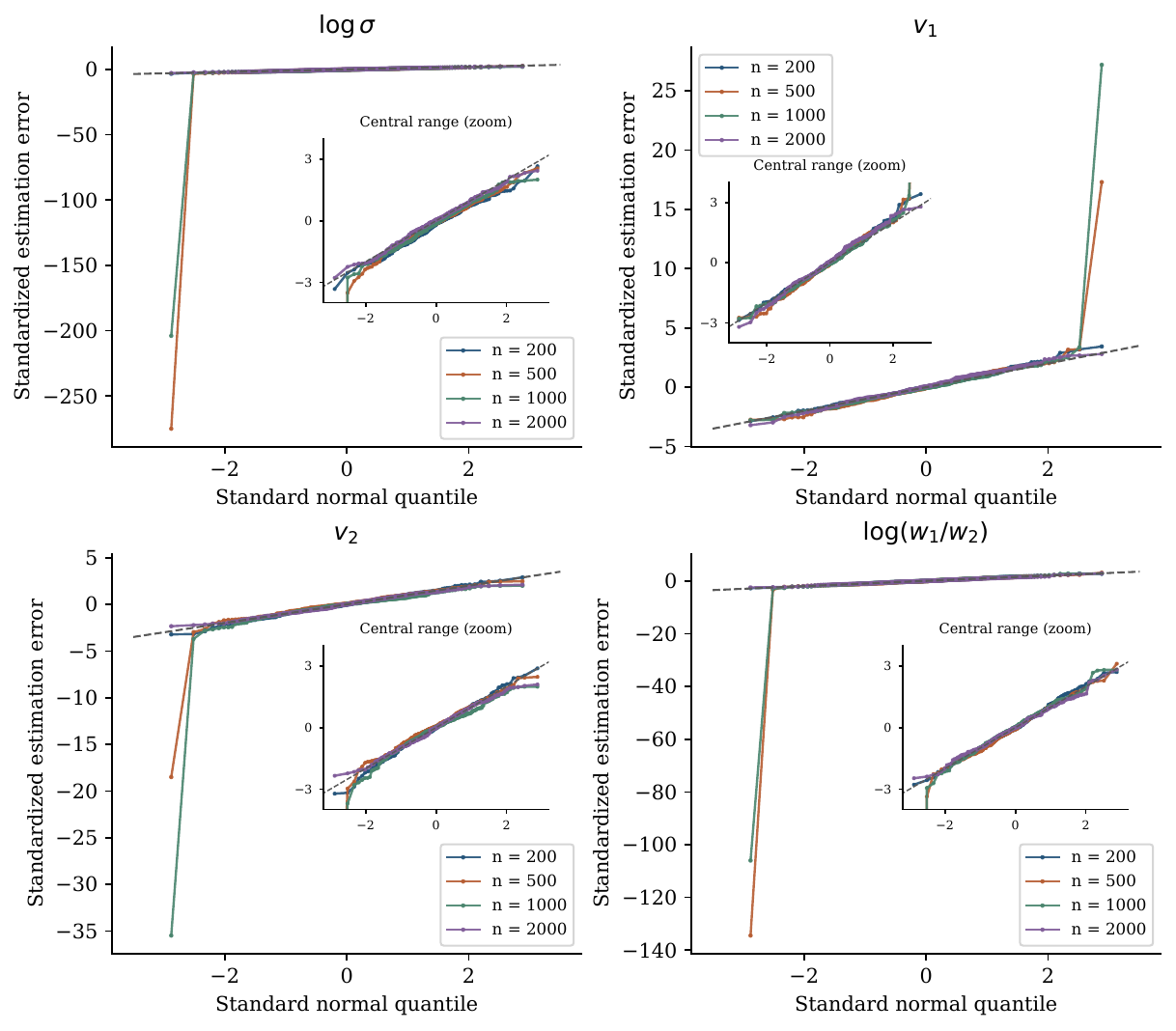}
			\caption{Marginal QQ plots of parameter estimates standardized by their Fisher-information standard errors across $n \in \{200,500,1000,2000\}$. Each subplot corresponds to a selected coordinate of the parameter vector. The closer the empirical quantiles lie to the $45^\circ$ line, the stronger the agreement with the Gaussian benchmark.}
			\label{fig:qq-overlay}
		\end{figure}
		
		\paragraph{Asymptotic Normality.}
		We next investigate the large-sample distributional behavior of the proposed estimator. The synthetic data are generated under the following one-dimensional setting: we fix the number of clusters $k=2$, fuzziness parameter $m=2.0$, and scale parameter $\sigma_0=0.9$. The two true cluster centers are
		\[
		v_1=-2,
		\quad v_2=2,
		\]
		and cluster weights are set to $\bw=(0.35,0.65)$.
		Given these parameters, we simulate independent draws from the cluster-based density by numerically inverting its cumulative distribution function on a grid of $200{,}001$ points. For each sample size $n\in\{200,500,1000,2000\}$, we generate $250$ independent datasets from the data-generating mechanism described above, each using a different random seed. On each replicate we compute the aligned estimator using Algorithm~\ref{alg:emfuzzy}, where label switching is resolved as in the consistency experiment.
		
		To assess asymptotic normality, we perform the following normalization procedure. First, for each $n$, we center the estimated parameter vectors at the truth $\theta_0$. Second, we divide each centered coordinate by its asymptotic standard error from Theorem~\ref{thm:mle-an} with $m$ fixed, $\{[I_{\eta\eta}(\theta_0)^{-1}]_{jj}/n\}^{1/2}$, where $I_{\eta\eta}(\theta_0)$ is the fixed-$m$ Fisher information at the true parameter in the coordinates $(\log\sigma, v_1, v_2, \log(w_1/w_2))$, computed from 750,000 independent draws. If Theorem~\ref{thm:mle-an} holds, each standardized coordinate should be approximately standard normal. 
		
		Figure~\ref{fig:qq-overlay} displays marginal quantile--quantile (QQ) plots for selected coordinates of the standardized parameter estimates, overlaid across the four sample sizes. Each subplot compares the empirical quantiles to the theoretical standard normal quantiles. The central ranges (insets) lie close to the $45^\circ$ line at every $n$. The extreme points at $n=500$ and $n=1000$ come from two fits, one at each of these sample sizes, in which Algorithm~\ref{alg:emfuzzy} converged to a degenerate solution with $\hat\sigma<0.002$ and one weight below $2\times10^{-7}$. Both fits passed the numerical convergence checks, but their negative log-likelihoods exceed that of the true parameter on the same integration cloud by about 101 and 185. This implementation omits the bounds $\sigma\ge\sigma_{\min}$ and $w_j\ge\varepsilon_w$ of~\eqref{eq:regular-parameter-set}; imposing them, as in the ADNI analysis, or using additional starts would exclude these two solutions. Both fits are retained in Figures~\ref{fig:con} and~\ref{fig:qq-overlay}.
		
		\begin{figure}
			\centering
			\includegraphics[width=\linewidth]{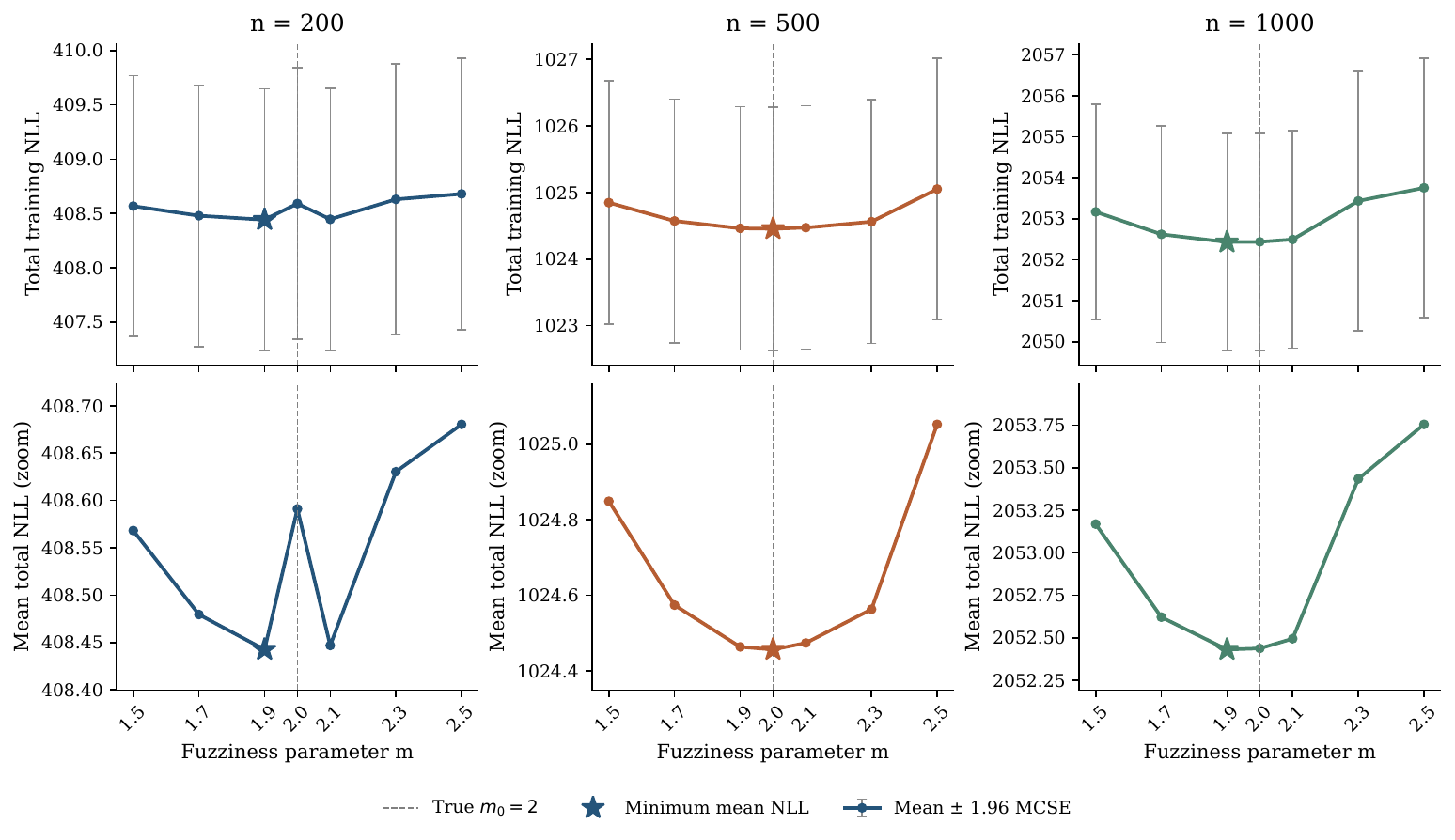}
			\caption{Selection of the fuzziness parameter $m$ via approximate trainNLL. For each $m\in\{1.5,1.7,1.9,2.0,2.1,2.3,2.5\}$ we fit $(\sigma, \bV, w)$ and select the best $m$ using the negative log-likelihood. A star marks the minimum of the mean curve, and the vertical line marks $m_0=2$. The lower panels enlarge the mean curves, without error bars. The spike at $m=2.0$ for $n=200$ comes from one degenerate fit (see text).}
			\label{fig:m-trainNLL}
		\end{figure}
		
		\paragraph{Estimating the fuzziness parameter $m$.}
		In addition to estimating $(\sigma, \bV, \bw)$, we also propose a procedure that estimates the parameter $m$ from data generated under the same one-dimensional mechanism ($k=2$, $\sigma_0=0.9$, $v_{0,1}=-2$, $v_{0,2}=2$, $\bw_0=(0.35,0.65)$, and $m_0=2$). For each sample size $n\in\{200,500,1000\}$, we generate $R=200$ independent datasets. For each dataset, we perform a grid search over $\mathcal M=\{1.5,1.7,1.9,2.0,2.1,2.3,2.5\}$. Parameters $(\sigma, \bV, \bw)$ are estimated by Algorithm~\ref{alg:emfuzzy}.
		
		To evaluate the likelihood for model comparison across $m$, we use importance sampling with a data-adaptive proposal. We refer the reader to Section~\ref{sec:EM-like} for detailed explanations. Concretely, for each dataset we fit a $2$-component Gaussian mixture to the observed $\bx_{1:n}$ and combine it with a Student-$t_5$ component, assigning mixture weights $0.9$ and $0.1$, respectively. We draw $M=\max\{2000,\lceil n^{3/2}\rceil\}$ proposal points. The same proposal samples and initialization seeds are reused across all $m\in\mathcal M$. The total training negative log-likelihood (trainNLL) for $m$ is then computed with these shared proposal samples. We summarize the results across the $R$ independent datasets by the mean trainNLL with $\pm1.96$ Monte Carlo standard-error bars, where the Monte Carlo standard error is the sample standard deviation divided by $\sqrt{R}$. Figure~\ref{fig:m-trainNLL} displays these curves, where lower values indicate better fit and the error bars describe the Monte Carlo precision of the estimated mean. The lower panels enlarge the same mean curves. For each dataset, we select $\hat m$ as the grid value with the smallest trainNLL and retain the corresponding parameter estimates. Individual selections are variable: $\hat m=2$ is chosen in 3.0\%, 9.0\%, and 9.5\% of datasets at $n=200$, $500$, and $1000$, while the two grid endpoints together are chosen in 72.5\%, 60.5\%, and 33.5\%. Paired differences $\mathrm{NLL}(m)-\mathrm{NLL}(2)$ within each dataset separate optimization failures from weak identification. One dataset at each $n$ contains degenerate fits of the kind described above: at $m=2.0$, $2.3$, and $2.5$ for $n=200$, which produce the spike at $m=2.0$ in Figure~\ref{fig:m-trainNLL}; at $m=2.5$ for $n=500$; and at $m=2.3$ and $2.5$ for $n=1000$. None of these fits is selected as $\hat m$. Excluding these three datasets, the mean paired difference is smallest at $m=2.0$ for $n=200$ and $500$ and at $m=1.9$ for $n=1000$ ($-0.006$, MCSE $0.016$), and its largest value over the grid is $0.13$, $0.39$, and $0.73$, respectively. For 65\%, 36\%, and 20\% of these datasets, the NLL varies by less than one unit across the whole grid. The spread of $\hat m$ therefore reflects weak identification of $m$ at these sample sizes rather than optimization failure, and identification improves as $n$ grows.
		
		\par
\begingroup
\subsection{Comparison across geometry, imbalance, and overlap}
\label{sec:revision-benchmark}

We compare WFCM with FCM, polynomial FCM, $k$-means, equilibrium $k$-means \citepS{he2024imbalanced}, HDBSCAN \citepS{campello2013density}, Ward hierarchical clustering, spectral clustering \citepS{ng2002spectral}, and a full-covariance GMM. The reported comparison crosses two Gaussian geometries, size ratios $1{:}1$, $4{:}1$, and $10{:}1$, and three overlap levels: 18 cells, each with 100 independent datasets of $n=600$ observations in two dimensions. All methods use the same datasets. Spherical clusters have identity covariance and means separated by $4$, $3$, or $2$. Anisotropic clusters apply $T=R_{\pi/6}\operatorname{diag}(2.5,0.4)R_{\pi/6}^{\top}$ to both means and observations, where $R_\alpha$ is planar rotation. Sample counts follow the prescribed ratio, rounded to total 600.

Except for HDBSCAN, methods receive $k=2$. WFCM uses $m=2$, $M=12000$, a defensive proposal as described in Section~\ref{sec:revision-computation}, and ten starts. FCM uses $m=2$ and polynomial FCM uses the quadratic fuzzifier with $\beta=0.5$. Equilibrium $k$-means uses the implementation's distance-variance choice of its tuning parameter. HDBSCAN uses minimum cluster size and minimum samples of 12; spectral clustering uses a 15-nearest-neighbor graph. Randomized methods use ten starts where supported, with a maximum of 450 iterations for iterative fits. These settings were fixed in advance. These studies fit WFCM by multistart L-BFGS-B on a fixed-cloud approximate likelihood instead of Algorithm~\ref{alg:emfuzzy}.

Table~\ref{tab:revision-benchmark} reports ARI, accuracy after optimal one-to-one label matching, minority-class recall, and prototype-location RMSE. For methods without native prototypes, we use within-cluster means. RMSE is $\{k^{-1}\sum_j\|\hat v_{\pi(j)}-\mu_j\|^2\}^{1/2}$ after matching, where $\mu_j$ is the generating component mean. HDBSCAN noise is included in ARI and counted as incorrect for accuracy and recall; center error is defined only when exactly two clusters are returned. For a balanced cell, the prespecified first class supplies the recall target.

WFCM's mean ARI exceeds FCM's for spherical and anisotropic clusters, but the full-covariance GMM has higher ARI in both. On anisotropic data, WFCM and GMM have similar prototype RMSE despite different ARI, illustrating that the two metrics address different targets. Figure~\ref{fig:revision-benchmark} shows the cellwise results: imbalance and overlap can sharply reduce WFCM's performance. Higher ARI need not coincide with higher minority recall.

The same protocol was also run on a nonconvex two-moons geometry (nine imbalance--noise cells, 100 datasets each). Averaged over these cells, mean ARI was 0.094 (MCSE 0.002) for WFCM, 0.129 for FCM, 0.134 for $k$-means and polynomial FCM, 0.136 for equilibrium $k$-means, 0.161 for the GMM, 0.196 for Ward clustering, 0.294 for spectral clustering, and 0.608 for HDBSCAN. WFCM therefore performed worst of the nine methods in this geometry, consistent with its centroid-based, approximately isotropic structure. The dataset-level results are included in the code repository.

\begin{table}[p]
\centering\small\singlespacing
\caption{Clustering performance by geometry: equally weighted means over nine imbalance--overlap cells, with MCSEs in parentheses.}
\label{tab:revision-benchmark}
\setlength{\tabcolsep}{5pt}
\begin{tabular}{lcccc}
\toprule
Method & ARI & Accuracy & Minority recall & Prototype RMSE\\
\midrule
\multicolumn{5}{l}{\textit{Spherical}}\\
WFCM & 0.495 (0.003) & 0.870 (0.002) & 0.653 (0.007) & 0.505 (0.007)\\
FCM & 0.468 (0.001) & 0.829 (0.001) & 0.955 (0.001) & 0.570 (0.003)\\
Polynomial FCM & 0.536 (0.002) & 0.858 (0.001) & 0.951 (0.001) & 0.400 (0.004)\\
$k$-means & 0.543 (0.002) & 0.861 (0.001) & 0.950 (0.001) & 0.385 (0.004)\\
Equilibrium $k$-means & 0.592 (0.002) & 0.881 (0.001) & 0.942 (0.001) & 0.288 (0.003)\\
HDBSCAN & 0.173 (0.005) & 0.417 (0.006) & 0.224 (0.004) & 0.862 (0.024)\\
Ward hierarchical & 0.544 (0.005) & 0.858 (0.003) & 0.860 (0.005) & 0.417 (0.009)\\
Spectral & 0.469 (0.003) & 0.823 (0.001) & 0.951 (0.001) & 0.522 (0.006)\\
Full-covariance GMM & 0.638 (0.002) & 0.900 (0.001) & 0.918 (0.002) & 0.378 (0.004)\\
\midrule
\multicolumn{5}{l}{\textit{Anisotropic}}\\
WFCM & 0.520 (0.004) & 0.847 (0.002) & 0.847 (0.005) & 0.946 (0.016)\\
FCM & 0.424 (0.002) & 0.814 (0.001) & 0.936 (0.001) & 1.152 (0.007)\\
Polynomial FCM & 0.443 (0.002) & 0.823 (0.001) & 0.934 (0.001) & 1.019 (0.008)\\
$k$-means & 0.443 (0.002) & 0.823 (0.001) & 0.934 (0.001) & 1.019 (0.009)\\
Equilibrium $k$-means & 0.471 (0.002) & 0.835 (0.001) & 0.929 (0.001) & 0.877 (0.009)\\
HDBSCAN & 0.310 (0.005) & 0.601 (0.006) & 0.308 (0.004) & 1.333 (0.053)\\
Ward hierarchical & 0.429 (0.006) & 0.810 (0.003) & 0.847 (0.007) & 1.162 (0.026)\\
Spectral & 0.339 (0.002) & 0.768 (0.001) & 0.943 (0.001) & 1.460 (0.008)\\
Full-covariance GMM & 0.595 (0.002) & 0.882 (0.001) & 0.923 (0.001) & 0.948 (0.011)\\
\bottomrule
\end{tabular}
\par\smallskip
\begin{minipage}{\linewidth}\footnotesize
Each method has 900 valid fits per geometry. Geometry-level MCSEs combine within-cell mean variances. HDBSCAN returns exactly two clusters in 623/900 spherical and 523/900 anisotropic datasets; its prototype RMSE is conditional on those fits. Other methods receive $k=2$. Noise observations remain in ARI and count as incorrect for accuracy and recall.
\end{minipage}
\end{table}

\begin{figure}[p]
\centering
\includegraphics[width=\linewidth]{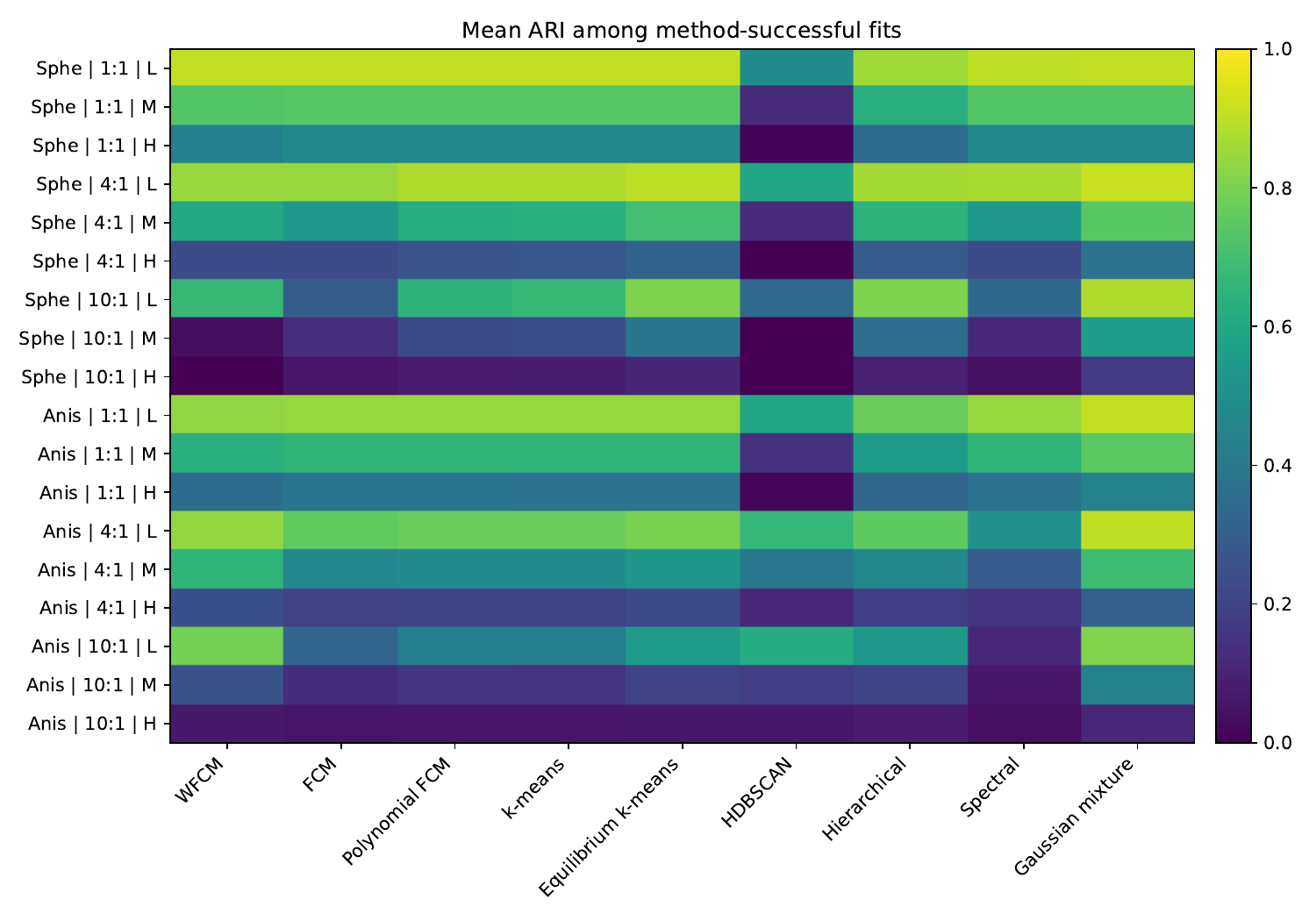}
\caption{Mean ARI over 100 independent datasets in each geometry--imbalance--overlap cell. Sphe and Anis denote spherical and anisotropic designs; L, M, and H denote low, moderate, and high overlap.}
\label{fig:revision-benchmark}
\end{figure}
\clearpage

\subsection{Selecting the number of clusters before inference}
\label{sec:revision-selection}

We generate 250 independent datasets at each $n\in\{150,300\}$ from the one-dimensional WFCM density with centers $(-2,2)$, weights $(0.3,0.7)$, $\sigma=1$, and $m=2$, using numerical inverse-CDF sampling. Candidate fits use $k\in\{2,3,4\}$, three starts, $M=\max(4000,12n)$, and the normalized $\mathrm{WXBI}_{\mathrm{rel}}$ from Section~4.3. The fuzzifier remains fixed. We construct nominal 95\% marginal center intervals from model Fisher information, estimated with 50,000 draws; centers are aligned to the truth. If $k=2$ is selected, the fixed-$k$ and selected analyses reuse the same fitted endpoint and intervals.

Table~\ref{tab:revision-selection} separates three quantities: fixed-$k$ coverage, coverage conditional on selecting $k=2$, and the joint select-and-cover probability. The last is the average of the two center-coverage indicators multiplied by $\mathbf 1\{\hat k=2\}$; a wrong $k$ or invalid fit contributes zero. WXBI chooses $k=3$ or $4$ in 40.4\% of datasets at $n=150$ and 19.6\% at $n=300$. Consequently, near-nominal coverage among correct selections coexists with much lower joint select-and-cover rates.

\begin{table}[htbp]
\centering\small\singlespacing
\caption{Selection and marginal center coverage, in percent (MCSE).}
\label{tab:revision-selection}
\setlength{\tabcolsep}{5pt}
\begin{tabular}{rccccc}
\toprule
$n$ & Selected $k=2,3,4$ & $\Pr(\hat k=2)$ & Fixed $k=2$ & Conditional & Select-and-cover\\
\midrule
150 & 149, 34, 67 & 59.6 (3.1) & 93.8 (1.2) & 95.0 (1.3) & 56.6 (3.1)\\
300 & 201, 14, 35 & 80.4 (2.5) & 95.6 (1.0) & 96.0 (1.0) & 77.2 (2.5)\\
\bottomrule
\end{tabular}
\par\smallskip
\begin{minipage}{\linewidth}\footnotesize
Each row uses 250 datasets, all with a complete candidate grid and valid fixed-$k$ intervals. Conditional coverage uses only the 149 or 201 correct selections; selection probabilities and select-and-cover use all 250 datasets.
\end{minipage}
\end{table}
\clearpage

\subsection{Computational sensitivity}
\label{sec:revision-computation}

We examine the complete fitting cost, including proposal fitting, importance draws, and three-start likelihood optimization (at most 450 iterations per start). The data are drawn from a three-component Gaussian mixture with probabilities $(0.15,0.30,0.55)$ and first-two-coordinate means $(-3.4,-1.8)$, $(0,3)$, and $(3.4,-1.5)$. Component standard deviations in these coordinates are $(1.25,0.60)$, $(0.65,1.15)$, and $(1.05,0.70)$, rotated by $-0.45$, $0$, and $0.45$ radians. Additional coordinates are independent $N(0,0.75^2)$. Center error here is measured relative to the generating mixture means. All fits use $k=3$ and $m=2$. Each configuration uses 20 independent datasets; proposal and $M$ comparisons are paired within a dataset. Timing was collected with one worker on macOS/ARM64 (Python 3.12.3).

The dimension experiment uses $n=500$, $M=10000$, and $d\in\{2,10,25,50\}$. We compare full- and diagonal-covariance GMM proposals with a defensive proposal comprising 90\% full GMM and 10\% broad multivariate $t_5$. For normalized importance weights $a_s$, we report $\mathrm{ESS}=1/\sum_s a_s^2$ and $\max_s a_s$. Figure~\ref{fig:revision-computation} shows that defensive-proposal ESS/$M$ decreases from 66.8\% to 2.1\% between $d=2$ and $50$; at $d=50$, the diagonal proposal achieves 17.1\%. Center-coordinate RMSE at $d=50$ is 0.385 for the defensive proposal, 0.272 for the diagonal GMM, and 0.463 for the full GMM. The choice of proposal therefore matters increasingly as dimension grows.

Table~\ref{tab:revision-computation} varies $M/n$ at $d=10$. Fits use independent clouds and are evaluated on a common independent scoring cloud of size $\max(50000,40n)$; the $M/n=80$ endpoint serves as the numerical reference. Increasing $M$ reduces the scored NLL difference and center displacement at additional computational cost. In a separate sample-size experiment with $M=20n$, median runtime increases from 0.307 to 3.827 seconds as $n$ increases from 250 to 4000. All 460 fits passed the numerical convergence checks.

\begin{figure}[p]
\centering
\includegraphics[width=\linewidth]{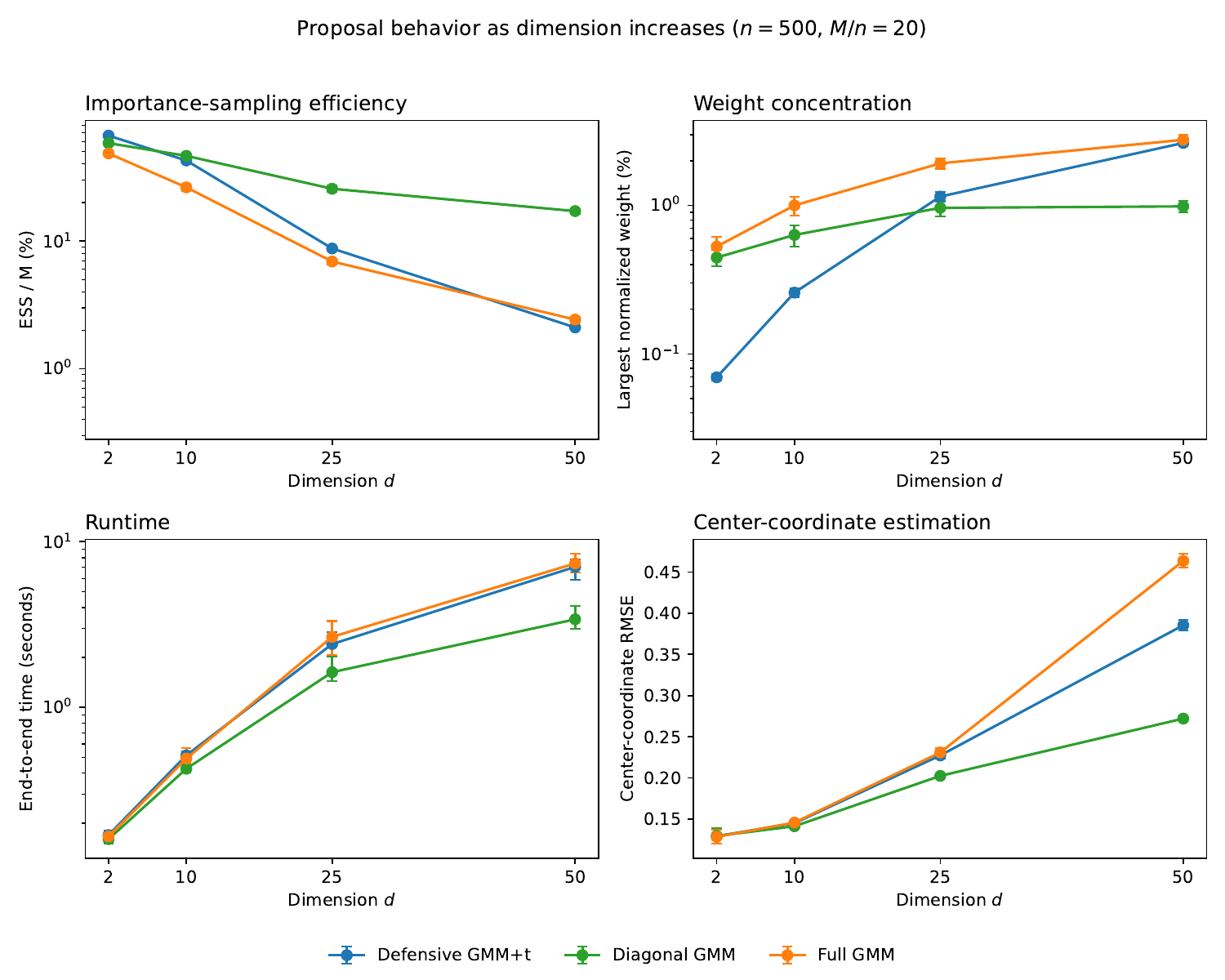}
\caption{Importance-sampling efficiency, weight concentration, runtime, and center-coordinate RMSE as dimension and proposal vary. Means have $\pm1$ MCSE bars; runtime shows medians and interquartile ranges. Each configuration uses 20 independent datasets. Center-coordinate RMSE divides the matched squared error by $kd$, whereas Table~\ref{tab:revision-benchmark} divides by $k$.}
\label{fig:revision-computation}
\end{figure}

\begin{table}[p]
\centering\small\singlespacing
\caption{Monte Carlo budget and sample-size sensitivity at $d=10$ with the defensive proposal.}
\label{tab:revision-computation}
\setlength{\tabcolsep}{6pt}
\begin{tabular}{rrccc}
\toprule
\multicolumn{5}{l}{(a) Monte Carlo budget: 20 paired datasets per $n$}\\
$n$ & $M/n$ & $\Delta$ average NLL & Center displacement & Seconds [IQR]\\
\midrule
250 & 1  & 0.8791 (0.1895) & 0.3060 (0.0226) & 0.144 [0.128, 0.188]\\
250 & 5  & 0.1121 (0.0161) & 0.1349 (0.0098) & 0.152 [0.137, 0.193]\\
250 & 20 & 0.0132 (0.0022) & 0.0641 (0.0038) & 0.281 [0.268, 0.311]\\
250 & 80 & reference & reference & 0.945 [0.867, 1.011]\\
1000 & 1  & 0.0556 (0.0061) & 0.0958 (0.0060) & 0.164 [0.146, 0.186]\\
1000 & 5  & 0.0068 (0.0006) & 0.0359 (0.0013) & 0.297 [0.276, 0.319]\\
1000 & 20 & 0.0011 (0.0001) & 0.0182 (0.0007) & 0.930 [0.887, 0.971]\\
1000 & 80 & reference & reference & 3.645 [3.329, 3.780]\\
\midrule
\multicolumn{5}{l}{(b) Sample-size scaling: $M=20n$, 20 datasets per $n$}\\
$n$ & $M$ & ESS/$M$ (\%) & Center-coordinate RMSE & Seconds [IQR]\\
\midrule
250 & 5000 & 28.37 (1.07) & 0.172 (0.007) & 0.307 [0.284, 0.376]\\
1000 & 20000 & 52.39 (0.43) & 0.128 (0.002) & 0.918 [0.860, 0.985]\\
4000 & 80000 & 58.76 (0.25) & 0.121 (0.001) & 3.827 [3.710, 3.927]\\
\bottomrule
\end{tabular}
\par\smallskip
\begin{minipage}{\linewidth}\footnotesize
Parentheses contain MCSEs; runtime is the median [first, third quartile]. In (a), NLL differences and center displacements are relative to the $M/n=80$ endpoint on the same dataset; all NLLs use the common independent scoring cloud. Displacement is matched center-coordinate RMSE between fitted endpoints. In (b), RMSE is relative to the generating means.
\end{minipage}
\end{table}

\clearpage
\endgroup

		\begingroup
\subsection{Bootstrap and Fisher center intervals}
\label{sec:revision-coverage}

We generate 200 datasets at each $n\in\{200,500\}$ from the one-dimensional WFCM density with centers $(-2,2)$, weights $(0.35,0.65)$, $\sigma=0.9$, and $m=2$, using numerical inverse-CDF sampling.

At fixed $k=2,m=2$, nominal 95\% model-Fisher intervals use inverse information in the free fixed-$m$ coordinates, divided by $n$, with information estimated from 50,000 independent fitted-density draws. This benchmark omits finite-IS covariance. Normal-approximation intervals follow Algorithm~\ref{alg:inference}, using 199 parametric bootstrap samples from the fitted WFCM density. Bootstrap centers are aligned to the observed fit, and each refit rebuilds the proposal and importance-sampling cloud. Each refit uses a single start at the observed fit; the observed fit uses three seeded starts. Fits use $M=\max(4000,20n)$ importance samples. All 400 observed fits and 79,600 parametric bootstrap refits passed the numerical checks.

In the upper panel of Table~\ref{tab:revision-coverage}, normal-bootstrap and Fisher coverage is 93.5\% and 92.2\% at $n=200$, and 96.2\% and 95.8\% at $n=500$. Normal-bootstrap intervals are 5.2\% and 3.1\% longer, respectively, and require more computation.

As a supplementary comparison, the lower panel reports nonparametric percentile intervals under the same design and fitting settings. Each of the 199 bootstrap samples resamples the observed data with replacement, and the interval is $[Q_{0.025}(\hat\theta_j^*),Q_{0.975}(\hat\theta_j^*)]$, where $Q_q$ denotes the empirical $q$-quantile of the aligned estimates. All 400 observed fits and 79,600 nonparametric refits passed the numerical checks.

\begin{table}[htbp]
\centering\small\singlespacing
\caption{Nominal 95\% marginal center intervals using Algorithm~\ref{alg:emfuzzy}: means and MCSEs.}
\label{tab:revision-coverage}
\setlength{\tabcolsep}{5pt}
\begin{tabular}{rlccc}
\toprule
$n$ & Method & Coverage (\%) & Length & CPU time (s)\\
\midrule
\multicolumn{5}{l}{\textit{Parametric normal bootstrap comparison}}\\
200 & Normal bootstrap & 93.5 (1.2) & 0.540 (0.006) & 67.739 (0.581)\\
200 & Model Fisher & 92.2 (1.3) & 0.513 (0.005) & 0.979 (0.014)\\
500 & Normal bootstrap & 96.2 (0.9) & 0.338 (0.002) & 111.119 (0.539)\\
500 & Model Fisher & 95.8 (1.0) & 0.328 (0.002) & 1.846 (0.023)\\
\midrule
\multicolumn{5}{l}{\textit{Supplementary nonparametric percentile comparison}}\\
200 & Percentile bootstrap & 94.0 (1.2) & 0.543 (0.008) & 49.654 (0.748)\\
200 & Model Fisher & 92.2 (1.3) & 0.513 (0.005) & 0.979 (0.014)\\
500 & Percentile bootstrap & 96.0 (1.0) & 0.340 (0.003) & 92.221 (0.931)\\
500 & Model Fisher & 95.8 (1.0) & 0.328 (0.002) & 1.846 (0.023)\\
\bottomrule
\end{tabular}

\par\smallskip
\begin{minipage}{\linewidth}\footnotesize
Each sample size has 200 planned datasets. Coverage and length average the two aligned centers within each dataset; MCSEs use datasets as independent units. Both panels report CPU seconds, including the observed-data fit.
\end{minipage}
\end{table}

\subsection{Null rejection and power}
\label{sec:revision-lrt}

At fixed $m=2$ and nominal level 0.05, the null is $N(0,1)$; alternatives have WFCM centers $(-\Delta/2,\Delta/2)$, weights $(1/2,1/2)$, $\sigma=1/\sqrt2$, and $\Delta\in\{1,2,3\}$. Each $(n,\Delta)$ cell uses 150 datasets. The one-center null is Gaussian and fitted analytically. Two-center fits use $M=\max(4000,15n)$ importance samples and three starts including the embedded null, with at most 450 iterations per start. The fitted alternatives are evaluated using 100,001-point quadrature with local refinement.

Writing $\ell_0$ for the fitted-null log-likelihood and $\ell_1^{\rm score}$ for the numerically evaluated fitted-alternative log-likelihood, we retain the null fit as a candidate and compute
\[
T=2\{\max(\ell_1^{\rm score},\ell_0)-\ell_0\}.
\]
Thus, $T=0$ when numerical fitting yields an alternative score below the null log-likelihood. Such cases occurred frequently: under the null, $T=0$ in 60 of 150 datasets at $n=200$, 72 of 150 at $n=500$, and 296 of 600 in the confirmation run, and in every cell about 42--47\% of bootstrap statistics were zero.

For each dataset, we draw 199 Gaussian bootstrap samples from the fitted null and repeat the same fitting and scoring procedure. The bootstrap p-value is
\[
p=\frac{1+\sum_{b=1}^{B_v}\mathbf1\{T_b^*\ge T\}}{1+B_v},
\]
where $B_v$ is the number of bootstrap refits satisfying the convergence criteria. All 1,200 analyses had at least 197 valid refits out of 199, with 238,766 valid refits out of 238,800 overall.

Table~\ref{tab:revision-lrt} reports the empirical null rejection rates and power of this procedure. The null rejection rate at $n=500$ exceeds the nominal level. Following this result, we pre-registered a protocol for, and then completed, an independent confirmation run using 600 datasets and the same procedure. Its result is reported separately alongside the primary result.

\begin{table}[htbp]
\centering\small\singlespacing
\caption{Null rejection and power at nominal 5\% (MCSE in percentage points).}
\label{tab:revision-lrt}
\setlength{\tabcolsep}{9pt}
\begin{tabular}{rrrr}
\toprule
$n$ & $\Delta$ & Datasets & Rejection (\%)\\
\midrule
200 & 0 & 150 & 4.7 (1.7)\\
200 & 1 & 150 & 39.3 (4.0)\\
200 & 2 & 150 & 99.3 (0.7)\\
200 & 3 & 150 & 100.0 (0.0)\\
500 & 0 & 150 & 9.3 (2.4)\\
500 & 1 & 150 & 76.7 (3.5)\\
500 & 2 & 150 & 100.0 (0.0)\\
500 & 3 & 150 & 100.0 (0.0)\\
\midrule
\multicolumn{4}{l}{Independent null confirmation}\\
500 & 0 & 600 & 3.83 (0.78)\\
\bottomrule
\end{tabular}
\par\smallskip
\begin{minipage}{\linewidth}\footnotesize
When $\Delta=0$, the data follow $N(0,1)$, and the rejection rate estimates the type I error rate. When $\Delta>0$, the two centers are separated by $\Delta$, and the rejection rate estimates power. Results are reported as rejection percentages, with Monte Carlo standard errors in parentheses.
\end{minipage}
\end{table}
\clearpage
\endgroup

		\clearpage
		\section{Derivation of the WFCM Density from the FCM Loss}
		\label{app:density-derivation}
		
		We retain the algebraic connection between the WFCM loss and the density, while distinguishing this identity from the additional modeling choice and the parameter-dependent normalizer.
		
		\subsection*{Step 1. Per-observation decomposition of the WFCM loss}
		
		The WFCM loss with fuzzifier $m > 1$ and cluster weights $w_j > 0$ is
		\begin{equation}
			\label{eq:wfcm-loss}
			J_m^{\mathrm{WFCM}}(V,w)
			= \sum_{i=1}^{n} \sum_{j=1}^{k} u_{ij}^{m}\, w_j\, \|x_i - v_j\|^2,
		\end{equation}
		where the optimal memberships satisfy
		\begin{equation}
			\label{eq:membership-appendix}
			u_{ij} = \frac{(w_j\,\|x_i - v_j\|^2)^{-1/(m-1)}}
			{\displaystyle\sum_{l=1}^{k}(w_l\,\|x_i - v_l\|^2)^{-1/(m-1)}}.
		\end{equation}
		Let $d_{ij} = \|x_i - v_j\|$ and $S_i = \sum_{l=1}^k (w_l d_{il}^2)^{-1/(m-1)}$.
		Substituting \eqref{eq:membership-appendix} into the inner sum of \eqref{eq:wfcm-loss}:
		\begin{align}
			\sum_{j=1}^{k} u_{ij}^m\,w_j d_{ij}^2
			&= \sum_{j=1}^{k} \frac{(w_j d_{ij}^2)^{-m/(m-1)}}{S_i^m}\cdot w_j d_{ij}^2
			\notag \\
			&= \frac{1}{S_i^m} \sum_{j=1}^{k} (w_j d_{ij}^2)^{-m/(m-1)+1}
			\notag \\
			&= \frac{1}{S_i^m} \sum_{j=1}^{k} (w_j d_{ij}^2)^{-1/(m-1)}
			\notag \\
			&= \frac{S_i}{S_i^m}
			= S_i^{-(m-1)}.
			\label{eq:per-obs-simplification}
		\end{align}
		Here the third equality uses $-m/(m-1)+1 = -1/(m-1)$. At a zero distance, the minimum loss contribution is zero; the expression is interpreted by continuity, with membership supported on the coincident centers.
		Therefore the WFCM loss decomposes \emph{exactly} into a sum of per-observation
		terms:
		\begin{equation}
			\label{eq:loss-decomposed}
			J_m^{\mathrm{WFCM}}(V,w)
			= \sum_{i=1}^{n} \varphi(x_i;\,V,w,m),
			\qquad
			\varphi(x;\,V,w,m)
			= \left[\sum_{j=1}^{k}
			\bigl(w_j\,\|x - v_j\|^2\bigr)^{-1/(m-1)}
			\right]^{-(m-1)}.
		\end{equation}
		
		\begingroup
		\subsection*{Step 2. A Gibbs density with energy $\varphi$}
		For $\sigma>0$, let $\mathcal E(x;\theta)=\sigma^{-2}\varphi(x;V,w,m)$. Choosing Lebesgue measure and the exponential kernel gives
		\begin{equation}\label{eq:density-derivation}
			f(x;\theta)=C(\theta)e^{-\mathcal E(x;\theta)},\qquad
			C(\theta)^{-1}=\int_{\mathbb R^d}e^{-\mathcal E(x;\theta)}\,dx.
		\end{equation}
		This normalization is unique for the chosen kernel, but the choice of kernel is an additional modeling assumption. For positive weights, put $A=\sum_j w_j^{-1/(m-1)}$ and $R=\max_j\|v_j\|$. If $\|x\|>2R$, then
		\[
		\varphi(x;V,w,m)\geq A^{-(m-1)}\min_j\|x-v_j\|^2
		\geq \tfrac14 A^{-(m-1)}\|x\|^2.
		\]
		The kernel is bounded by one on bounded sets and has an integrable Gaussian tail, so the normalizer is finite and positive. This gives precisely~\eqref{eq:WFCM-density}.
		
		\subsection*{Step 3. The likelihood includes the normalizer}
		For independent observations, the identity is
		\begin{equation}\label{eq:nll-equals-loss}
			-\ell_n(\theta)=-n\log C(\theta)
			+\sigma^{-2}J_m^{\mathrm{WFCM}}(V,w).
		\end{equation}
		The first term generally varies with $V,w,m,\sigma$. Consequently, WFCM loss minimization and likelihood maximization are different estimation problems.
		
		\begin{remark}
			The working density permits likelihood-based inference under the assumptions of the corresponding results, while the memberships remain geometric quantities. Nonparametric resampling and model-based Fisher information have distinct validity conditions, as discussed in Remark~\ref{rem:density-derivation}.
		\end{remark}
		\endgroup
		
		\section{Related Literature}\label{sec:sup_related-literature}
		
		Clustering is a core problem in statistics and machine learning that aims to group observations with high within-cluster similarity and low between-cluster similarity \citepS{jain2010data,duda2006pattern,basu2008constrained,ball1965isodata}. Classical approaches include hard partitioning methods such as $k$-means and hierarchical clustering, as well as probabilistic model-based clustering \citepS{jain1988algorithms,meilua2006uniqueness}.
		
		Soft clustering assigns each observation a vector of membership weights. Two prominent paradigms are mixture-model methods (e.g., Gaussian mixture models, GMMs) and distance-based fuzzy clustering. GMMs posit a finite mixture of Gaussians and estimate parameters by maximum likelihood, typically via EM \citepS{dempster1977maximum,mclachlan2000finite}, enabling likelihood-based uncertainty quantification but requiring covariance estimation and distributional assumptions that can be fragile in high-dimensional or small-sample regimes. In contrast, fuzzy $c$-means (FCM) \citepS{bezdek1984fcm} is a distance-based alternative that optimizes a distance-based objective with a fuzziness parameter controlling softness, and has been widely used in applications \citepS{pham1999adaptive,hathaway2006extending}. Numerous extensions improve robustness and flexibility, including adaptive/metric variants, regularized formulations, spatially constrained and entropy-penalized versions, and kernel/possibilistic approaches \citepS{gustafson1979fuzzy,gath1989unsupervised,fan2003suppressed,ji2011modified,zhang2003clustering,zhang2004novel}.
		
		Despite extensive algorithmic development, formal statistical inference for optimization-based fuzzy clustering remains limited. The FCM objective is not derived from a probability model, so fitted memberships and centers carry no standard errors and hypotheses about cluster parameters cannot be tested. This gap has been approached from several directions, none of which addresses it within the fuzzy clustering framework itself.
		Overlapping and mixed-membership models \citepS{banerjee2005model,xie2013overlapping} allow an observation to belong to several clusters, as fuzzy memberships do, but specify the generative mechanism at the outset rather than deriving it from a clustering objective; the resulting memberships need not coincide with those an FCM practitioner would obtain.
		Post-clustering, or selective, inference \citepS{taylor2015statistical,gao2024selective,yun2023selective,zhu2026functional} takes the opposite route, retaining the clustering algorithm and correcting the subsequent inference for the fact that hypotheses were formed after seeing the data. This is complementary to our aim rather than an alternative: selective inference conditions on a realized hard partition and asks whether the resulting groups differ, whereas we specify a normalized working density for the observations, permitting likelihood-based inference for its parameters $(V,w,\sigma)$ at fixed $k,m$ under the stated conditions.
		Bayesian nonparametric formulations, including Dirichlet process mixtures \citepS{antoniak1974mixtures} and the Indian buffet process for overlapping memberships \citepS{griffiths2011indian}, estimate cluster structure and quantify uncertainty jointly through the posterior, and are the closest existing answer to the inferential question. Our approach differs in its starting point. We construct a normalized working density motivated by the weighted FCM objective. Because the normalizing constant depends on the model parameters, the resulting likelihood-based estimator generally differs from the loss-minimizing WFCM estimator; the likelihood inference developed here concerns the former.
		
		Cluster-size imbalance has been addressed through several algorithmic modifications to the standard FCM objective. \citetS{dave1991characterization} introduced noise clustering, and \citetS{krishnapuram1993possibilistic} introduced possibilistic $c$-means, which removes the partition-sum constraint and down-weights outliers; however, these methods do not specifically target minority-cluster dominance, and possibilistic $c$-means requires careful initialization to avoid cluster coincidence. Methods aimed directly at cluster-size imbalance include cluster-size-insensitive FCM \citepS{noordam2002multivariate}, which rescales memberships by estimated relative cluster sizes so that large clusters exert less pull on small-cluster prototypes, and size-insensitive integrity-based FCM \citepS{lin2014size}, which modifies this rescaling using a measure of cluster integrity. Like the polynomial fuzzifier below, these are modifications of the FCM iterations without an explicit normalized density; in WFCM, the cluster weights are instead parameters of a working likelihood, estimated by maximum likelihood together with the centers and scale. \citetS{klawonn2003alternative} and \citetS{winkler2011fuzzy} proposed the \emph{polynomial fuzzifier}: the power penalty $u^m$ is replaced by a polynomial function of $u$, changing the membership and prototype updates. This is an optimization-based construction, for which uncertainty could be assessed by resampling or estimating equations. In contrast, our WFCM introduces cluster-specific weights $w_j$ that enter the density $f(\mathbf{x};\,V,w,m,\sigma)$ explicitly (Section~\ref{sec:method}), enabling maximum-likelihood estimation with identifiable parameters and provable asymptotic guarantees.
		\citetS{gustafson1979fuzzy} proposed adaptive-metric FCM (Gustafson-Kessel) with cluster-specific covariance matrices, which handles anisotropic geometry but introduces $O(kd^2)$ additional parameters and loses the simplicity of WFCM's centroid-based updates. \citetS{hathaway2006extending} addressed large-scale FCM via streaming extensions but without inferential guarantees.
		
		A parallel line of work addresses heterogeneous feature relevance by incorporating per-feature weights into the clustering objective. \citetS{huang2005automated} proposed variable-weighted $k$-means, which alternates between hard assignments, center updates, and updates of feature weights raised to a power. \citetS{jing2007entropy} developed an entropy-weighted subspace clustering formulation with cluster-specific feature weights, in which an entropy penalty keeps the weights from concentrating on a few features. Within the FCM framework specifically, feature relevance has been incorporated via adaptive distance metrics \citepS{gustafson1979fuzzy} and via regularized objectives that penalize uniform weight vectors \citepS{pedrycz2005knowledge}.
		
		Our WFCM instead places weights on \emph{clusters} rather than features, targeting cluster-size imbalance rather than feature irrelevance; the two mechanisms are complementary and could in principle be combined. 
		Several authors have sought probabilistic or likelihood-based interpretations of fuzzy clustering objectives. \citetS{gath1989unsupervised} proposed an unsupervised optimal fuzzy clustering algorithm whose objective is derived from a maximum-likelihood criterion under Gaussian cluster models with cluster-specific covariances, making it an early example of likelihood-driven fuzzy partitioning. \citetS{miyamoto1997fuzzy} interpreted the fuzzifier $m$ of standard FCM as a regularization of hard $c$-means and proposed an entropy-regularized alternative, with penalty weight $\lambda>0$, whose memberships take an exponential form, $u_{ij}\propto\exp(-\|\bx_i-\bv_j\|^2/\lambda)$, linking fuzzy clustering to a maximum-entropy formulation. These contributions establish that a probabilistic reading of fuzzy clustering is possible, but neither builds a normalized density for the observations from the fuzzy clustering objective nor provides asymptotic inference for the resulting estimators, which is the gap that motivates our construction in Section~\ref{sec:method}.
		
		Finally, cluster validity indices are commonly used for model selection in fuzzy clustering. The Xie--Beni index (XBI) \citepS{xie1991new} measures a compactness--separation trade-off and is frequently used to select the number of clusters and the fuzziness parameter. In this paper, we use a heuristic weighted Xie--Beni index adapted to our weighted fuzzy $c$-means (WFCM) framework.
		
		\begingroup
\section{Theory for the Approximate Likelihood Estimator}\label{app:approx-mle}
This appendix supplies the assumptions and proof of Theorem~\ref{thm:approx-transfer}. We distinguish approximation accuracy from the ability of a local algorithm to find the relevant solution.

\subsection{Conditional importance-sampling assumptions}
Let $\mathcal D_n$ be the $\sigma$-field generated by $X_1,\ldots,X_n$. The proposal $\hat p_n$ is $\mathcal D_n$-measurable, positive wherever the kernel or its derivatives are nonzero, and $Y_1,\ldots,Y_{M_n}$ are conditionally i.i.d. from $\hat p_n$. Write $g_\vartheta=g_{T(\vartheta)}$. The fitted proposal is not differentiated with respect to $\vartheta$.

\paragraph{Uniform approximation condition (U).}
For orders $a=0,1,2$, assume
\begin{equation}\label{eq:is-uniform-condition}
 \sup_{\vartheta\in\Xi}
 \left\|\frac1{M_n}\sum_{r=1}^{M_n}
 \frac{D^a g_\vartheta(Y_r)}{\hat p_n(Y_r)}-D^a Z(\vartheta)\right\|
 =O_p(M_n^{-1/2}).
\end{equation}
Derivatives are taken on an open regular extension of the parameter set. For consistency alone only the value condition, with $o_p(1)$ error, is needed. The lower bound on $Z$ and the derivative envelopes in Appendix~\ref{app:normality} give, by differentiation of the logarithm,
\begin{equation}\label{eq:criterion-is-errors}
 \Delta_{a,n}:=\sup_{\vartheta\in\Xi}
 \|D^a\widehat Q_n(\vartheta)-D^aQ_n(\vartheta)\|
 =O_p(M_n^{-1/2}),\qquad a=0,1,2.
\end{equation}

Condition (U) is a uniform empirical-process requirement over the parameter set. The following lemmas give a sufficient proposal construction and verify the derivative, moment, and entropy bounds explicitly.
\begin{lemma}[Uniform H\"older envelopes]\label{lem:is-holder}
Let $s=kd+k+1$ be the dimension of the free coordinates and choose
$0<\alpha<\min\{1,2/(m_{\max}-1)\}$. There are constants $C,c>0$ such that
$\Gamma(x)=C(1+\|x\|^6)e^{-c\|x\|^2}$ satisfies, for every multi-index $a$ with $|a|\leq2$ and every $\vartheta,\vartheta'\in\Xi$,
\begin{align}
 |D^ag_\vartheta(x)|&\leq \Gamma(x),\label{eq:is-holder-envelope}\\
 |D^ag_\vartheta(x)-D^ag_{\vartheta'}(x)|
 &\leq \Gamma(x)\|\vartheta-\vartheta'\|^\alpha.\label{eq:is-holder-increment}
\end{align}
The derivatives at centers are their continuous extensions. Constants may depend on the fixed parameter bounds and $\alpha$, but not on $x$ or the parameters.
\end{lemma}

\begin{proof}
Choose $\alpha<\beta<\min\{1,2/(m_{\max}-1)\}$. Enlarge the parameter bounds slightly, preserving positive weights and scale, $m>1$, and a positive center-separation margin, so that $q=2/(m-1)\geq q_0>\beta$ throughout the enlarged set. Any sufficiently short segment joining two points of $\Xi$ lies in this regular extension. It need not lie in $\Xi$ itself.

Near one center, write $r=\|x-v_j\|$ and use \eqref{eq:mle-center-factorization}. With
\[
 F(z,t)=\frac{(1+z)^{-t}-1}{z},\qquad F(0,t)=-t,
\]
the remainder has the exact representation
\[
 R_\vartheta(x):=E_\vartheta(x)-A_jr^2
 =A_jK_\vartheta(x)r^{2+q}
       F\{K_\vartheta(x)r^q,t\}.
\]
The other centers remain separated from $x$, so $K$ and its first three parameter derivatives are bounded in this neighborhood. The same is true of the derivatives of $F$ on the relevant compact $(z,t)$ set. Differentiating a radial power with respect to a center lowers its power by at most one; differentiating with respect to $m$ introduces at most one factor $\log r$. The product and chain rules consequently give, for $0<r<r_0\leq1$,
\begin{equation}\label{eq:is-remainder-derivatives}
 \max_{|a|=j}|D^aR_\vartheta(x)|
 \leq C r^{2+q_0-j}(1+|\log r|^3)
 \leq C_\beta r^{2+\beta-j},\qquad j=0,1,2,3.
\end{equation}
For clarity, the radial bound used here is
$|\partial_v^b\partial_m^l r^{\lambda(m)}|
\leq C r^{\lambda(m)-|b|}(1+|\log r|^l)$ for $|b|+l\leq3$, applied to the powers in the displayed factorization. Extra factors $r^q$ from differentiating $F$ only improve the power. The last inequality in \eqref{eq:is-remainder-derivatives} follows from
$\sup_{0<r\leq1}r^{q_0-\beta}(1+|\log r|^3)<\infty$.

Put $e=\|\vartheta-\vartheta'\|$ and first take $e$ sufficiently small. Shrink the center neighborhood so that the factorization and bounds remain valid along the joining segment. If $r\leq2e$, then $\|x-v'_j\|\leq3e$; the second-derivative bound in \eqref{eq:is-remainder-derivatives} bounds the two remainder Hessians by $Ce^\beta$. If $r>2e$, the center distance along the joining segment is at least $r/2$. Integration of the third-derivative bound along this segment gives a Hessian difference at most
\[
 Ce r^{\beta-1}\leq C'e^\beta.
\]
The Hessian of $A_j\|x-v_j\|^2$ is Lipschitz on bounded sets. These bounds prove the required H\"older control for $D^2E$ near moving centers, including when an endpoint is a center. The bounded first and second derivatives similarly give Lipschitz control for $E$ and $DE$. Away from the centers on a bounded $x$ set, all third derivatives are bounded and ordinary mean-value bounds apply. A finite cover gives uniform constants. For parameter pairs farther apart, the bounded envelopes give the increment bound after enlarging the constant. Since $\beta>\alpha$, the resulting bounds imply exponent $\alpha$ on bounded parameter sets. Only derivatives away from centers were taken to order three; no third derivative at a center is required.

Finally, for large $\|x\|$, the tail factorization in Appendix~\ref{app:normality} gives $E_\vartheta(x)\geq c\|x\|^2$ and
$|D^aE_\vartheta(x)|\leq C(1+\|x\|^2)$ through order three. Thus
$|D^ag_\vartheta(x)|\leq C(1+\|x\|^6)e^{-c\|x\|^2}$ through order three in the tails. Mean-value bounds there give \eqref{eq:is-holder-increment} for nearby parameter pairs; the envelope handles the remaining pairs. On bounded $x$ sets, the identities
$Dg=-gDE$ and
$D^2g=g\{DE(DE)^{\mathsf T}-D^2E\}$ transfer the established bounds to $g$. Increasing $C$ on that bounded set yields a single envelope $\Gamma$ for both \eqref{eq:is-holder-envelope} and \eqref{eq:is-holder-increment}.
\end{proof}

\begin{lemma}[A defensive proposal implies (U)]\label{lem:is-defensive}
Let $p_0$ be a fixed $d$-variate Student $t$ density with positive degrees of freedom and positive definite scale matrix. Suppose, almost surely, that
\begin{equation}\label{eq:is-defensive-lower-bound}
 \hat p_n(x)\geq\epsilon_{\rm IS}p_0(x),\qquad
 0<\epsilon_{\rm IS}<1,
\end{equation}
where $\epsilon_{\rm IS}$ does not depend on $n$. Conditional on $\mathcal D_n$, draw $M_n$ independent points from $\hat p_n$, with $M_n\to\infty$. Then (U) holds for orders zero, one, and two. In particular, \eqref{eq:is-defensive-lower-bound} holds for
\[
 \hat p_n=(1-\epsilon_{\rm IS})\hat p_{n,\rm GMM}
            +\epsilon_{\rm IS}p_0
\]
with any data-measurable fitted GMM that is a proper density. No compactness restriction on its fitted means or covariances is needed for this lemma.
\end{lemma}

\begin{proof}
Condition on the data and abbreviate $p=\hat p_n$ and $M=M_n$. For a scalar derivative component, set
\[
 f_{\vartheta,p}(x)=\frac{D^ag_\vartheta(x)}{p(x)},\qquad
 \mathcal F_{a,p}=\{f_{\vartheta,p}:\vartheta\in\Xi\},\qquad
 F_p(x)=\frac{\Gamma(x)}{p(x)}.
\]
Here $P_p$ denotes integration against $p$. Lemma~\ref{lem:is-holder} gives an envelope $F_p$ and the pointwise increment bound
\begin{equation}\label{eq:is-weighted-holder}
 |f_{\vartheta,p}(x)-f_{\vartheta',p}(x)|
 \leq F_p(x)\|\vartheta-\vartheta'\|^\alpha.
\end{equation}
For every fixed $\eta\geq0$,
\begin{equation}\label{eq:is-uniform-moment-bound}
 P_pF_p^{2+\eta}
 =\int\frac{\Gamma(x)^{2+\eta}}{p(x)^{1+\eta}}\,dx
 \leq\epsilon_{\rm IS}^{-(1+\eta)}
       \int\frac{\Gamma(x)^{2+\eta}}{p_0(x)^{1+\eta}}\,dx
 =:K_\eta<\infty.
\end{equation}
Indeed, if $p_0$ has $\nu>0$ degrees of freedom, then
$p_0(x)\geq c_0(1+\|x\|)^{-(d+\nu)}$; the integrand is bounded by a polynomial times a Gaussian tail. The constants $K_\eta$ are deterministic and uniform in $n$ and the fitted GMM.

We next give the entropy bound. The bounded set $\Xi\subset\mathbb R^s$ has a Euclidean $u$-net with at most $C u^{-s}$ points for $0<u\leq1$. By \eqref{eq:is-weighted-holder}, the brackets
\[
 [f_{\vartheta_j,p}-u^\alpha F_p,\,
   f_{\vartheta_j,p}+u^\alpha F_p]
\]
cover $\mathcal F_{a,p}$ and have $L^2(p)$ width $2u^\alpha\|F_p\|_{P_p,2}$. Hence, uniformly in $p$, for $0<\varepsilon\leq1$,
\begin{align}
 N_{[]}\{\varepsilon\|F_p\|_{P_p,2},\mathcal F_{a,p},L^2(p)\}
 &\leq C\varepsilon^{-s/\alpha},\label{eq:is-entropy-bound}\\
 \int_0^1\sqrt{\log(C\varepsilon^{-s/\alpha})}\,d\varepsilon
 &<\infty.\nonumber
\end{align}
The same pointwise brackets work with any empirical $L^2$ norm. Continuity in $\vartheta$ makes each supremum equal to one over a countable dense subset of $\Xi$, so these suprema are measurable.

For completeness, the following chaining argument turns these bounds into the required rate uniformly over the fitted proposal. Let $P_M$ be the empirical measure of the cloud, let $\varepsilon_1,\ldots,\varepsilon_M$ be independent Rademacher signs, and define
\[
 Z_\vartheta=M^{-1/2}\sum_{r=1}^M
      \varepsilon_r f_{\vartheta,p}(Y_r),\qquad
 A_M=(P_MF_p^2)^{1/2}.
\]
Conditional on the cloud, $\cosh z\leq e^{z^2/2}$ gives
\[
 \mathbb E_\varepsilon e^{\lambda(Z_\vartheta-Z_{\vartheta'})}
 \leq\exp\left\{\frac{\lambda^2}{2}
       P_M(f_{\vartheta,p}-f_{\vartheta',p})^2\right\},
\]
and the variance proxy is at most
$A_M^2\|\vartheta-\vartheta'\|^{2\alpha}$. Use parameter nets of radius $2^{-l/\alpha}$, with cardinality at most $C2^{ls/\alpha}$, and choose a net point $\pi_l\vartheta$ within that radius of each $\vartheta$. For the increments $Z_{\pi_l\vartheta}-Z_{\pi_{l-1}\vartheta}$, the variance proxies are at most $C A_M^2 2^{-2l}$ and the number of possible pairs is at most $C2^{2ls/\alpha}$. The exponential bound and a union bound therefore bound the expected maximum absolute increment at level $l$ by
$C A_M2^{-l}\sqrt{1+l}$. The initial finite net contributes at most $C A_M$. Telescoping the nets is valid by sample-path continuity, and thus
\begin{equation}\label{eq:is-chaining-bound}
 \mathbb E_\varepsilon\sup_{\vartheta\in\Xi}|Z_\vartheta|
 \leq C A_M\left\{1+\sum_{l=1}^{\infty}2^{-l}\sqrt{1+l}\right\}
 \leq C' A_M.
\end{equation}
Symmetrization using an independent cloud, followed by Jensen's inequality, now gives
\begin{align}
 &\mathbb E\left[\sup_{\vartheta\in\Xi}
    \left|\sqrt M(P_M-P_p)f_{\vartheta,p}\right|
       \,\middle|\,\mathcal D_n\right]\nonumber\\
 &\hspace{1cm}\leq2\mathbb E\left[
      \mathbb E_\varepsilon\sup_{\vartheta\in\Xi}|Z_\vartheta|
          \,\middle|\,\mathcal D_n\right]
 \leq C\|F_p\|_{P_p,2}\leq C\sqrt{K_0}.\label{eq:is-conditional-maximal-bound}
\end{align}
Since $P_pf_{\vartheta,p}=\int D^ag_\vartheta(x)\,dx=D^aZ(\vartheta)$, conditional Markov's inequality proves (U). There are only finitely many derivative components through order two, so the same conclusion holds for their vector and matrix norms.
\end{proof}

The proof also covers a data-adaptive Student component with fixed degrees of freedom, bounded location, and scale eigenvalues bounded above and away from zero on events with probability tending to one. On those events its density has a common polynomial lower bound, so the same constants apply; the probabilities of the complementary events are added to the probability bounds. This establishes (U); the covariance convergence in (MC) below is a separate condition. The defensive component is what secures these importance moments.

\paragraph{Conditional Monte Carlo limit condition (MC).}
For the distributional conclusion, define
\begin{equation}\label{eq:mc-score-covariance}
\begin{split}
 \psi(x)&=\nabla E_{\vartheta_0}(x)+\nabla\log Z(\vartheta_0),\\
 \xi_n(y)&=\frac{\nabla g_{\vartheta_0}(y)-g_{\vartheta_0}(y)
                  \nabla\log Z(\vartheta_0)}{Z(\vartheta_0)\hat p_n(y)}
          =-\frac{f_{\vartheta_0}(y)}{\hat p_n(y)}\psi(y),\\
 \Omega_n&=\mathbb E\{\xi_n(Y)\xi_n(Y)^{\mathsf T}\mid\mathcal D_n\}
           \xrightarrow{p}\Omega_*.
\end{split}
\end{equation}
Assume the limit $\Omega_*$ is deterministic and, for every $\epsilon>0$, the conditional Lindeberg condition
\[
 \mathbb E\{\|\xi_n(Y)\|^2
   1(\|\xi_n(Y)\|>\epsilon\sqrt{M_n})\mid\mathcal D_n\}
 \xrightarrow{p}0
\]
holds. A conditional $(2+\epsilon)$ moment bounded in probability is sufficient for the latter. The conditional mean of $\xi_n$ is exactly zero. Conditional independence and these conditions give
$M_n^{-1/2}\sum_r\xi_n(Y_r)\Rightarrow\mathcal N(0,\Omega_*)$ with conditional characteristic functions converging in probability to the nonrandom Gaussian characteristic function. Multiplying these characteristic functions by any bounded function of the data shows that this limit is asymptotically independent of the data score. Thus fitting the proposal to the same data does not require an additional independent pilot sample. A regular fitted proposal converging to a deterministic limit, with the above uniform moment bounds, is one way to ensure covariance convergence.

\subsection{Consistency and error transfer}
The numerical output must belong to the stated compact feasible set (or do so on events with probability tending to one). The grid changes only $m$, so rounding $m$ preserves feasibility. From Appendix~\ref{sec:pftheory} and (U),
\[
 \sup_{\vartheta\in\Xi}|\widehat Q_n(\vartheta)-Q(\vartheta)|=o_p(1),
 \qquad Q(\vartheta)=\mathbb E_{\theta_0}\{-\log f_\vartheta(X)\}.
\]
If $h_n\to0$, uniform continuity of $Q$ implies that its infimum on $\Xi_{h_n}$ tends to $Q(\vartheta_0)$. With $r_n=o_p(1)$, the output is therefore within $o_p(1)$ of the minimum population value. The strict orbit gap established in Appendix~\ref{sec:pftheory} gives consistency. In particular, data-adaptive box constraints or unconstrained multistart routines must separately be shown to satisfy this feasible-set and optimization requirement.

Let $\bar\vartheta_n$ be an exact continuous minimizer of $\widehat Q_n$ over $\Xi$. The same argument makes it consistent. After matching labels, both $\bar\vartheta_n$ and $\hat\vartheta_n$ lie in an interior convex neighborhood of $\vartheta_0$ with probability tending to one. The local Hessian law and \eqref{eq:criterion-is-errors} bound the eigenvalues of both criterion Hessians above and away from zero there. Since $\nabla Q_n(\hat\vartheta_n)=0$ and $\nabla\widehat Q_n(\bar\vartheta_n)=0$, an integral Taylor expansion gives
\begin{equation}\label{eq:continuous-is-distance}
 \|\bar\vartheta_n-\hat\vartheta_n\|\leq C\Delta_{1,n}
 =O_p(M_n^{-1/2}).
\end{equation}
Round only the $m$ coordinate of $\bar\vartheta_n$ to a grid value, obtaining $\bar\vartheta_n^g$. The zero gradient at the continuous minimizer and the upper Hessian bound imply
\[
 \widehat Q_n(\bar\vartheta_n^g)-\widehat Q_n(\bar\vartheta_n)=O_p(h_n^2).
\]
Consequently, by the definition of $r_n$ and the lower local Hessian bound,
\[
 c\|\tilde\vartheta_n-\bar\vartheta_n\|^2
 \leq\widehat Q_n(\tilde\vartheta_n)-\widehat Q_n(\bar\vartheta_n)
 \leq O_p(h_n^2)+r_n.
\]
Consistency justifies using the same local neighborhood for these points. Combining the last two bounds proves \eqref{eq:approx-transfer-rate}. The comparison is with the continuous minimizer; one cannot presume that a discrete grid minimizer is stationary or unique.

\subsection{First-order limit and Monte Carlo variance}
At $\vartheta_0$, a ratio expansion of $\nabla\log\widehat Z_n$ gives
\begin{equation}\label{eq:is-score-expansion}
 \nabla\widehat Q_n(\vartheta_0)
 =\frac1n\sum_i\psi(X_i)+\frac1{M_n}\sum_r\xi_n(Y_r)+O_p(M_n^{-1}).
\end{equation}
For example, the first-order ratio term is
$Z_0^{-1}\{(\nabla\widehat Z_n-\nabla Z_0)
-(\widehat Z_n-Z_0)\nabla Z_0/Z_0\}$; its product remainder has the stated order by (U). Put $H=\nabla^2Q(\vartheta_0)=I_r$ and $J=\operatorname{Var}_{\theta_0}\psi(X)=I_r$. If $n/M_n=O(1)$, the ratio remainder is $o_p(n^{-1/2})$. Taylor expansion at the consistent continuous minimizer, followed by the grid and numerical bound, yields
\begin{equation}\label{eq:approx-linear-representation}
\begin{split}
 \sqrt n(\tilde\vartheta_n-\vartheta_0)
 ={}&-H^{-1}\left\{\frac1{\sqrt n}\sum_i\psi(X_i)
       +\sqrt{\frac n{M_n}}\frac1{\sqrt{M_n}}\sum_r\xi_n(Y_r)\right\}+R_n,\\
 \|R_n\|={}&O_p(\sqrt n h_n+\sqrt{nr_n})+o_p(1).
\end{split}
\end{equation}
The data central limit theorem and (MC) give covariance
$H^{-1}(J+\rho\Omega_*)H^{-1}$ when $n/M_n\to\rho$ and the displayed grid and optimization remainder vanishes. This proves \eqref{eq:approx-transfer-clt}. For $M_n/n\to\infty$, \eqref{eq:approx-transfer-rate} already gives exact-MLE equivalence without needing (MC). In contrast, $M_n=cn$ with fixed $c>0$ generally retains $c^{-1}H^{-1}\Omega_*H^{-1}$ in the limiting covariance. A fixed $M_n$ does not satisfy the consistency conditions.

\paragraph{A gradient stopping rule after continuous refinement.}
If an output is already known to converge to the correct interior basin and all coordinates, including $m$, are continuously refined, let
$e_n=\|\nabla\widehat Q_n(\tilde\vartheta_n)\|$.
The same positive local curvature gives
$\|\tilde\vartheta_n-\bar\vartheta_n\|\leq Ce_n$ with probability tending to one. Thus $e_n=o_p(n^{-1/2})$ replaces the grid and value-gap requirements for first-order transfer. For the value-gap formulation, $r_n$ uses the \emph{average} negative log-likelihood; $r_n=o_p(n^{-1})$ is equivalently an $o_p(1)$ gap on the total negative log-likelihood scale.

\subsection{Fixed finite grids and fixed fuzziness}
Let $\mathcal G$ be fixed and finite, and retain the value part of (U), $M_n\to\infty$, and $r_n=o_p(1)$ relative to $\Xi_{\mathcal G}$. If $m_0\in\mathcal G$, compactness and identifiability give a positive population gap between the $m_0$ slice and the union of the other slices. The uniform criterion convergence therefore implies
\[
 \Pr(\tilde m_n=m_0)\longrightarrow1,\qquad
 d_{\rm lab}\{T(\tilde\vartheta_n),\theta_0\}\xrightarrow{p}0.
\]
On this event, write $\vartheta=(\eta,m_0)$ and repeat the continuous-minimizer argument in $\eta$. Under the derivative and (MC) conditions, $n/M_n\to\rho<\infty$, and $r_n=o_p(n^{-1})$,
\[
 \sqrt n(\tilde\eta_n-\eta_0)\Rightarrow
 \mathcal N\{0,H_\eta^{-1}(J_\eta+\rho\Omega_\eta)H_\eta^{-1}\},
\]
where $H_\eta=J_\eta=I_{\eta\eta}$ and $\Omega_\eta$ is the active-coordinate block of $\Omega_*$. The selected $m$ coordinate has a degenerate limit, not the joint continuous-$m$ normal law. Fixing the true $m$ in advance is the special case of a one-point grid.

If $m_0\notin\mathcal G$, the same uniform convergence instead gives convergence to the set of minimizers of $Q$ over $\Xi_{\mathcal G}$. Only if that restricted minimizer is unique modulo labels can a single restricted target be asserted. In general, neither $m$ nor the other coordinates need converge to their unrestricted true values. This result concerns selection by the approximate likelihood.

\subsection{Misspecified working densities and implementation scope}
The approximation argument also applies when observations have a distribution $P$ outside the WFCM family. Suppose $Q_P(\vartheta)=P\{-\log f_\vartheta(X)\}$ has a unique interior minimizer $\vartheta^\dagger$ modulo labels, $P\|X\|^4<\infty$, and $H_P=P\nabla^2\{-\log f_{\vartheta^\dagger}(X)\}>0$. The moment assumption and the established envelopes give the criterion and local Hessian uniform laws and a finite score variance. Let $J_P=\operatorname{Var}_P\{\nabla[-\log f_{\vartheta^\dagger}(X)]\}$, and impose (U) and (MC) at this target. Repeating the separated-argmin, local-curvature, and score arguments gives the same approximation rates and, if $n/M_n\to\rho<\infty$, $h_n=o(n^{-1/2})$, and $r_n=o_p(n^{-1})$,
\[
 \sqrt n(\tilde\vartheta_n-\vartheta^\dagger)
 \Rightarrow\mathcal N\{0,H_P^{-1}(J_P+\rho\Omega_\dagger)H_P^{-1}\}.
\]
Interiority makes the $P$-mean score zero; the importance score has mean zero under its proposal by normalization of $f$, independently of correct specification. Thus the two score limits combine as before. Here $H_P$ need not equal $J_P$, so the sandwich covariance replaces the inverse model Fisher information.

Finally, software may parameterize the kernel by coefficients $a_j>0$, $\sum_ja_j=1$, and scale $s$, using $s^{-2}\{\sum_ja_j\|x-v_j\|^{-2/t}\}^{-t}$. Its equivalence to the paper's parameterization requires the joint transformation
\[
 w_j=\frac{a_j^{-t}}{\sum_la_l^{-t}},\qquad
 \sigma=\frac{s}{(\sum_la_l^{-t})^{1/2}}.
\]
The inverse is $a_j=w_j^{-1/t}/A$, $s=\sigma A^{t/2}$, where $A=\sum_jw_j^{-1/t}$. This is smooth on the regular interior, so covariance must be transformed jointly, including the $m$ coordinate when estimated. The data and normalizer must use the same energy. Fixed distance clipping, penalization, or a different scale convention changes the criterion unless separately shown to be negligible.

\paragraph{Initialization and parameter bounds in the reported fits.}
In the real-data analyses (Section \ref{sec:real-data-analysis} and Appendix \ref{app:brca}) and the one-dimensional simulations of Appendix \ref{sec:sup_simulation} (Figures~\ref{fig:con}--\ref{fig:m-trainNLL} and Table~\ref{tab:revision-coverage}), $\bm{\bV}^{(0)}$ consists of $k$ distinct observations drawn at random with a fixed seed, with $\sigma^{(0)}=1$ and $w_j^{(0)} = 1/k$, and the same starting values are used at every $m$ in the grid. We use one start for the single-cell and ADNI analyses, four starts for TCGA-BRCA, and three starts in the one-dimensional simulations, keeping the fit with the smallest numerical NLL at each $m$. The simulation studies in Sections~\ref{sec:revision-benchmark}--\ref{sec:revision-computation} and~\ref{sec:revision-lrt}, which optimize the approximate likelihood directly, instead start from $k$-means++ centroids with $w_j^{(0)} = 1/k$ and $\sigma^{(0)}$ set from the within-cluster dispersion of that preliminary clustering. The resulting numerical fit can depend on initialization. None of the numerical fits imposes the center-separation constraint of~\eqref{eq:regular-parameter-set}. The direct-likelihood fits bound the scale between $10^{-3}$ and $10^{3}$ times the data standard deviation, the centers to an enlarged data range, and the weights below by $10^{-4}$; Algorithm~\ref{alg:emfuzzy} fits impose no bounds except in the ADNI analysis ($\sigma\ge0.05$, $w_j\ge0.01$).

\paragraph{Importance-sampling proposals in the applications.}
The real-data analyses (Section~\ref{sec:real-data-analysis} and Appendix~\ref{app:brca}) use the defensive importance-sampling proposal
\begin{equation}\label{eq:defensive-proposal}
q(\mathbf{x})
= 0.2\sum_{j=1}^{k}\pi_j\,\mathcal{N}(\mathbf{x}\mid\boldsymbol{\mu}_j,\boldsymbol{\Sigma}_j)
+ \frac{0.7}{k}\sum_{j=1}^{k}\mathcal{N}(\mathbf{x}\mid\boldsymbol{\mu}_j,\bar{s}^2\mathbf{I}_d)
+ 0.1\, t_5\!\left(\mathbf{x}\mid\bar{\mathbf{x}},\,4\bar{s}^2\mathbf{I}_d\right),
\end{equation}
where the first term is a $k$-component GMM fitted to the data by maximum likelihood, with weights $\pi_j$, means $\boldsymbol{\mu}_j$, and full covariances $\boldsymbol{\Sigma}_j$ (the single-cell analysis uses a three-component GMM for every candidate $k$, so that one cloud serves all candidates), $\bar{\mathbf{x}}$ is the sample mean,
$\bar{s}^2$ is the average per-coordinate sample variance, and $t_5(\cdot\mid\bar{\mathbf{x}},4\bar{s}^2\mathbf{I}_d)$
denotes the multivariate Student-$t$ density with five degrees of freedom, location $\bar{\mathbf{x}}$, and scale matrix $4\bar{s}^2\mathbf{I}_d$.
The isotropic components cover the isotropic tails of the WFCM density~\eqref{eq:WFCM-density} in directions where the fitted covariances are thin.
The heavy-tailed $t$ component has a fixed positive weight, similar in form to the construction in Lemma~\ref{lem:is-defensive}. Importance weights use the full mixture density $q$.
Following Theorem~\ref{thm:approx-transfer}, the proposal and the integration cloud are held fixed throughout optimization,
including across candidate values of $m$ (and of $k$ in the single-cell analysis), and are rebuilt for each bootstrap refit.
Draws are independent for the single-cell ($M = 20{,}000$) and ADNI ($M = 500{,}000$ for the observed fit; $M = 100{,}000$ for bootstrap refits) analyses. For TCGA-BRCA we use $M = 2^{15} = 32{,}768$ scrambled Sobol points mapped through the mixture; Theorem~\ref{thm:approx-transfer} is stated for independent draws. In each application, the NLL ranking across the $m$ grid is checked on an independent cloud of $100{,}000$, $2{,}000{,}000$, and $131{,}072$ points, respectively.
\endgroup

		\section{Case Study: TCGA-BRCA Breast Cancer Subtype Clustering} \label{app:brca}
		
		\subsection{Dataset and Experimental Setup}
		\label{app:brca:data}
		
		We evaluate $k$-means, FCM, and WFCM on the TCGA-BRCA (The Cancer Genome
		Atlas Breast Invasive Carcinoma) dataset~\citepS{tcga2012}, a widely used
		benchmark resource for breast cancer molecular subtyping. Our analysis cohort comprises 819 patients,
		with one primary breast tumor sample retained per patient.
		Reference subtype labels were obtained from 
		PAM50 subtype annotations accompanying the downloaded TCGA-BRCA
		data~\citepS{Parker2009}.
		After excluding samples with missing PAM50 annotations or
		Normal-like assignments, we retained four intrinsic molecular
		subtypes: Basal-like (Basal, $n = 139$), HER2-enriched (Her2,
		$n = 67$), Luminal~A (LumA, $n = 421$), and Luminal~B
		(LumB, $n = 192$). These subtypes differ substantially in prognosis, treatment
		response, and underlying biology. Basal-like tumors are predominantly
		triple-negative with poor prognosis; Her2-enriched tumors exhibit ERBB2
		gene amplification and are targeted by anti-HER2 therapies; and the two
		luminal subtypes share estrogen receptor positivity but differ in
		proliferative index and clinical outcome~\citepS{Sorlie2003}. The
		discrimination of LumA from LumB is a clinically significant but
		expression-difficult task, as the two subtypes occupy overlapping regions
		of gene expression space, making them a natural stress test for clustering
		methods.
		
		We use the gene-level expression values as distributed by the UCSC Xena GDC hub, which are on the $\log_2(x+1)$ scale; no further transformation is applied. After removing genes with non-finite values or near-zero variance, we select the 5,000 genes with the largest variance across the 819 tumors (highly variable genes, HVGs). We apply PCA to these genes after centering, without scaling each gene to unit variance, and retain the first five principal components. The PC scores are used without further standardization for all three clustering methods. This choice of $N_{\mathrm{PC}} = 5$ retains 35.0\% of the total variance of the selected genes.
		
		A sensitivity analysis over $N_{\mathrm{PC}} \in \{3, 5, 10, 20\}$ is reported in Appendix~\ref{app:brca_sensitivity}. All three methods are configured with $K = 4$ clusters; For the five-PC analysis,  $m=1.50$ for FCM, and selected $m=2.008$ for WFCM based on the refined negative log-likelihood, each chosen from eight geometrically spaced values between 1.5 and 2.5. For FCM, $m$ is chosen by profiling the equal-weight working likelihood over the scale at the FCM centers; the selected value is the lower end of the grid. Clustering quality is assessed by Adjusted Rand Index (ARI), Normalized Mutual Information (NMI), per-subtype recall, and for the soft methods, mean maximum membership and normalized membership entropy.
		
		\subsection{Expression Space Overlap and Soft Assignment Uncertainty}
		\label{app:brca:uncertainty}
		
\begin{figure}[p]
			\includegraphics[width=\linewidth]{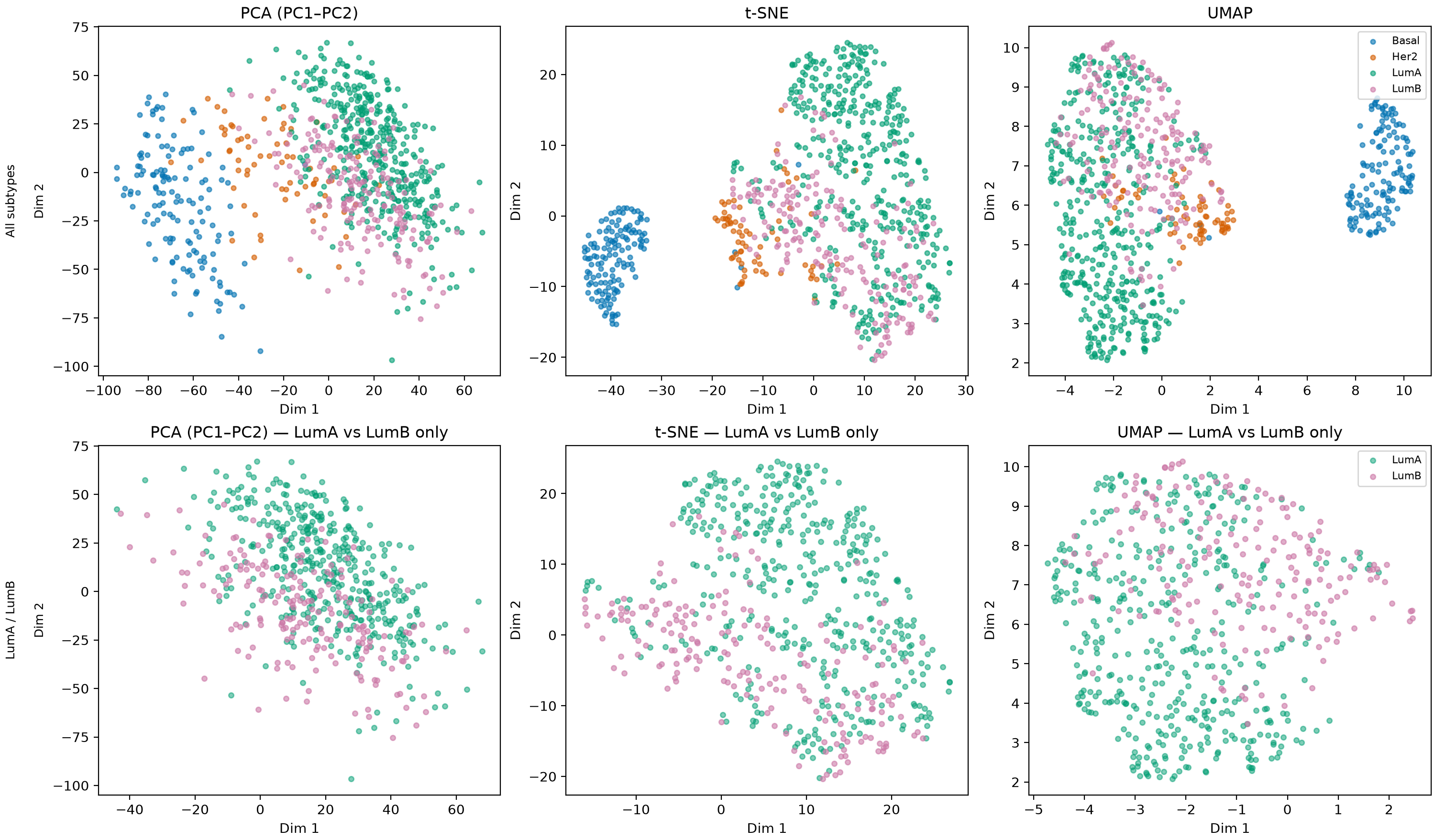}
			\caption{PAM50 breast cancer subtypes visualized using PCA, t-SNE and UMAP. The LumA and LumB subtypes exhibit substantial overlap in all three representations, illustrating the difficulty of separating these groups using low-dimensional visualizations.}
			\label{fig:embedding_overlap}
		\end{figure}

		Before comparing methods, we characterize the intrinsic geometry of
		the classification problem. Figure~\ref{fig:embedding_overlap} displays t-SNE and UMAP
		projections~\citepS{vanDerMaaten2008,mcinnes2018umap} of all 819 samples colored by PAM50
		subtype. While Basal-like tumors occupy a well-separated region of
		expression space, the Her2, LumA, and LumB subtypes form a continuous,
		partially overlapping manifold with no sharp natural boundary.
		
\begin{figure}[p]
			\includegraphics[width=\linewidth]{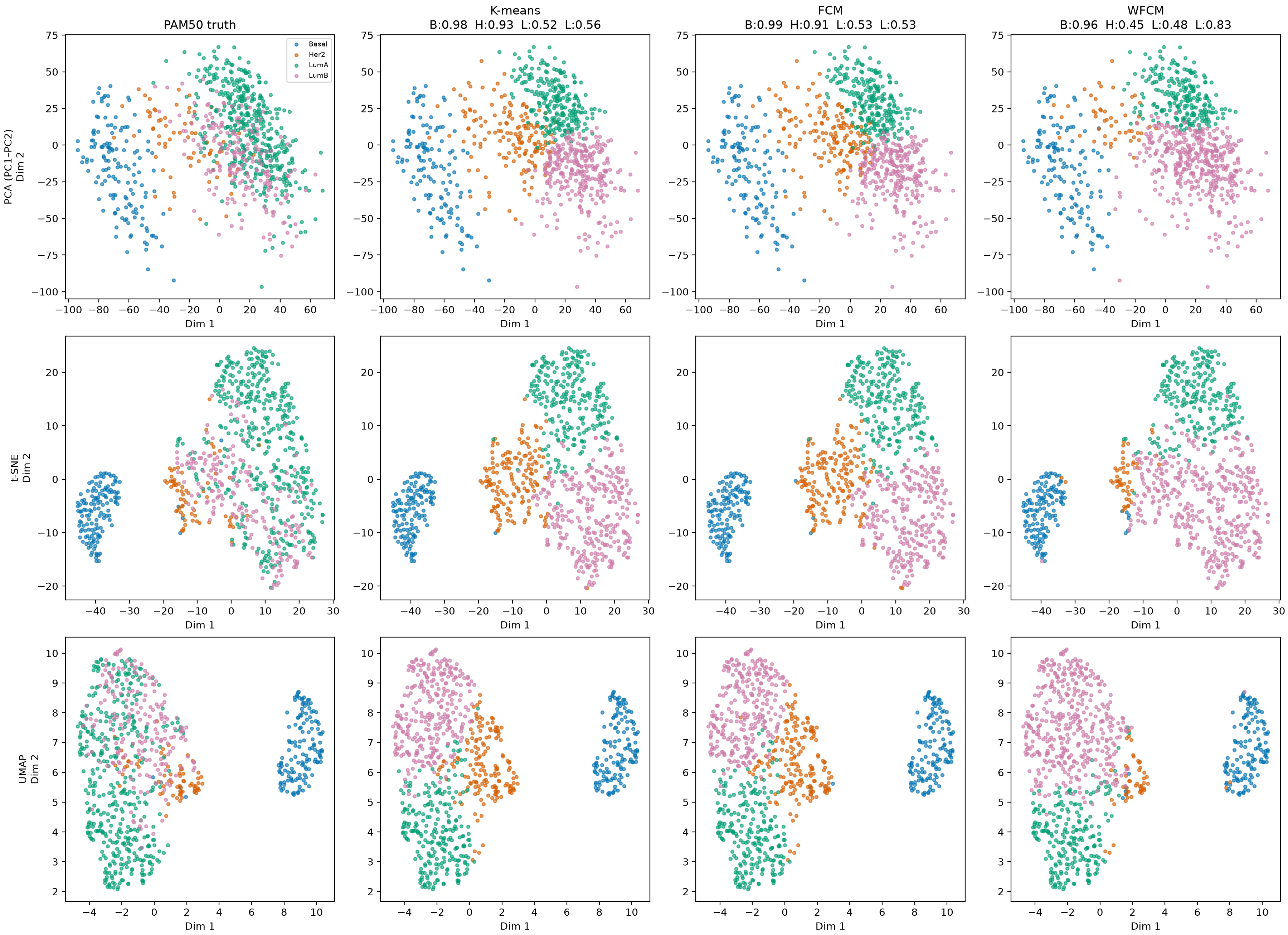}
			\caption{Hard cluster assignments in alternative embeddings. WFCM has higher LumB recall (0.83 vs 0.56 for $k$-means) but lower HER2-enriched recall.}
			\label{fig:cluster_assignments}
		\end{figure}

		Figure~\ref{fig:cluster_assignments} shows hard cluster assignments across PCA (PC1--PC2, top row), t-SNE (middle row) and UMAP (bottom row) embeddings. The leftmost column reproduces the PAM50 ground-truth labels as a reference; the remaining three columns show the hard cluster assignments of $k$-means, FCM, and WFCM, with per-subtype recall values reported in each column header. Basal-like tumors (blue) are consistently well-separated and correctly recovered by all three methods. The luminal region, by contrast, forms a continuous, partially overlapping manifold in all three embeddings, making LumA--LumB discrimination the central challenge. $k$-means and FCM both assign a large proportion of true LumB patients (pink) to the LumA cluster (green) and Her2 cluster (orange), yielding LumB recalls of $0.56$ and $0.53$ respectively. WFCM identifies a substantially larger proportion of LumB samples, achieving a LumB recall of $0.83$ while maintaining comparable Basal and LumA recall, but with much lower HER2-enriched recall ($0.45$ versus $0.91$--$0.93$).
		Figures~\ref{fig:cluster_assignments} and~\ref{fig:lumB_membership} provide complementary PCA, t-SNE, and UMAP visualizations of the fitted cluster assignments and soft memberships. The consistent patterns across three representations further illustrate WFCM's improved identification of LumB samples relative to conventional FCM and $k$-means.
		
\begin{figure}[p]
			\includegraphics[width=\linewidth]{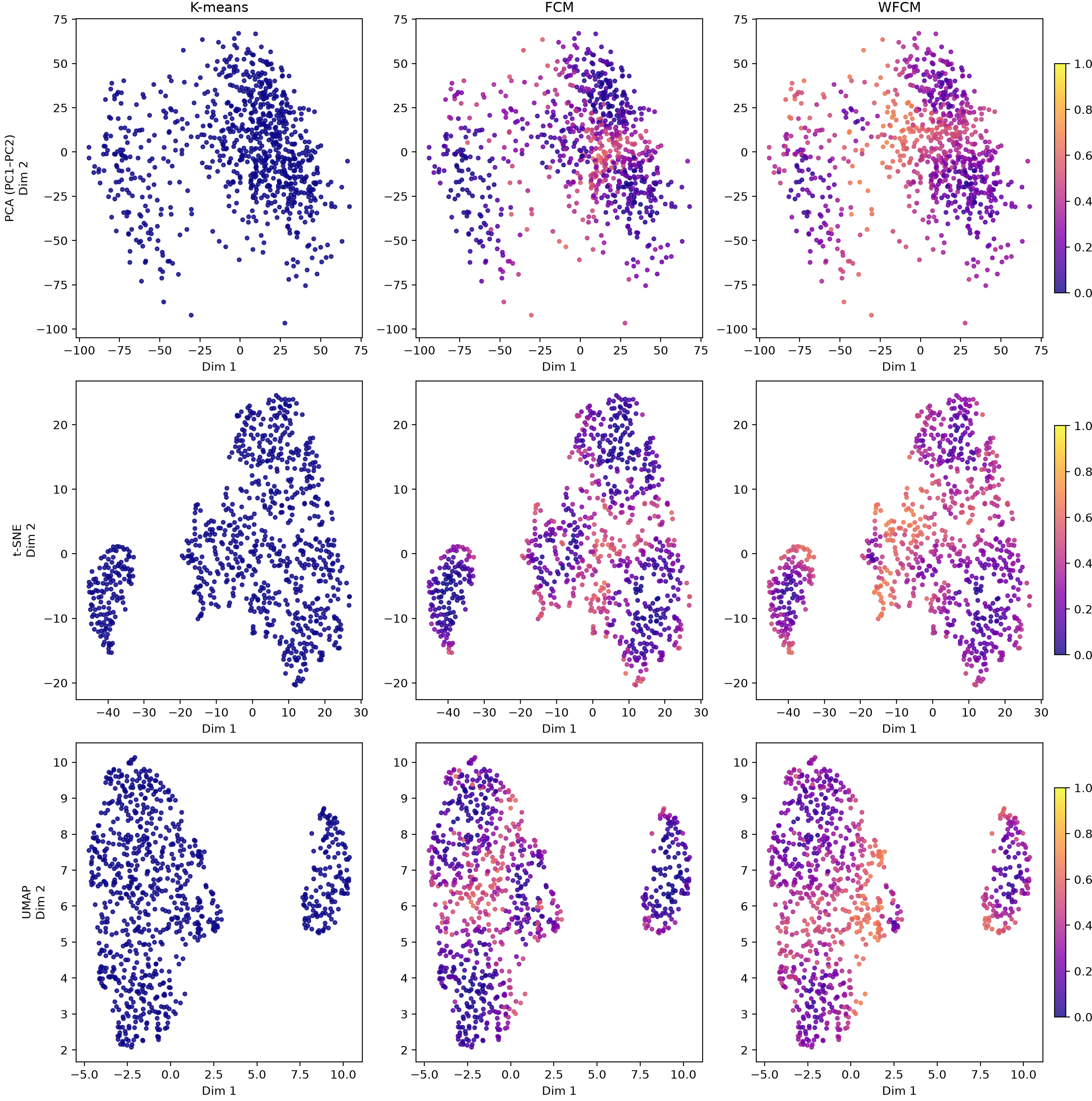}
			\caption{Membership uncertainty (1 – max membership) in each embedding. Both soft methods place uncertainty mainly at the LumA/LumB interface; $k$-means cannot express uncertainty.}
			\label{fig:uncertainty_embedding}
		\end{figure}

		Figure~\ref{fig:uncertainty_embedding} compares per-sample uncertainty across all three methods. It is defined as $1 - \max_k\, u_{ik}$, where $u_{ik}$ is the membership of sample $i$ in cluster $k$. Because $k$-means produces hard
		assignments ($u_{ik} \in \{0,1\}$), uncertainty is identically zero for
		all samples; the corresponding panel appears uniformly dark under the
		plasma colormap. FCM and WFCM both reveal non-trivial uncertainty
		concentrated in the LumA--LumB boundary region of t-SNE and UMAP space.
		The two soft methods place uncertainty in similar regions; they differ
		in how membership is distributed there, as quantified in
		Section~\ref{app:brca:membership}.
		
		\subsection{Luminal Subtype Discrimination: Membership Concentration and Performance Metrics}
		\label{app:brca:membership}
		
\begin{figure}[p]
			\includegraphics[width=\linewidth]{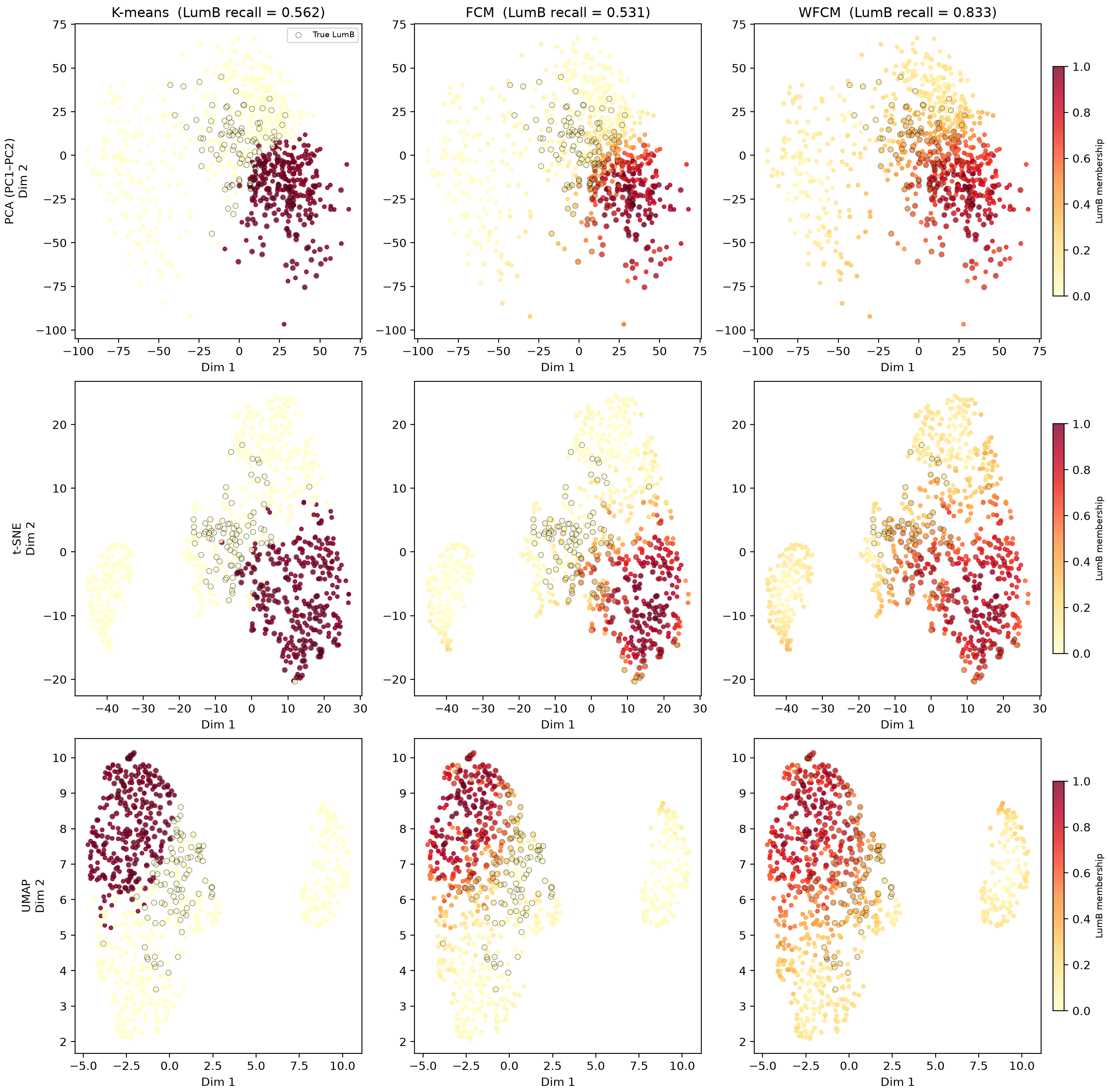}
			\caption{LumB soft membership across embeddings. WFCM assigns high LumB membership to true LumB patients that $k$-means and FCM misclassify as LumA.}
			\label{fig:lumB_membership}
		\end{figure}

		Figure~\ref{fig:lumB_membership} compares LumB membership across all three methods in PCA (PC1--PC2, top row), t-SNE (middle row), and UMAP (bottom row) embeddings. Each point is colored by its LumB
		membership value $u_{i,\text{LumB}} \in [0,1]$ (dark red $= 1.0$, yellow $= 0.0$); open circles identify ground-truth LumB patients. In the $k$-means panels, LumB membership is binary by construction: a substantial fraction of true LumB patients (open circles) located in the luminal overlap region receive zero LumB membership, appearing yellow despite being genuine positives. FCM assigns graded membership values throughout both embeddings but produces diffuse, low-to-moderate $u_{i,\text{LumB}}$ for true LumB patients in the LumA--LumB boundary zone, indicating that soft assignment alone is insufficient to resolve the ambiguity. WFCM, by contrast, assigns substantially higher LumB membership (darker red) to the true LumB patients that $k$-means and FCM misclassify or under-score, while correctly maintaining low membership for true LumA patients in the same spatial region. Similar patterns in the PCA and t-SNE visualizations further illustrate the differences in LumB membership assignments among the three methods.
		
		\begin{figure}[p]
			\includegraphics[width=\linewidth]{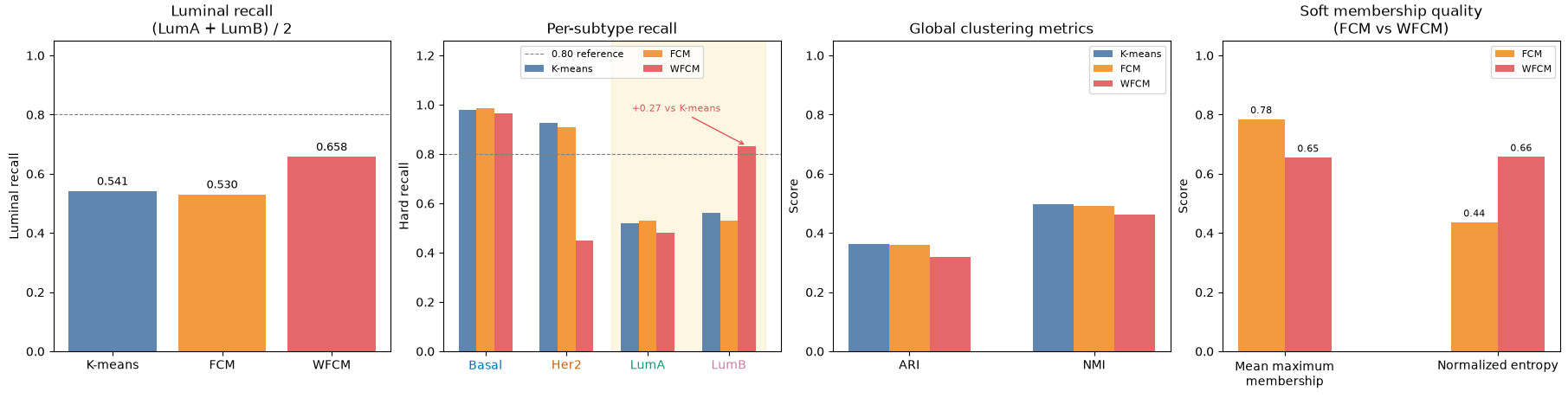}
			\caption{Per-subtype, luminal, global, and soft-membership metrics. WFCM has the highest luminal recall but the lowest ARI, NMI, and HER2-enriched recall; its soft memberships are less concentrated than FCM's (lower mean maximum membership, higher normalized entropy).}
			\label{fig:metrics}
		\end{figure}
		
		Figure~\ref{fig:metrics} summarizes the three methods across four
		complementary metrics. \textbf{Panel 1} reports luminal recall, defined
		as the mean of LumA and LumB per-subtype recall:
		\begin{equation}
			\text{Luminal Recall} = \frac{\text{Recall}_{\text{LumA}} +
				\text{Recall}_{\text{LumB}}}{2}.
		\end{equation}
		WFCM achieves a luminal recall of $0.658$, compared with $0.541$ for
		$k$-means and $0.530$ for FCM. This is a gain of $11.7$ percentage points over $k$-means and $12.8$ percentage points over FCM.
		
		\textbf{Panel 2} presents per-subtype recall across all four
		subtypes (Table~\ref{tab:brca} provides the corresponding
		numerical values). WFCM achieves substantially higher LumB
		recall than conventional FCM ($0.83$ vs.\ $0.53$),
		highlighting its ability to better identify this challenging
		luminal subtype. This improvement is accompanied by lower
		HER2-enriched recall ($0.45$ vs.\ $0.91$), illustrating
		a trade-off in subtype-specific recovery. These findings
		suggest that cluster-specific weighting can alter the balance
		of clustering performance across overlapping molecular subtypes, with a particularly pronounced benefit for LumB identification in this application.
		
		\textbf{Panel 3} summarizes overall agreement with the PAM50 labels.
		$k$-means and FCM achieved similar ARI values (both $0.361$), with NMI
		values of $0.496$ and $0.491$, respectively. WFCM yielded an ARI of
		$0.320$ and an NMI of $0.461$. The per-subtype results show a different
		pattern for LumB: WFCM recovered $83.3\%$ of LumB tumors, compared with
		$56.2\%$ for $k$-means and $53.1\%$ for FCM.
		
		\textbf{Panel 4} compares the concentration of the fitted memberships.
		WFCM had a lower mean maximum membership ($0.65$ versus $0.78$ for FCM)
		and higher mean normalized entropy ($0.66$ versus $0.44$). Both measures indicate that,
		on average, WFCM distributed membership more broadly across the four clusters.
		
		Compared with FCM and \(k\)-means, WFCM raises LumB recall at the cost of lower HER2-enriched recall and lower global agreement (ARI and NMI).
		
		\begin{table}[h]
			\centering
			\caption{Clustering performance on TCGA-BRCA ($n=819$, $K=4$).
				Per-subtype recall is the fraction of samples of a given PAM50 subtype
				assigned to its corresponding cluster after one-to-one matching.
				Luminal recall is the mean of LumA and LumB recalls.}
			\label{tab:brca}
			\small
			\begin{tabular}{lcccccccc}
				\toprule
				\textbf{Method} & \textbf{ARI} & \textbf{NMI}
				& \textbf{Basal} & \textbf{Her2} & \textbf{LumA} & \textbf{LumB}
				& \textbf{Luminal} & \textbf{Mean Max} \\
				& & & \textbf{Recall} & \textbf{Recall} & \textbf{Recall}
				& \textbf{Recall} & \textbf{Recall} & \textbf{Membership} \\
				\midrule
				$k$-means & 0.361 & 0.496 & 0.978 & 0.925 & 0.520 & 0.562 & 0.541 & $1.000^{*}$ \\
				FCM       & 0.361 & 0.491 & 0.986 & 0.910 & 0.530 & 0.531 & 0.530 & 0.784 \\
				WFCM      & 0.320 & 0.461 & 0.964 & 0.448 & 0.482 & 0.833 & 0.658 & 0.655 \\
				\bottomrule
			\end{tabular}
			\par\vspace{2pt}
			{\footnotesize $^{*}$For $k$-means, each sample has one hard assignment,
				so its maximum membership is $1$ by construction.}
		\end{table}
		
		\subsection{Sensitivity to the Number of Retained Principal Components}
		\label{app:brca_sensitivity}
		
		Table~\ref{tab:brca_pca_sensitivity} compares clustering performance
		across $N_{\mathrm{PC}}\in\{3,5,10,20\}$ using the same cohort, gene set,
		and fitting procedures. The number of clusters was held fixed, and
		the fuzzifier was reselected at each dimension for FCM and WFCM.
		LumB recall varied less for $k$-means ($0.547$--$0.562$) than for FCM
		($0.464$--$0.557$). WFCM achieved the highest LumB recall at each
		dimension, with values of $0.781$, $0.833$, $0.766$, and $0.677$ at
		three, five, ten, and twenty PCs, respectively. Its LumB recall peaked
		at five PCs and declined as more components were retained.
		
		\begin{table}[ht]
			\centering
			\caption{Sensitivity to the number of retained PCs in TCGA-BRCA.
				Luminal recall is the mean of LumA and LumB hard recalls.}
			\label{tab:brca_pca_sensitivity}
			\small
			\resizebox{\textwidth}{!}{%
				\begin{tabular}{cc*{9}{r}}
					\toprule
					$N_{\mathrm{PC}}$ & Var.\ expl.\ (\%) &
					\multicolumn{3}{c}{LumB recall} &
					\multicolumn{3}{c}{Luminal recall} &
					\multicolumn{3}{c}{ARI} \\
					\cmidrule(lr){3-5}\cmidrule(lr){6-8}\cmidrule(lr){9-11}
					& & $k$-means & FCM & WFCM
					& $k$-means & FCM & WFCM
					& $k$-means & FCM & WFCM \\
					\midrule
					3 & 28.9 & 0.557 & 0.536 & 0.781
					& 0.543 & 0.531 & 0.650
					& 0.340 & 0.332 & 0.240 \\
					5 & 35.0 & 0.562 & 0.531 & \textbf{0.833}
					& 0.541 & 0.530 & \textbf{0.658}
					& 0.361 & 0.361 & 0.320 \\
					10 & 44.1 & 0.547 & 0.464 & 0.766
					& 0.522 & 0.484 & 0.637
					& 0.356 & 0.341 & 0.340 \\
					20 & 52.5 & 0.547 & 0.557 & 0.677
					& 0.523 & 0.548 & 0.625
					& 0.354 & 0.297 & 0.294 \\
					\bottomrule
				\end{tabular}%
			}
		\end{table}
		
	\end{sloppypar}
	
	\bibliographystyleS{apacite}
	\bibliographyS{ref}
\end{document}